\documentclass[a4paper,11pt]{article}
\pdfoutput=1 % if your are submitting a pdflatex (i.e. if you have
\usepackage{jheppub} % for details on the use of the package, please
\usepackage[T1]{fontenc} % if needed
\usepackage{graphicx}
\usepackage[makeroom]{cancel}
\usepackage{makecell}
\usepackage{bm}
\usepackage{amssymb}
\usepackage{amsmath}
\usepackage{epsfig}
\usepackage{hyperref}
\usepackage{hhline}
\usepackage{multirow}
\usepackage[normalem]{ulem}
\usepackage{adjustbox}
\usepackage{subfig}
\usepackage{tikz}
\usepackage{epstopdf}
\usetikzlibrary{snakes}
\usepackage{slashed}
\usepackage{float}
\usepackage{url}
\usepackage{makecell}

\allowdisplaybreaks

\usepackage{empheq}
\usepackage[most]{tcolorbox}
\newtcbox{\mymath}[1][]{%
    nobeforeafter, math upper, tcbox raise base,
    enhanced, colframe=blue!30!black,
    colback=blue!30, boxrule=1pt,
    #1}

\usepackage[section]{placeins}

 \definecolor{jd}{rgb}{0.858, 0.188, 0.478}

\def\lapp{\mathrel{\rlap{\raise.5ex\hbox{$<$}}
                    {\lower.5ex\hbox{$\sim$}}}}
\def\gapp{\mathrel{\rlap{\raise.5ex\hbox{$>$}}
                    {\lower.5ex\hbox{$\sim$}}}}

\def\sb#1{{\bf \boldmath \textcolor{cyan}{{[MI:~#1]}}}}

{\newcommand{\lsim}{\mbox{\raisebox{-.6ex}{~$\stackrel{<}{\sim}$~}}}
{\newcommand{\gsim}{\mbox{\raisebox{-.6ex}{~$\stackrel{>}{\sim}$~}}}
\def\sb{\color{blue}}
\def\lam{\lambda}
\def\lSH{\lam_{SH}}
\def\lHP{\lam_{H\Phi}}
\def\lSP{\lam_{S\Phi}}
\def\lH{\lam_H}
\def\lS{\lam_S}
\def\lP{\lam_\Phi}
\def\lSM{\lam_{SM}}
\def\vh{v_h}
\def\vs{v_S}
\def\vP{v_\Phi}
\def\mP{m_\Phi}
\def\mx{m_X}
\def\gx{g_X}
\def\trh{T_\text{RH}}
\def\tew{T_\text{EW}}
\def\u1{U(1)_X}
\def\sm{s_\text{min}}

\def\tm{\theta_\text{min}}
\def\trh{T_\text{RH}}
\def\mpl{M_\text{Pl}}

\def\gev{\;\text{GeV}}
\def\tev{\;\text{TeV}}
\def\zBB{{\mathbbm Z}}
\def\z2{\zBB_2}
\newcommand{\nc}{\newcommand}
\nc{\nn}{\nonumber}
\nc{\beq}{\begin{equation}}  \nc{\eeq}{\end{equation}}
\nc{\bea}{\begin{eqnarray}}  \nc{\eea}{\end{eqnarray}}
\nc{\baa}{\begin{array}}     \nc{\eaa}{\end{array}}
\nc{\bit}{\begin{itemize}}   \nc{\eit}{\end{itemize}}
\nc{\ben}{\begin{enumerate}} \nc{\een}{\end{enumerate}}
\nc{\bce}{\begin{center}}    \nc{\ece}{\end{center}}
\nc{\bpm}{\begin{pmatrix}}   \nc{\epm}{\end{pmatrix}}
\nc{\bvt}{\begin{verbatim}}  \nc{\evt}{\end{verbatim}}
\newcommand{\mathsym}[1]{{}}

\nc{\hsp}{\hspace{0.4cm}}
\nc{\lsp}{\hspace{0.8cm}}
\nc{\Lsp}{\hspace{1.6cm}}
\nc{\what}{\widehat}
\nc{\LLsp}{\lsp\lsp}
\nc{\lra}{\longrightarrow}
\textwidth 15.55cm \textheight 22.5cm
\hoffset -1.6cm
\voffset -1cm

\title{Feebly coupled vector boson dark matter in effective theory}
\author[a]{Basabendu Barman,} 
\emailAdd{bb1988@iitg.ac.in}
\author[a]{Subhaditya Bhattacharya,}
\emailAdd{subhab@iitg.ac.in}
\author[b]{Bohdan Grzadkowski}
\emailAdd{bohdan.grzadkowski@fuw.edu.pl}
  \affiliation[a]{Department of Physics, Indian Institute of Technology Guwahati,\\ North Guwahati, Assam-781039, India.}
  \affiliation[b]{Faculty of Physics, University of Warsaw, Pasteura 5, 02-093 Warsaw, Poland}
\abstract{
A model of dark matter (DM) that communicates with the Standard Model (SM) 
exclusively through suppressed dimension five operator is discussed. 
The SM is augmented with a symmetry $U(1)_X \otimes Z_2$, where $U(1)_X$ is gauged and 
broken spontaneously by a very heavy decoupled scalar. 
The massive $U(1)_X$ vector boson ($X^\mu$) is stabilized being odd under 
unbroken $Z_2$ and therefore may contribute as the DM component of the universe. 
Dark sector field strength tensor $X^{\mu\nu}$ couples to the SM 
hypercharge tensor $B^{\mu\nu}$ via the presence of a heavier $Z_2$ odd real scalar $\Phi$, i.e. 
$1/\Lambda \; X^{\mu\nu}B_{\mu\nu}\Phi$, with $\Lambda$ being a scale of new physics.
The freeze-in production of the vector boson dark matter feebly coupled to the SM is
advocated in this analysis. Limitations of the so-called UV freeze-in mechanism that emerge when 
the maximum reheat temperature $T_\text{RH}$ drops down close to the scale of DM mass are discussed. 
The parameter space of the model consistent with the observed DM abundance is determined.
The model easily and naturally avoids both direct and indirect DM searches.
Possibility for detection at the Large Hadron Collider (LHC) is also considered.
A Stueckelberg formulation of the model is derived. 
}

\begin{document}

\maketitle
\flushbottom

\setcounter{footnote}{0}
\renewcommand*{\thefootnote}{\arabic{footnote}}
%%%%%%%%%%%%%%%%%%%%%%%%%%%%%%%%%%%%%%%%%%%%%%%%%%%%%%%%%%
%%%%%%%%%%%%%%%%%%%%%%%%%%%%%%%%%%%%%%%%%%%
\section{Introduction}
\label{sec:intro}
%%%%%%%%%%%%%%%%%%%%%%%%%%%%%%%%%%%%%%%%%%%

The existence of Dark Matter (DM) is motivated from different astrophysical observations like galaxy rotation curves~\cite{Zwicky:1933gu,Zwicky:1937zza,Rubin:1970zza}, bullet cluster~\cite{Clowe:2006eq}, 
gravitational lensing~\cite{Massey:2010hh}, and cosmological observations like anisotropies in Cosmic Microwave Background (CMB)~\cite{Hu:2001bc}(for a review, see, 
for example~\cite{Bertone:2004pz,Feng:2010gw}). However, we still do not know what DM actually is. 
DM as a fundamental particle has to be electromagnetic charge neutral and stable at the scale of universe life time. From satellite experiments like WMAP and 
PLANCK~\cite{Spergel:2006hy,Jarosik:2010iu,Hinshaw:2012aka,Ade:2013zuv,Aghanim:2018eyx}, that measure anisotropies in CMB, we learn that DM constitutes almost $85\%$ of the total matter content and $26.4\%$ of the total energy budget of the universe, often expressed in terms of relic density, which provides an important constraint to abide by. 
Since no Standard Model (SM) particle resembles the properties of a DM particle, many possibilities beyond the SM (BSM) have been formulated 
to explain the particle nature of the DM, as scalar, fermion or a vector boson stabilized by an additional symmetry $\mathcal{G}_{DM}$. 

Amongst different possibilities, the most popular one assumes DM to be present in thermal bath in early universe due to non-negligible coupling with the SM, which eventually 
freezes out to provide correct thermal relic as universe expands and cools down. Weekly Interacting Massive Particles (WIMP) belongs to such thermal relic category
and is widely studied due to its phenomenological richness~\cite{Kolb:1990vq,Jungman:1995df,Baer:2014eja,Arcadi:2017kky}. However, it is also viable to assume that DM is very weakly coupled to visible sector 
and therefore does not equilibrate to hot soup of SM particles in the early universe and gets produced via decay or annihilation of particles already in 
equilibrium. Such non-thermal DM production halts after the temperature of the bath drops smaller than DM mass and the yield freezes in to 
provide correct relic density, see for example, \cite{Hall:2009bx}. DM particles which freezes in are often called feebly interacting massive particle (FIMP) and easily evades the bounds from non-observation 
of DM in direct or collider searches. Such a DM is mainly studied in the analysis presented here.

Vector boson DM (VDM) candidate can only appear in models with extended gauge group, the simplest being an Abelian $U(1)$. 
Many possibilities of an Abelian VDM have been studied~\cite{Farzan:2012hh,Baek:2012se,Bian:2013wna,Choi:2013qra,Baek:2013dwa,Baek:2013fsa,Ko:2014gha,Baek:2014poa,Ko:2014loa,Duch:2015jta,Beniwal:2015sdl,Kamon:2017yfx,Duch:2017nbe,Arcadi:2017jqd,Baek:2018aru,YaserAyazi:2019caf,Duch:2017khv,Choi:2020kch,Choi:2020dec}, while non-Abelian extensions to adopt VDM are fewer~\cite{DIAZCRUZ2011264,DiazCruz:2010dc,Bhattacharya:2011tr,Farzan:2012kk,Barman:2017yzr,Fraser:2014yga,Barman:2018esi,Barman:2019lvm,Abe:2020mph}. 
The VDM can become massive after spontaneous symmetry breaking of the additional gauge group and often requires additional stabilizing symmetry $\mathcal{G}_{DM}$~\cite{DIAZCRUZ2011264,Baek:2012se}. 
The advantage of the non-Abelian realization of this scenario is that, in this case, 
there is no need to impose an extra symmetry by hand that provides stability of vector DM~\footnote{This statement is valid if no extra degrees of freedom charged under the dark gauge symmetry is present.}.
The main parameters that characterize VDM are DM mass and the portal quartic coupling that connects dark and visible sectors.
Therefore, the portal coupling crucially distinguishes the possibility of (i) DM freeze-out when the coupling is moderately weak~\cite{Jungman:1995df,Profumo:2017hqp,Arcadi:2017kky,Roszkowski:2017nbc} and (ii) freeze-in~\cite{Chu:2011be,Elahi:2014fsa,Co:2015pka,Bernal:2017kxu,Duch:2017khv,Heeba:2018wtf,Zakeri:2018hhe,Becker:2018rve,Biswas:2018aib,Hambye:2018dpi,Lebedev:2019ton,Chang:2019xva,Bernal:2019mhf,Chakraborti:2019ohe,Biswas:2019iqm} when the coupling is very tiny. Freeze-in possibilities have also been studied in the context of non-Abelian cases, for example in \cite{Barman:2019lvm}. Our goal in this paper is to realize the presence of a VDM coupled to the SM via effective theory.

Effective DM-SM operators  provide a model-independent framework to probe DM characteristics like relic density, direct search and collider search prospects. 
Such operators are usually written as $\mathcal{O}=\mathcal{O}_{SM}\mathcal{O}_{DM}$, where $\mathcal{O}_{SM}$ consists of SM fields and $\mathcal{O}_{DM}$ 
consists of additional DM fields (scalar, fermion or vector boson). The Lagrangian is assumed to be invariant under $\mathcal{G}_{SM} \times \mathcal{G}_{DM}$, 
where SM fields in $\mathcal{O}_{SM}$ transform only under SM gauge symmetry 
($\mathcal{G}_{SM}=SU(3)_c \times SU(2)_L \times U(1)_Y$) and neutral under $\mathcal{G}_{DM}$, 
while DM fields transform only under dark symmetry ($\mathcal{G}_{DM}$, often assumed to be $Z_2$) and are singlets under $\mathcal{G}_{SM}$. 
A heavy mediator is assumed to couple to both dark and visible sector weakly and the operators are expected to vanish when the 
mass of the heavy mediator goes to infinity following decoupling theorem.  A complete set of such operators have been written upto dimension six assuming $\mathcal{G}_{SM}$ to be SM 
gauge group~\cite{Duch:2014xda,Macias:2015cna} as well as  assuming $ \mathcal{G}_{SM}\sim U(1)_{EM}$ after spontaneous electroweak symmetry breaking \cite{Goodman:2010ku} keeping dark symmetry intact. 
Detailed phenomenological analysis assuming the DM to freeze-out have been carried out including the collider search prospects at Large Hadron Collider (LHC)~\cite{Goodman:2010ku,Fox:2011pm,Dreiner:2013vla,Busoni:2013lha,Busoni:2014sya,Abercrombie:2015wmb,Kumar:2015wya,Belyaev:2018pqr,10.3389/fphy.2019.00075}.   
   
Here, we elaborate a model where the dark sector is coupled to visible sector {\it only} via effective dimension five operator. We choose the 
simplest extension of the SM by Abelian $U(1)_X$ gauge group. The $U(1)_X$ vector boson is electromagnetic charge neutral and must be stable for becoming DM. The 
stability is guaranteed by imposing an additional $Z_2$ symmetry under which the dark vector boson is odd, then the kinetic mixing $X^{\mu\nu}B_{\mu\nu}$ is forbidden. 
However a direct connection between DM and the visible sector (SM) still could be introduced if an extra real scalar ($\Phi$) odd under the stabilizing symmetry is present.
Then an operator of mass dimension five, $ X^{\mu\nu}B_{\mu\nu} \Phi/\Lambda $, is allowed. 
For dimensional reasons the interaction must be suppressed by an unknown new physics (NP) scale $\Lambda$.
This operator has been listed in~\cite{Macias:2015cna} and a WIMP phenomenology has recently been 
performed  in~\cite{Fortuna:2020wwx}. It is worthy to mention here, even without $X^{\mu\nu}B_{\mu\nu}\Phi$ term, 
dark sector can couple to the SM, via the mixing of scalar boson (call it $S$) that breaks $U(1)_X$ 
and the Higgs doublet ($H$) via a portal term $|S|^2 |H|^2$ \cite{Duch:2017khv}. 
Here however, we will assume that the scalar $S$ is super heavy and decouples. 
In addition a quartic portal interaction of the scalar $\Phi$, $\Phi^2 |H|^2$ is also allowed by the symmetry.  
The coupling is relevant for $\Phi$ being in thermal equilibrium with the SM, however fails to produce vector boson DM without the dimension five operator.
It is important to note that in absence of the dimension five term, $\Phi$ becomes a stable DM candidate together with $X$, 
while the latter is completely decoupled from the SM in the limit of heavy $S$. 
With the presence of the higher dimension interaction term, $X$ becomes stable DM, given $m_\Phi>m_X$, as we assume here. 
We will show in sec.~\ref{sec:large-trh}, that even large portal coupling of $\Phi^2 |H|^2$ fails to contribute significantly to 
DM ($X$) production, compared to the $\Phi$ decay after Electroweak symmetry breaking (EWSB).
For the consistency of Effective Field Theory (EFT), the NP scale also requires to be larger than the maximum reheat temperature $\Lambda >T_\text{RH}$. Together, 
it is more appealing to assume that the VDM is feebly connected to the SM and it freezes-in. The paper analyzes such possibility in details. 
We also demonstrate the limitation of {\it UV} freeze-in which is advocated in context of effective operators \cite{Elahi:2014fsa}. We show when the reheat temperature 
comes closer to the DM mass scale ($m$) involved in production process with $T_\text{RH} \gtrsim m$, 
massive kinematics plays an important role and IR aspects are becoming relevant.

It is worth noticing that owing to feeble DM-SM interaction to account for correct relic density in FIMP like models, 
the possibility of detecting such DM candidates 
at direct or collider searches is limited. However, if one has an extended dark sector, like we have $\Phi$ having same $Z_2$ symmetry as of VDM ($X^\mu$), 
there can still be a possibility. We comment on seeing mono-X (where X stands 
for jet, photon, $W,Z$ or $H$) plus missing energy signature in this framework at the upcoming run of Large Hadron Collider (LHC). 

Finally, let's comment on the so called small scale cosmological problems. 
Even though comparison of the standard cosmological model, i.e. the $\Lambda$CDM model, with observations is very successful on scales 
larger than galaxies, the model has some difficulties at sub-galaxy scales, predicting, via computer simulations, too many dwarf galaxies 
("missing-satellites problem") and/or too much dark matter (``core-cusp problem'') in central regions of galaxies. 
Among possible solutions of this ``small scale crisis" are e.g. models of strongly interacting DM, for VDM see e.g. \cite{Duch:2017khv}. 
However there exist also a simpler explanation of the crises. Namely, there has been an extensive investigation recently 
of the possibility that a realistic treatment of baryonic physics in simulations, 
such as supernovae feedback, stellar winds, etc. can eliminate the tension (see \cite{Stafford:2020ppr} and references therein). 
Therefore in this work the issue of small scale problems has not been addressed.

The paper is arranged as follows. Sec.~\ref{sec:intro} contains an introduction to VDM models. In Sec.~\ref{Model} the model considered here is described and its Stueckelberg formulation specified.  
Sec.~\ref{freeze-in} discusses properties of the Boltzmann equation relevant for the DM production.
Sec.~\ref{sec:relic-dm} contains our findings for the DM abundance via the freeze-in and shows regions of the parameter space consistent with the observed DM abundance. 
In Sec.~\ref{sec:collider} we comment on experimental constraints and collider signatures of the model. 
Sec.~\ref{sec:summ} shows summary and conclusions. In Appendices~\ref{parameters_relations}-\ref{sec:app-amp-bewsb} we collect useful formulae.

%%%%%%%%%%%%%%%%%%%%%%%%%%%%%%%%%%%%%%%%%%%%%%%%%%%%%%%
\section{The Model}
\label{Model}
%%%%%%%%%%%%%%%%%%%%%%%%%%%%%%%%%%%%%%%%%%%%%%%%%%%%%%%
The minimal VDM model contains a $U(1)_X$ gauge boson denoted here by $X_\mu$. In order to enable direct interactions between $X_\mu$ and the SM one also requires presence of a real scalar $\Phi$. 
Both of them should be odd under a $Z_2$ which stabilises DM candidate, i.e. the vector boson ($m_X < m_\Phi$). Therefore the symmetry group of the model is ${\cal G} = SU(3)_C \times SU(2)_L \times U(1)_Y \times U(1)_X \times Z_2$. In order to generate a mass for the dark gauge boson we also introduce a complex scalar $S$ charged under $U(1)_X$, which acquires a vacuum expectation value to break $U(1)_X$ spontaneously. The $Z_2$ transformation acts on these fields as follows:
\beq
Z_2: ~~X_\mu \to -X_\mu, \lsp S \to S^\star, \lsp \Phi\to -\Phi\,.
\label{z2_sym}
\eeq 
%Note here, that the transformation of $Z_2: S \to S^{*}$ is automatically guaranteed from the invariance under the kinetic piece $ |D_\mu^{X} S|^2$, when $Z_2:  X_\mu \to -X_\mu$.

The quantum numbers under $SU(3)_c\times SU(2)_L\times U(1)_Y\times Z_2$ of the new fields are tabulated in Tab.~\ref{tab:particles}.

\begin{table}[htb!]
\begin{center}
\begin{tabular}{|c|c|c|c|c|c|c|c|c|c|c|}
\hline
Fields &$SU(3)_c$ & $SU(2)_L$ & $U(1)_Y$ & $Z_2$ \\ [0.5ex] 
\hline
\hline
$\Phi$ & 1 & 1 & 0 & $-\Phi$\\
$X$ & 1 & 1 & 0 & $-X$\\
$S$ & 1 & 1 & 0 & $S^{*}$\\
\hline
\hline
\end{tabular}
\end{center}
\caption {Charges of the new particles under $SU(3)_c\times SU(2)_L\times U(1)_Y\times Z_2$ symmetry.}\vspace{0.3cm} 
\label{tab:particles}
\end{table}

With these fields and the charges, we can write the renormalizable $SU(3)_c\times SU(2)_L\times U(1)_Y\times Z_2$ invariant scalar potential as:
\bea
V(H,S,\Phi) & = & -\mu_H^2 |H|^2 - \mu_S^2 |S|^2 + \mu_\Phi^2 \Phi^2 \\
&& + \lambda_H |H|^4+\lambda_S |S|^4 +\lambda_\Phi \Phi^4 + \lambda_{H\Phi}|H|^2 \Phi^2 + \lambda_{S\Phi}|S|^2 \Phi^2 +\lambda_{SH}|H|^2 |S|^2\,.\nn
\label{eq:pot}
\eea
The total renormalizable Lagrangian then reads:
\beq
{\cal L}_\text{tot}=-\frac14 X_{\mu\nu} X^{\mu\nu} + |D_\mu^{X} S|^2 + (D_\mu^{SM} H)^\dagger (D^{SM\,\mu} H)  + 
\frac12 \partial_\mu \Phi \partial^\mu \Phi - V(H,S,\Phi) + {\cal L}_\text{SM}\,,
\label{lag_ini}
\eeq
where $H$ is the $SU(2)_L$ SM Higgs doublet and $D_\mu^{SM}$ is the SM covariant derivative. The $X_\mu$ field tensor and corresponding covariant derivative are defined as  
\beq
X_{\mu\nu}=\partial _\mu X_\nu-\partial _\nu X_\mu; ~~D_\mu^{X} S=\partial_\mu S-ig_X S X_\mu;
\eeq
where $g_X$ denotes $U(1)_X$ gauge coupling. 

Below we are going to investigate limit of the above model when the mass of one of the physical scalars contained in the spectrum becomes very large. 
We expect to reproduce a version of the Stueckelberg model coupled to the extra scalar $\Phi$ and the SM Higgs doublet $H$. The goal is to determine, among the degrees of freedom of the considered model, 
the Stueckelberg scalar introduced in order to make the Stueckelberg Lagrangian gauge symmetric. Some subtleties of the limiting procedure will be addressed.

%%%%%%%%%%%%%%%%%%%%%%%%%%%%%%
\subsection{Positivity criteria}
%%%%%%%%%%%%%%%%%%%%%%%%%%%%%%%

In order to formulate conditions for asymptotic positivity (for large field strengths)  of the potential in Eq.~\eqref{eq:pot} we shall first write down the matrix of quartic couplings in the basis: $|S|^2, \Phi^2, |H|^2$:

\bea
W\equiv
\quad
\begin{pmatrix} 
\lambda_S & \frac{\lambda_{S\Phi}}{2} & \frac{\lambda_{SH}}{2} \\
\frac{\lambda_{S\Phi}}{2} & \lambda_{\Phi} & \frac{\lambda_{H\Phi}}{2}\\
\frac{\lambda_{SH}}{2} & \frac{\lambda_{H\Phi}}{2} & \lambda_H,
\end{pmatrix} \,.
\eea
Now, a scalar potential biquadratic in fields is bounded from below if the matrix $W$ is co-positive~\cite{Kannike:2012pe}. Thus, the vacuum stability conditions for the potential in Eq.~\eqref{eq:pot} are given by the Sylvester criteria for the co-positivity of $W$~\cite{Kannike:2012pe,Kannike:2016fmd}:

\bea
\lambda_{S} > 0,~~\lambda_\Phi > 0,~~\lambda_H > 0,
\eea
also,
\bea
&\tilde{\lambda}_{S\Phi}\equiv\lambda_{S\Phi}+2(\lambda_{S} \lambda_\Phi)^{1/2} > 0,& \label{posit_SP}\\
&\tilde{\lambda}_{H\Phi}\equiv\lambda_{H\Phi} + 2(\lambda_\Phi\lambda_H)^{1/2} > 0,& \label{posit_PH}\\
&\tilde{\lambda}_{SH}\equiv\lambda_{SH}+ 2(\lambda_{S}\lambda_H)^{1/2} > 0,& \label{posit_SH} \\
&\lambda_\Phi^{1/2}\lambda_{SH}+\lambda_S^{1/2}\lambda_{H\Phi}+\lambda_H^{1/2}\lambda_{S\Phi}+
2(\lambda_{S}\lambda_\Phi \lambda_H)^{1/2}+2(\tilde{\lambda}_{S\Phi}\tilde{\lambda}_{H\Phi}\tilde{\lambda}_{SH})^{1/2} > 0.&
\label{posit_com}
\eea
We emphasize that these are necessary and sufficient conditions for vacuum stability~\cite{Kannike:2012pe}.

%%%%%%%%%%%%%%%%%%%%%%%%%%%
\subsection{Minimization conditions and spontaneous symmetry breaking}
%%%%%%%%%%%%%%%%%%%%%%%%%%%

We parametrize the scalar fields as follows:
\bea
H=
\begin{pmatrix} 
\phi^+ \\
\frac{h+i \phi^0}{\sqrt{2}}
\end{pmatrix}\,,
\lsp S=\frac{\rho}{\sqrt{2}}e^{i \sigma_S/\vs}\,.
\label{parametrization}
\eea
The extrema conditions for the potential in Eq.~\eqref{eq:pot} read 
\bea
\left.\frac{\partial V}{\partial h}\right|_{h=v_h}=0,
~~\left.\frac{\partial V}{\partial \rho}
\right|_{\rho=v_S} =0,~~\left.\frac{\partial V}{\partial \Phi}\right|_{\Phi=v_\Phi} = 0.
\label{min_con}
\eea
Hereafter we assume $\mu_H^2, \mu_S^2, \mu_\Phi^2>0$ in order to generate proper symmetry breaking.

We will require that the above conditions are satisfied by non-zero vacuum expectation values (vevs) of $ \langle H \rangle =v_h\neq 0$ and $\langle S \rangle=v_S\neq 0$, while for $\Phi$ we require zero-vev; $\langle \Phi \rangle = v_\Phi=0$. 

The following relations are implied by the minimization conditions (\ref{min_con}):
\bea
v_h(2 \lam_H v_h^2+\lSH\vs^2-2\mu_H^2)&=&0\nn\\
\vs(2\lam_S\vs^2+\lSH\vh^2-2\mu_S^2)&=&0 \label{eq:mincond} \\
v_\Phi(2\mu_\Phi^2+4\lam_\Phi v_\Phi^2 + \lHP\vh^2+\lSP\vs^2)&=&0\nn
\eea
We will therefore expand $H$ and $S$ around the non-zero vevs as follows
\bea
H=
\begin{pmatrix} 
\phi^+ \\
\frac{h+\vh+i \phi^0}{\sqrt{2}}
\end{pmatrix}\,,
\lsp S=\frac{\rho + \vs}{\sqrt{2}}e^{i \sigma_S/\vs}\,,
\label{param1}
\eea
where we have used the same notation for the fluctuations around the vacuum as earlier for the initial fields. In the expression above $\sigma_S$ is the Goldstone boson that constitutes the longitudinal component of the 
$X_\mu$, while the SM Goldstone bosons are $\phi^{\pm,0}$. Note that there is no potential for $\sigma_S$. We have adopted a Cartesian parametrization for the doublet $H$ together with a polar 
parametrization for the complex singlet $S$. The purpose was to find out the degree of freedom that corresponds to the Stueckelberg scalar, it will be discussed in details shortly. 
\begin{table}[h]
\bce
\begin{tabular}{||c|c|c|c|c||} 
 \hline
 \# & $\vh^2$ & $\vs^2$ & $\vP^2$ & $V|_\text{extr}$ \\ 
 \hline\hline
 1 & 0 & 0 & 0 & 0 \\ 
 \hline
 2 & 0 & $\frac{\mu_S^2}{\lambda_S}$ & 0 & $-\frac{1}{4} \frac{\mu_S^4}{\lambda_S}$ \\
 \hline
 3 & $\frac{\mu_H^2}{\lambda_H}$ & 0 & 0 & $-\frac{1}{4} \frac{\mu_H^4}{\lambda_H}$ \\ 
 \hline
 4 & $\frac{2\left(2\lambda_{S}\mu_H^2-\lambda_{SH}\mu_{S}^2\right)}{4\lambda_H\lambda_{S}-\lambda_{SH}^2}$ & $\frac{2\left(2\lambda_H\mu_{S}^2-\lambda_{SH}\mu_H^2\right)}{4\lambda_H\lambda_{S}-\lambda_{SH}^2}$ & 0 & $-\frac{\lambda_S\mu_H^4-\lambda_{SH}\mu_H^2\mu_S^2+\lambda_H\mu_S^4}{4\lambda_H\lambda_{S}-\lambda_{SH}^2}$ \\ 
 \hline
\end{tabular}
\caption{The table shows possible extrema with $\vP=0$ and corresponding values of the potential \ref{eq:pot}.}
\label{pot_points}
\ece
\end{table}
In Table~\ref{pot_points} we list all possible extrema that satisfy (\ref{eq:mincond}) for $\vP=0$ together with corresponding values of the potential.
There may exist three other extrema with $\vP\neq 0$, however for the stability of $\Phi$ we are going to choose parameters that ensure $\vP = 0$.
We are going to find conditions that guarantee the solution \#4 to be the global minimum.
First we must make sure that $\vP=0$ is the only possible vev for $\Phi$, for that purpose we will assume that for given quartic couplings we adjust $\mu_\Phi^2$ such that $2\mu_\Phi^2+4\lam_\Phi v_\Phi^2 + \lHP\vh^2+\lSP\vs^2>0$, then indeed $\vP=0$ is the only solution of (\ref{eq:mincond}).

Next, it turns out that 
\bea
V_4-V_2&=&-\frac{(-2\lambda_S \mu_H^2+\lambda_{SH}\mu_S^2)^2}{4\lambda_S(4\lambda_H\lambda_{S}-\lambda_{SH}^2)} \label{glob42}
\\
V_4-V_3&=&-\frac{(-2\lambda_H \mu_S^2+\lambda_{SH}\mu_H^2)^2}{4\lambda_H(4\lambda_H\lambda_{S}-\lambda_{SH}^2)}\,. \label{glob43}
\eea
As it will be seen shortly we assume $4\lambda_H\lambda_{S}-\lambda_{SH}^2>0$ in order to ensure positivity of masses squared, in addition $\lambda_{S,H}>0$ for the positivity of the potential,
therefore hereby we have shown that the extremum \#4 is the deepest one, and it must be the global minimum regardless what is the nature of solutions \#1, \#2 and \#3 (refer to tabl.~\ref{pot_points}).  

The mass matrix squared corresponding to the solution \#4 for physical degrees of freedom expressed in the basis $\{h,s,\Phi\}$ reads:
\bea
\mathcal{M}^2 = \left(
\begin{array}{ccc}
 2 v_h^2 \lambda_H & v_h v_S \lambda_{SH} & 0 \\
 v_h v_S \lambda_{SH} & 2 v_S^2 \lambda_{S} & 0 \\
 0 & 0 & 2 \mu_\Phi^2+\lHP\vh^2+\lSP \vs^2 \\
\end{array}
\right),
\eea
where it is clearly seen that only $\{h,\rho\}$ mixes (as they get non-zero vevs) while $\Phi$ (the (3,3) element of the matrix) only receives contribution proportional to the vev of the other two fields. The eigenvalues of the mass matrix read:
\bea
m_\pm^2 &=& \lambda_H v_h^2+\lambda_{S} v_S^2\pm\sqrt{(\lambda_H v_h^2+\lambda_{S} v_S^2)^2-(\vh \vs)^2(4 \lS \lH - \lSH^2)}, 
\label{eigen_val_h}\\
m_{\Phi}^2 &=& 2 \mu_\Phi ^2+\lambda_{H\Phi} v_h^2+\lambda_{S\Phi} v_S^2\,. 
\label{eigen_val_P}
\eea
Hereafter we will adopt the convention that $h_1$ is always the $125\gev$ SM-like Higgs boson discovered in 2012 at the LHC. 
Therefore $m_1=m_\pm$ and $m_2=m_\mp$ for $h_1$ heavier (upper sign) or lighter (lower sign) than $h_2$. Hereafter we are going to consider the case of very heavy $h_2$, i.e. $m_2 \gg m_1$. As it is seen from (\ref{eigen_val_h}), 
for quartic couplings not exceeding perturbative limits $\sim 4\pi$, heavy $h_2$ requires large $v_S$, i.e. $v_S \gg \vh$. It is clear that the presence of a minimum at the extremum \#4 requires:
\beq
4 \lS \lH - \lSH^2 > 0 \lsp \text{and} \lsp 2 \mu_\Phi^2+\lHP\vh^2+\lSP\vs^2 > 0\,.
\label{loc_min_con}
\eeq
The first condition above together with the potential positivity condition (\ref{posit_SH})  implies
$\lSH < 2 \sqrt{\lS \lH}$. Note also that $4 \lS \lH - \lSH^2 > 0$ guarantees positivity of $\vh^2$ and $\vs^2$, see tabl.~\ref{pot_points}. 

Now we can now rotate the weak basis to get the mass basis via:
\bea
\quad
\begin{pmatrix} 
h_1 \\
h_2 \\
\Phi
\end{pmatrix} = \mathcal{R}^{-1}\quad
\begin{pmatrix} 
h \\
s \\
\Phi
\end{pmatrix},
\eea
where $\mathcal{R}$ is the Euler rotation matrix of the form:
\bea
\mathcal{R}=
\quad
\begin{pmatrix} 
\cos\alpha & -\sin\alpha & 0 \\
\sin\alpha & \cos\alpha & 0 \\
0 & 0 & 1 
\end{pmatrix}.
\eea
The mixing angle $\alpha$ is determined by the entries of the mass matrix as follows\footnote{See \cite{Duch:2015jta,Duch:2015cxa} for a detailed discussion of the $H-S$ system.}:
\beq
\sin 2\alpha =\frac{\text{sign}(\lSM-\lH)2{\cal M}_{12}^2}{\sqrt{({\cal M}_{11}^2-{\cal M}_{22}^2)^2+4({\cal M}_{12}^2)^2}}
\lsp
\cos 2\alpha =\frac{\text{sign}(\lSM-\lH)({\cal M}_{11}^2-{\cal M}_{22}^2)}{\sqrt{({\cal M}_{11}^2-{\cal M}_{22}^2)^2+4({\cal M}_{12}^2)^2}}
\label{mix_ang}
\eeq

The potential (\ref{eq:pot}) has 9 real parameters:
\bea
\{\mu_{H,S},\mu_\Phi,\lambda_{H,S,\Phi},\lambda_{H\Phi},\lambda_{S\Phi},\lambda_{SH}\}.
\eea
Amongst these, $\mu_{H,S}$ can be replaced by the vevs $v_h$ and $v_{S}$ following Eq.~\eqref{eq:mincond}. 
Adopting (\ref{eigen_val_h}) for $m_1=m_-$ one can express $\vs$ through $\{m_1^2=2\lambda_{SM} v_h^2,\lambda_H,\lambda_{S},\lambda_{SH}\}$ as follows:
\bea
v_S^2=v_h^2 \frac{4\lambda_{SM} (\lambda_H-\lambda_{SM})}{4 \lambda_{S} (\lambda_H-\lambda_{SM})-\lambda_{SH}^2}.
\label{eq:vsvh}
\eea
This reduces the number of free parameters in the theory to seven:

\beq
\{\mu_\Phi,\lambda_{H,S,\Phi},\lambda_{SH},\lambda_{H\Phi},\lambda_{S\Phi}\}. \nonumber
\eeq

All other useful relations of the parameters in the scalar potential have been furnished further in Appendix ~\ref{parameters_relations}.

%%%%%%%%%%%%%%%%%%%%%%%%%%%%%%% 
\subsection{Decoupling limit}
%%%%%%%%%%%%%%%%%%%%%%%%%%%%%%%

%which in turn needs $4 \lS \lH - \lSH^2 \to 0$ together with $2 \lH \mu_S^2-\lSH \mu_H^2 \neq 0$.

Here we would like to explore the {\it decoupling} limit of a very heavy new scalar ($h_2$). From (\ref{eigen_val_h}) it is clear that the limit $m_2 \to \infty$ requires $\vs \to \infty$. In order to do that, it is useful to define:
\beq
\Delta\equiv 4 \lS (\lH-\lSM) - \lSH^2 
\eeq
From (\ref{eq:vsvh}) we find:
\beq
\vs^2=\vh^2\frac{4\lSM(\lH-\lSM)}{\Delta},
\eeq
from where we see that large $\vs^2$ corresponds to $\Delta \to 0$.

From (\ref{m2_use}) we obtain:
\beq
m_2^2=\vh^2\frac{8\lS\lSM(\lH-\lSM)}{\Delta} + {\cal O}(\Delta^0).
\eeq

So, clearly $\Delta\to 0$ implies $m_2\to \infty$ unless $\lH=\lSM$.

Now we can investigate the behavior of the mixing angle for $\Delta\approx 0^+$, it is easy to see that:
\beq
\sin 2\alpha =\text{sign}(\lSM-\lH)\left(\frac{\Delta}{\lS \lSM}\right)^{1/2}+ {\cal O}(\Delta^{3/2})\,,
\eeq
so it is evident that $\alpha \to 0$ as $\Delta \to 0$ ($m_2\to \infty$). From now on we shall use the following set of parameters:~\footnote{We consider the case $m_2>m_1$, so $\lH>\lSM$.}
\beq
(m_1,m_2,\vh,\lS,\lH) \lsp\text{and}\lsp (m_\Phi,\lP,\lHP,\lSP).
\label{eq:parameters}
\eeq

Then $\vs^2$, $\lSH$, $\sin 2 \alpha$ and mass parameters could be calculated and expanded in powers of $m_2$ as follows:
\bea
\vs^2 &=& \frac{m_2^2}{2\lS}+ \frac{\lSM-\lH}{\lS}\vh^2\label{eq:vs2dec}\\
\lSH^2 &=& 4\lS\left[\lH-\frac{\lSM m_2^2}{m_2^2+2(\lSM-\lH)\vh^2}\right] = 4\lS(\lH-\lSM) + {\cal O}\left(\frac{1}{m_2^2}\right)\label{eq:lamSHdec}\\
\sin 2\alpha &=& -2\sqrt{2(\lH-\lSM)}\frac{\vh}{m_2} + {\cal O}\left(\frac{1}{m_2^3}\right)\label{eq:s2adec}\\
\mu_\Phi^2 &=& \frac12 \left[\mP^2 + \vh^2\left(\frac{\lSP(\lH-\lSM)}{\lS} - \lHP\right)
-\frac{\lSP}{2\lS}m_2^2\right]\\
\mu_H^2 &=& \frac12 \left(2\lH \vh^2+\lSH \vs^2\right) = \left(\frac{\lH-\lSM}{4\lS}\right)^{1/2} m_2^2 + {\cal O}\left(\frac{1}{m_2^0}\right)\\
\mu_S^2 &=& \frac12 \left(2\lS \vs^2+\lSH \vh^2\right) = \frac{m_2^2}{2} + {\cal O}\left(\frac{1}{m_2^0}\right).
\eea
Also note:
\beq
4\lS(\lH-\lSM)-\lSH^2=8\lS \lSM(\lH-\lSM)\frac{\vh^2}{m_2^2} + {\cal O}\left(\frac{1}{m_2^4}\right)\,.
\eeq

With all these relations amongst different parameters of the scalar potential, assuming large $m_2$, we are now going to construct an effective residual theory in the limit of large $m_2$. Note that then $\sin 2\alpha \to 0$, such that 
\bea
h_1&=&\phantom{-}\cos\alpha \; h + \sin\alpha \; \rho \longrightarrow h\,\\
h_2&=&-\sin\alpha \; h + \cos\alpha \; \rho \longrightarrow \rho \,.
\eea
Therefore all we need to do is to expand the Lagrangian for the SM supplemented by $S$, $X_\mu$ and $\Phi$ around the vacuum adopting the parametrization (\ref{param1}) and drop the $h_2 \leftrightarrow \rho $ and rename $h_1$ by $h$. It turns out that the resulting effective Lagrangian reads:
\beq
{\cal L}_\text{lim}=-\frac14 X_{\mu\nu} X^{\mu\nu} + |D_\mu^{X} S|^2 + 
 (D_\mu^{SM} H)^\dagger (D^{SM\,\mu} H)  + \frac12 \partial_\mu \Phi \partial^\mu \Phi - V_\text{lim}(h,\Phi)\,,
\label{eff_lag}
\eeq
where the potential is independent of $\sigma_S$ and given by:
\bea
V_\text{lim}(h,\Phi)&=&
\frac12 m_h^2 h^2 + \lH \vh h^3 + \frac14 \lH h^4 + \\
&&+\frac12 m_\Phi^2 \Phi^2 + \lHP \vh h \Phi^2 + \lP \Phi^4 + \frac12 \lHP h^2 \Phi^2  
+ \text{const.}\nn
\label{eff_pot}
\eea
where $m_h=m_1$. 

The kinetic terms  in the limit $m_2\to \infty$ should be written after expanding around the vacuum and decoupling/removing $\rho$ as follows: 
\bea\begin{split}
&|D_\mu S|^2 = \frac12 (\mx X_\mu-\partial_\mu \sigma_S) (\mx X^\mu-\partial^\mu \sigma_S) \\&
(D_\mu^{SM} H)^\dagger (D^{SM\,\mu} H)= \frac12 \partial_\mu h \partial^\mu h +\cdots,     
    \end{split}
\label{cov_der}
\eea
where the ellipsis contain all the interaction terms. 

%%%%%%%%%%%%%%%%%%%%%%%%%%%%%%% 
\subsection{Stueckelberg Lagrangian in decoupling limit}
%%%%%%%%%%%%%%%%%%%%%%%%%%%%%%%

One can easily notice that the effective Lagrangian (\ref{eff_lag}) coincides with the standard form of the Stueckelberg Lagrangian invariant under the following transformation:
\bea
X_\mu &\to& X_\mu^\prime = X_\mu+\partial_\mu\theta\nn\\
\sigma_S &\to& \sigma_S ^\prime = \sigma_S+\mx\theta\\
\Phi &\to& \Phi^\prime = \Phi\nn
\label{gauge_trans}
\eea
In other words we have just proven that in the limit $m_2\to \infty$ the theory defined by the Lagrangian (\ref{lag_ini}) reduces to the Stueckelberg Lagrangian. 

In addition our model is invariant under the $Z_2$:
\beq
X_\mu \to -X_\mu,\lsp \sigma_S\to -\sigma_S, \lsp \Phi \to -\Phi 
\label{ztwo}
\eeq

There are various comments here in order. First, note that the Stueckelberg scalar is just the Goldstone boson $\sigma_S$. To see this the polar parametrization of $S$ adopted in (\ref{param1}) was crucial. A consequence of that was also the disappearance of the potential for $\sigma_S$. 
\begin{figure}[htb!]
$$
\includegraphics[scale=0.25]{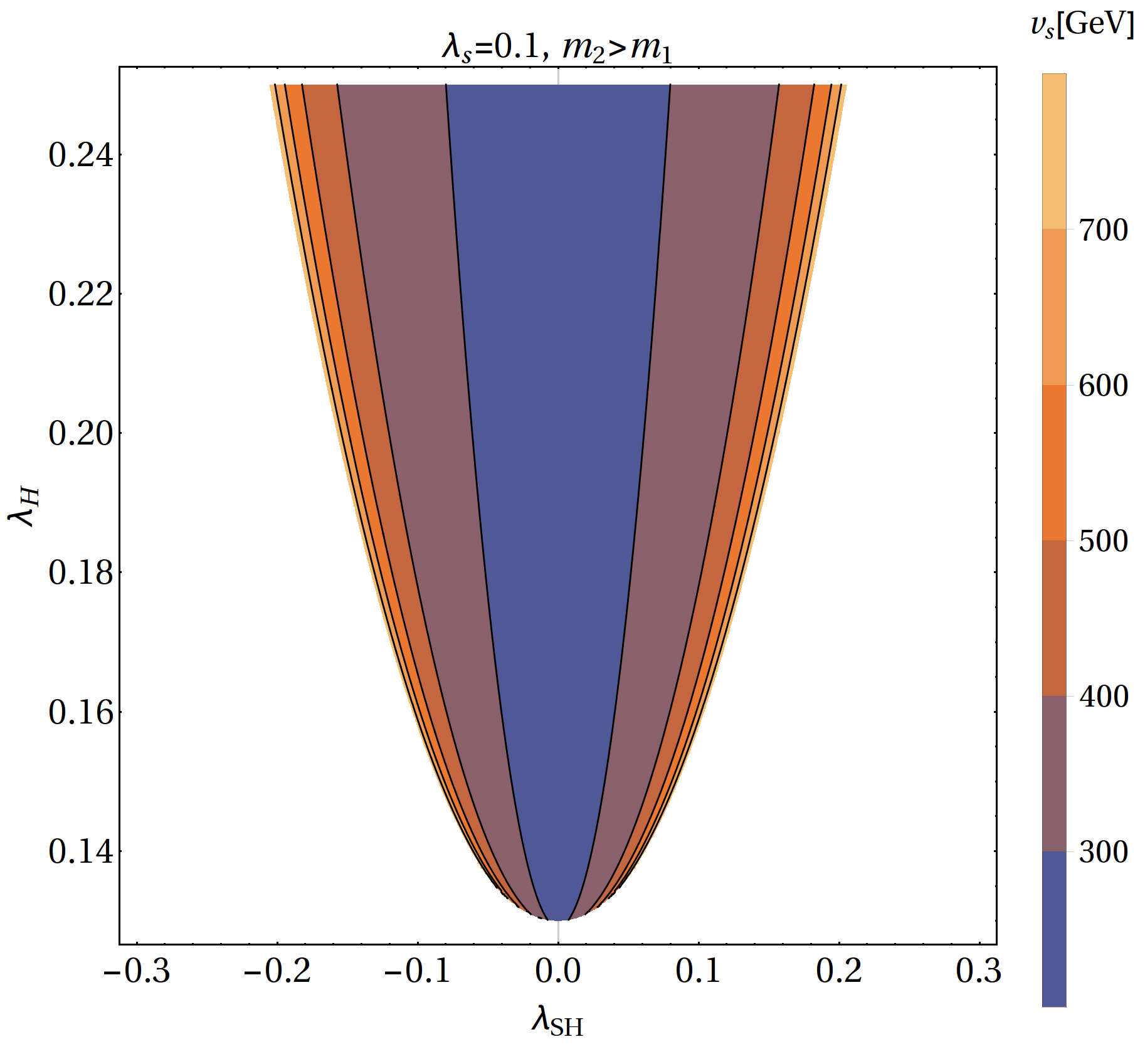}
\includegraphics[scale=0.25]{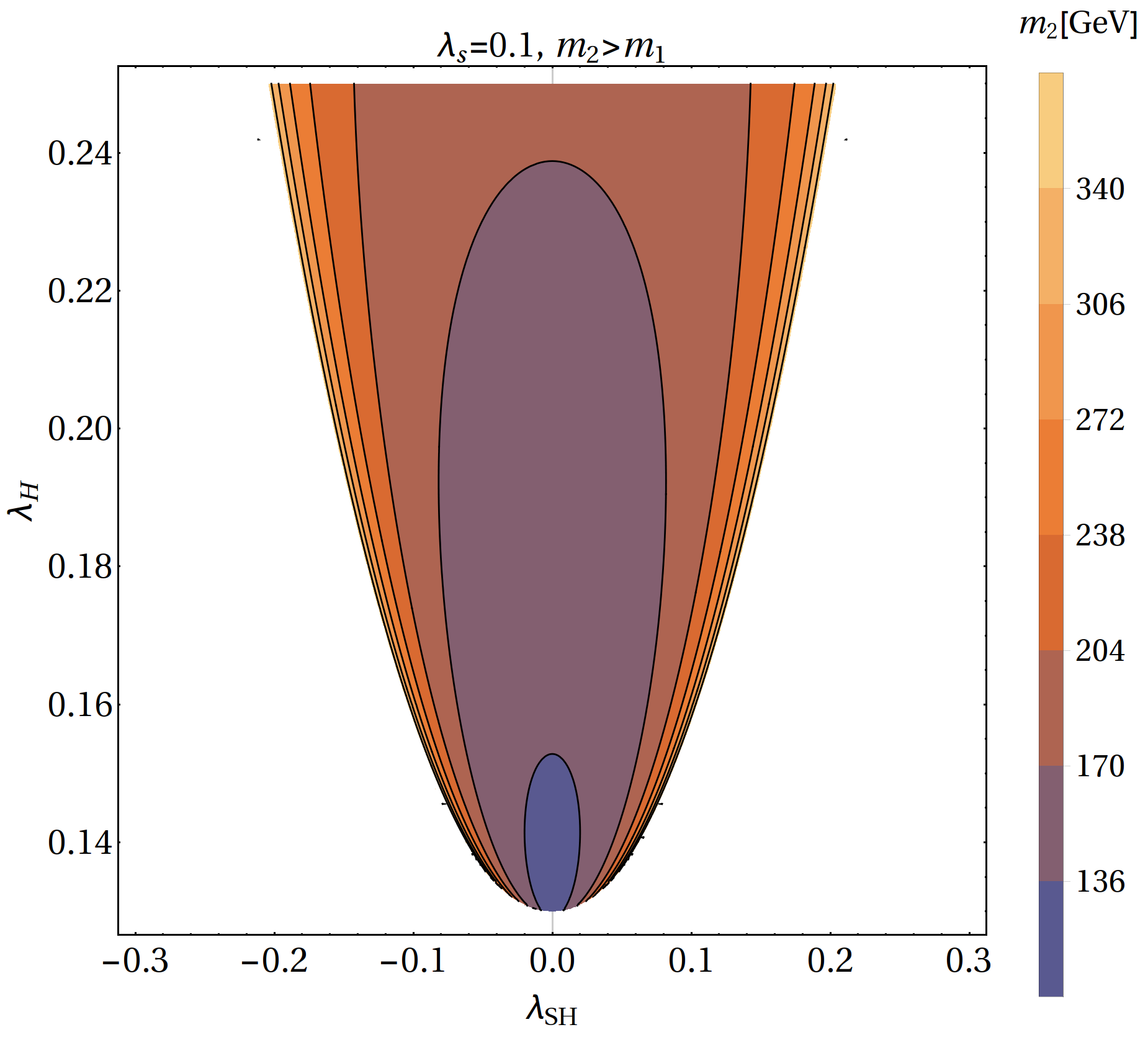}
$$\vspace{5mm}
$$
\includegraphics[scale=0.3]{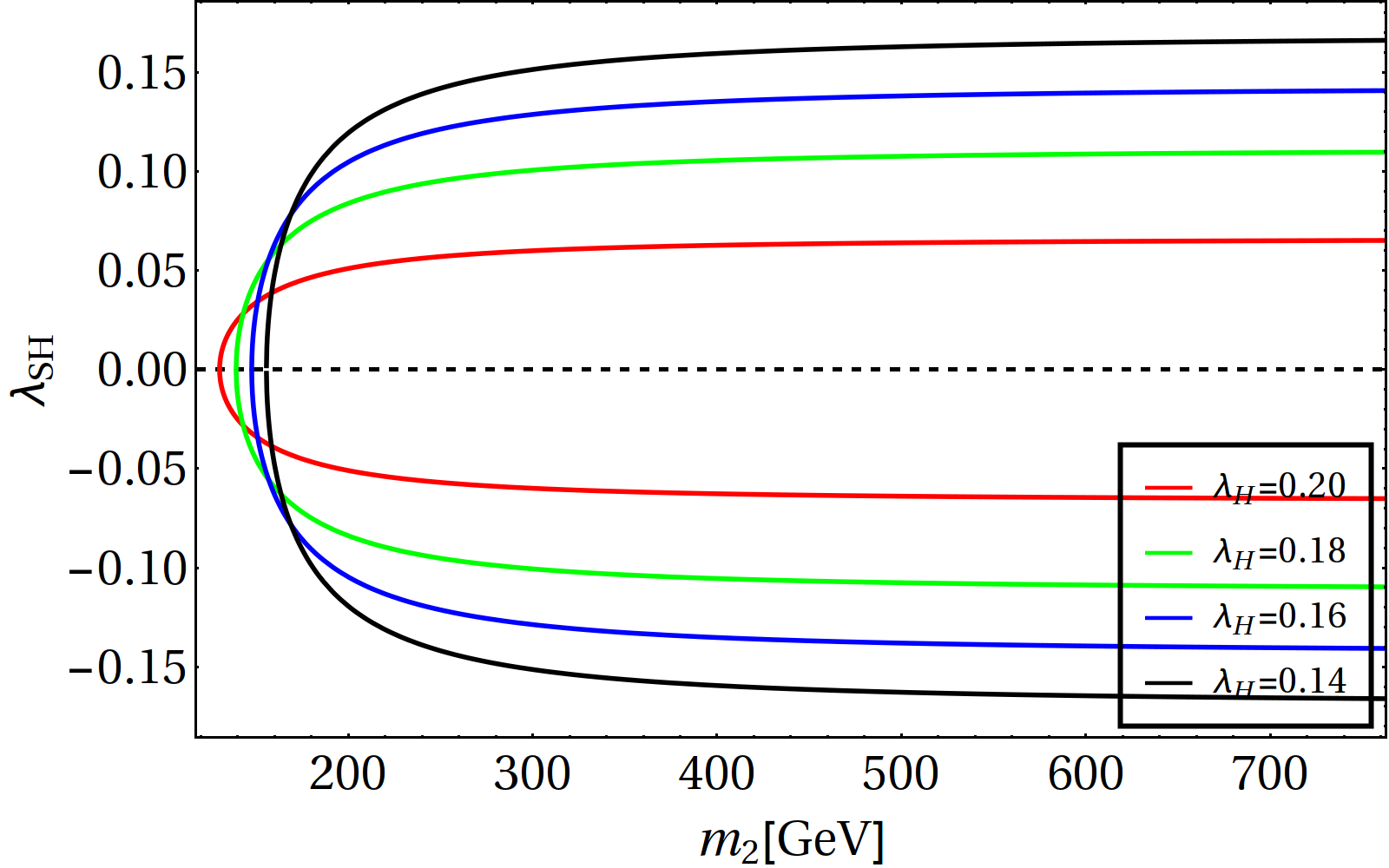}~~
\includegraphics[scale=0.3]{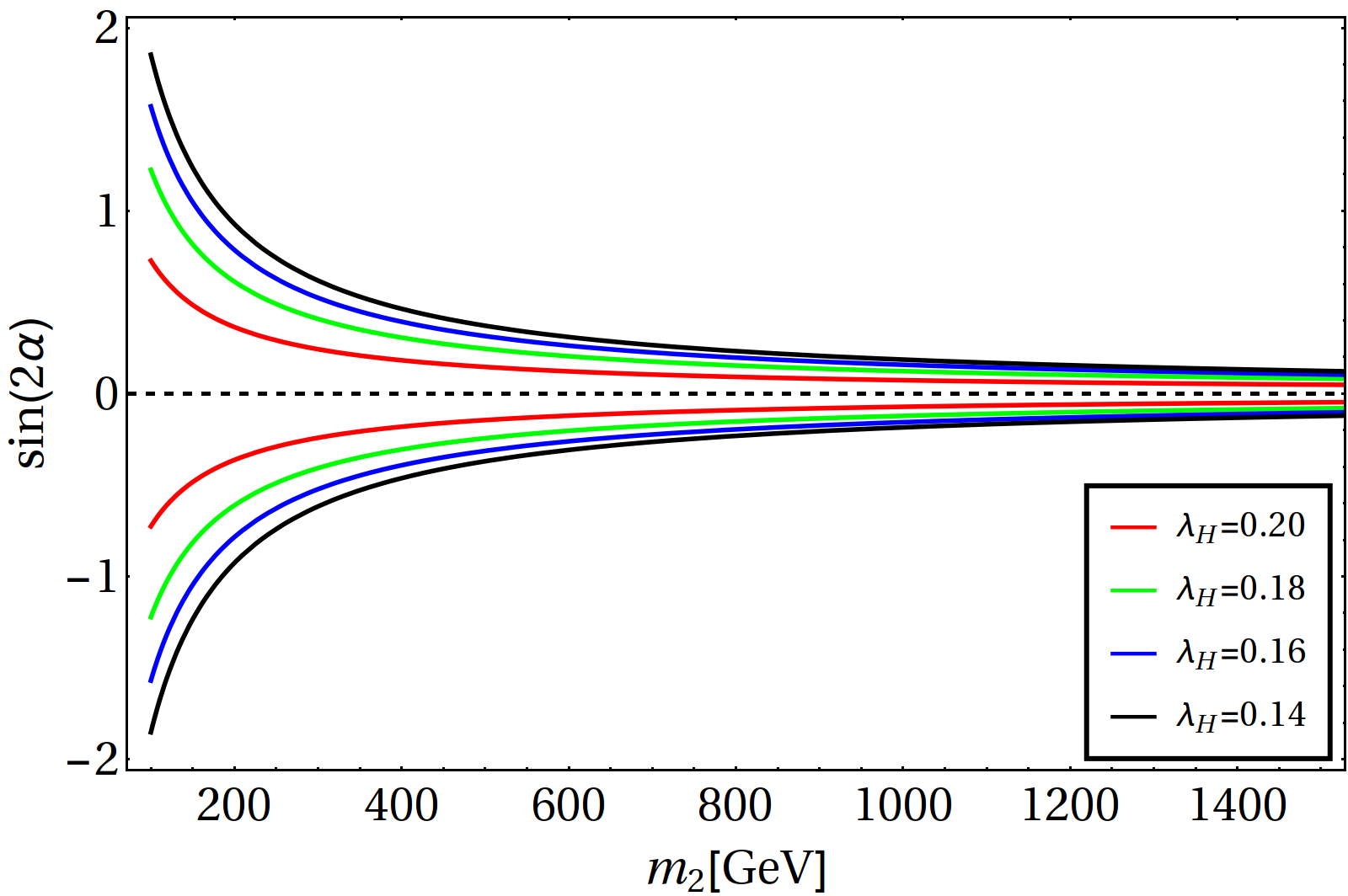}
$$
\caption{Top: The contours show allowed range of the vev $v_S$ and mass of the decoupled heavy scalar $m_2$ as a function of $\lam_H$ and $\lambda_{SH}$ for fixed $\lambda_S=0.1$.  Bottom left: $\lam_{SH}$ as a functions of $m_2$ for fixed values of $\lam_H$. Bottom right: $\sin (2\alpha)$ as a functions of $m_2$ for fixed values of $\lam_H$. Decoupling of the dark sector is clearly seen in the limit of large $m_2$.}
\label{fig:plot1}
\vspace{0.2 cm}
\end{figure}

Now let's define a current, $j_\mu\equiv (\mx X_\mu-\partial_\mu \sigma_S)$. Then note that the following potentially relevant term, $j_\mu \partial^\mu\Phi$, is invariant under (\ref{gauge_trans}) and therefore could be added to the standard Stueckelberg Lagrangian. 
However, it turns out that this operator could be omitted. It has been shown in sec.~5 of ref.~\cite{Duch:2014yma} 
that terms $\propto j_\mu \partial^\mu \Phi$ could be removed from the Lagrangian by field redefinition: a shift of $B$ and rescaling of $\Phi$. Therefore hereafter $j_\mu \partial^\mu \Phi$
will be ignored. Alternatively one could also follow arguments of ref.~\cite{Ruegg:2003ps}, where the author argues that the operator $\propto j_\mu \partial^\mu \Phi $ does not contribute to the $S$-matrix elements between physical states. 

Since our model is gauge invariant, the quantization requires fixing a gauge. We adopt the following gauge fixing term
\beq
{\cal L}_{gf}=-\frac{1}{2\xi}\left(\partial^\mu X_\mu+ \xi \mx \sigma_S \right)^2  
\label{gf}
\eeq
The advantage of the adopted gauge fixing is that it cancels mixing between $\partial^\mu X_\mu$ and $\sigma_S$. Expanding the Lagrangian one obtains eventually
\bea
{\cal L} &=& {\cal L}_\text{lim}+{\cal L}_{gf} = -\frac14 X_{\mu\nu}X^{\mu\nu} +  \frac{\mx^2}{2} X_\mu X^\mu -
\frac{1}{2\xi}\left(\partial^\mu X_\mu\right)^2
+\frac12 \partial_\mu \sigma_S \partial^\mu \sigma_S - \frac12 \xi\mx^2 \sigma_S^2 + \nn \\
&&
+ \frac12 \partial_\mu \Phi \partial^\mu \Phi + \frac12 \partial_\mu h \partial^\mu h - V_\text{lim}(h,\Phi) + \cdots
%+\frac{c}{\Lambda} B_{\mu\nu} X^{\mu\nu} \Phi + \frac{\tilde c}{\Lambda} B_{\mu\nu} \tilde{X}^{\mu\nu} \Phi
\,.  
\label{lag_fin}
\eea
In order to gauge-fix away $\sigma_S$ one must adopt the unitary gauge, which corresponds to $\xi \to \infty$. 
In the Stueckelberg formalism, in the unitary gauge, in presence of $\Phi$, one could expect presence of a term like $X_\mu X^\mu \Phi^2$ since it is allowed by symmetries.
However, it turns out that the operator $X_\mu X^\mu \Phi^2$ may only originate from dimension-6 term $|D_\mu S|^2\Phi^2$ and therefore must be suppressed by $1/\Lambda^2$. This is the only gauge invariant way to generate an operator $\propto X_\mu X^\mu \Phi^2$. It explains why the operator $X_\mu X^\mu\Phi^2$ can not appear as an unsuppressed dimension-4 operator, even though naively it could be added within the Stueckelberg strategy.

The decoupling limit of the scalar potential consistent with all the theoretical constraints is illustrated in Fig.~\ref{fig:plot1}. In the top left panel, we show the allowed region in the $(\lH,\lSH)$ plane for $m_2>m_1$ where the colors varying from blue to yellow show larger $v_s$. Top right panel shows the same but coloring is with respect to $m_2$. Both these top panels show the decoupling limit at the outer edge of the allowed parameter space. In bottom left panel, we show 
$\lSH$ as a function of $m_2$ at fixed values of $\lambda_H$. Similarly the bottom right figure shows $\sin(2\alpha)$ as a function of $m_2$, the convergence to zero-mixing angle is clearly shown. 

Since the mixing angle vanishes in the limit $m_2\to \infty$, i.e. $\sin(2\alpha)\propto \vh/m_2$, therefore the dimension-4 interaction between dark vector and the SM disappears. Note also that since $\lSP$ could be negative, therefore for $\lSP<0$ one can always adjust $\mu_\Phi^2$ so that the $\Phi$ mass squared remains at the weak scale even if $\vs$ grows. On the other hand, for  $\lSP>0$, to keep $m_\Phi$ at the weak scale $\lSP$ must behave as $\lSP \propto (\vh/\vs)^2 \propto (\vh/m_2)^2$.
%It compensates  contains a term $\propto \lSP \vs^2$, therefore for consistency some parameter tuning is needed to hold $m_\Phi$ at the weak mass scale, e.g. $\lSP \propto (\vh/\vs)^2 \propto (\vh/m_2)^2$. 
In addition, since we want to retain the vector boson mass, $\mx=\gx \vs$, of the order of weak scale, therefore it is necessary that the gauge coupling diminishes as $\gx \propto (\vh/\vs)^2 \propto (\vh/m_2)^2$. 
Note also that, since for large $m_2$ the value of the potential at the extremum \#~4 diverges as $\sim -m_2^4/(16 \lambda_S)$ therefore in order to avoid instability while preserving perturbativity we must limit ourself to large, but finite, values of $m_2$.

Summarizing, to reach the Stueckelberg limit starting from the Lagrangian (\ref{lag_ini}) a carefully adjusted trajectory in the parameter space must be adopted. An important consequence of approaching the $m_2\to \infty$ limit is the decoupling of the dark sector from the SM at dimension-4 operators by $\sin(2\alpha) \to 0$ and decoupling of $X_\mu$ from $S$ by $\gx \to 0$. It should also be recalled that to avoid instability of the potential $m_2$ must be finite (although can be large). 

%%%%%%%%%%%%%%%%%%%%%%%%%%%%%%% 
\subsection{Higher dimensional operator to connect DM and SM sectors}
%%%%%%%%%%%%%%%%%%%%%%%%%%%%%%%

Note that the coupling $\lHP$, which parametrizes the quartic portal interactions $\Phi^2|H|^2$, remains unsuppressed in the decoupling limit.  
Besides CP-violating operator $X_{\mu\nu}\tilde{X}^{\mu\nu}$~\footnote{That is irrelevant for DM phenomenology considered in this paper.}
this is the only renormalizable (dim-4) communication between the dark and visible sectors in the limit. Leading corrections to this communication will be provided by dim-5 operators
that are invariant under  transformations from ${\cal G} = SU(3)_C \times SU(2)_L \times U(1)_Y \times U(1)_X \times Z_2$ and under Lorentz transformations and which are made of the scalar $\Phi$, the vector $X_\mu$ and possibly combinations of SM fields. It is easy to see that there are only two non-trivial operators that satisfy required symmetry conditions~\footnote{Another dim-5 operator would require a presence of right-handed neutrinos $\nu_R$. This option will not be pursued hereafter.}:
\beq
{\cal L}_\text{dim-5}= \frac{c}{\Lambda} B_{\mu\nu} X^{\mu\nu} \Phi + \frac{\tilde c}{\Lambda} B_{\mu\nu} \tilde{X}^{\mu\nu} \Phi.
\label{dim5}
\eeq
This operator has already been introduced in~\cite{Macias:2015cna}, where it was mentioned that such an operator can be generated at the tree level only via antisymmetric tensor mediators. 
Finally, one can then write down the complete Lagrangian as:
\bea
{\cal L}_\text{tot}&=&{\cal L}_\text{lim} + {\cal L}_{gf} + {\cal L}_\text{dim-5} \nn\\
&=& -\frac14 X_{\mu\nu}X^{\mu\nu} +  \frac{\mx^2}{2} X_\mu X^\mu - \frac{1}{2\xi}\left(\partial^\mu X_\mu\right)^2
+\frac12 \partial_\mu \sigma_S \partial^\mu \sigma_S - \frac12 \xi\mx^2 \sigma_S^2 + \nn \\
&& + \frac12 \partial_\mu \Phi \partial^\mu \Phi + \frac12 \partial_\mu h \partial^\mu h -\left\{\frac12 m_h^2 h^2 + \lH \vh h^3 + \frac14 \lH h^4 + \right. \label{Ltot} \\
&& \left. +\frac12 m_\Phi^2 \Phi^2 + \lHP \vh h \Phi^2 + \lP \Phi^4 + \frac12 \lHP h^2 \Phi^2  
+ \text{const.} \right\} + \cdots + \nn \\
&& + \frac{c}{\Lambda} B_{\mu\nu} X^{\mu\nu} \Phi + \frac{\tilde c}{\Lambda} B_{\mu\nu} \tilde{X}^{\mu\nu} \Phi \nn\,,
\eea
where ellipsis denote interactions of the SM Higgs boson $h$ with other SM components that are not relevant here. We note here that we necessarily assume 
$\Lambda > m_2$, otherwise higher dimensional operators (neglected in this work) would appear in the scalar potential. We also adopt the following notations hereafter : 
$\alpha(\tilde{\alpha})=\frac{c (\tilde{c})}{\Lambda}; ~~\beta=\frac{\tilde{\alpha}}{\alpha}$.

%%%%%%%%%%%%%%%%%%%%%%%%%%%%%%%% 
%\subsection{Remarks}
%%%%%%%%%%%%%%%%%%%%%%%%%%%%%%%%
%\bit
%\item The solution \%7 is indeed a global minimum in the decoupling limit, $m_2 \gg \vh$.
%\item The mixing angle vanishes  in the limit, $\alpha\propto \vh/m_2$, so the interaction between dark vectors and SM sectors disappears. Note that, for a fixed $m_X=g_X \vs$, in the decoupling limit, also $\lim_{m_2\to \infty} g_X=0$. 
%\item The couplings $\Phi\Phi h_1 h_1$ and $\Phi\Phi h_1$ remain unsuppressed in the decoupling limit, i.e. they are $\propto \lHP$. This is the only renormalizable communication between the dark and visible sectors.
%\item $m_2$ must be large but finite, otherwise the potential becomes unstable, i.e.unbounded from below.
%\item In the Stueckelberg formalism, one could expect presence of couplings like $X_\mu X^\mu \Phi^2$ which are allowed by symmetries, however they originate from dim~6 operators (to be verified) so we can neglect them.
%\item I think we should assume $\Lambda > m_2$, otherwise we should include higher dimensional operators in the potential.
%\item In order to preclude massive production of $XX$ pairs by decays of heavy $h_2$ we must assume that $m_2>\trh$, so that the number density of $h_2$ is suppressed. 
%\eit

% %%%%%%%%%%%%%%%%%%%%%%%%%%%%%%%%%
\section{DM yield via freeze-in}
\label{freeze-in}
% %%%%%%%%%%%%%%%%%%%%%%%%%%%%%%%%

It is clear from the proceeding section that the couplings $\Phi\Phi h_1 h_1$ and $\Phi\Phi h_1$ remain unsuppressed in the decoupling limit that we are exercising here. 
These interactions are $\propto \lHP$, which is not suppressed.  This is the only renormalizable communication between the dark and visible sectors. 
So, it is quite likely to assume that in the early universe $\Phi$ is abundant being in equilibrium with the SM (i.e. with $h$).
Since $\Phi$ is the next lightest $Z_2$-odd dark component its decays and annihilations may produce DM (i.e. $X$) non-thermally.
This the mechanism (freeze-in) we will investigate hereafter. First, in this section, we will derive Boltzmann equations governing $X$  production in the early universe. We will also discuss applicability of neglecting various masses while calculating the amplitudes for decays and annihilations. Before going into the details, we would like to clarify that in the following sections, in order to satisfy the EFT limit and also have a successful freeze-in, we will adopt the following hierarchy amongst the scales and masses involved in the model:
\bea
\Lambda\gsim m_2 \gsim T_{RH} > m_\Phi > m_X,
\label{eq:hierarchy}
\eea
where $\trh$ is the reheating temperature at which inflaton decay products thermalize.

%%%%%%%%%%%%%%%%%%%%%%%%%%%%%%%%%
\subsection{DM production via decay and annihilation processes}
\label{sec:dm-prod}
%%%%%%%%%%%%%%%%%%%%%%%%%%%%%%%%%

\begin{figure}[htb!]
$$
\includegraphics[scale=0.4]{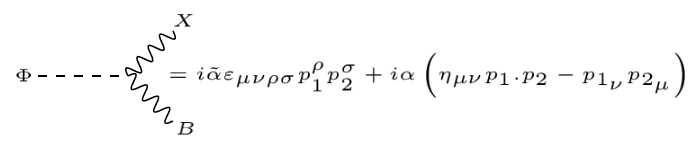}
$$
\caption{Decay of $\Phi \to XB$ before EWSB and $\Phi \to X \gamma(Z)$ after EWSB, which contributes to the freeze-in production of $X$. The vertex factor shown here is the one for $\Phi \to XB$.}
\label{fig:phi-decay}
\end{figure}

\begin{figure}[htb!]
$$
\includegraphics[scale=0.35]{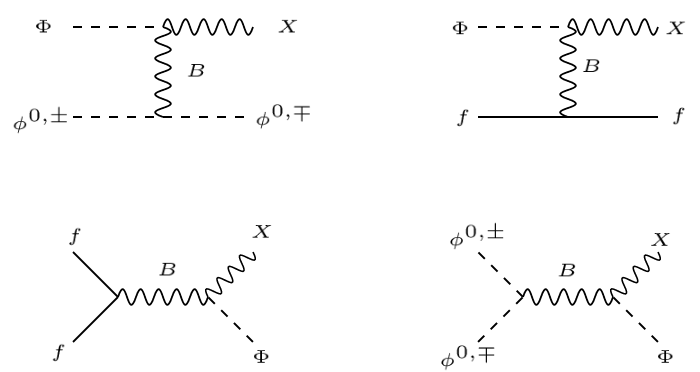}
$$
\caption{Annihilations via $t$-channel (top) and $s$-channel (bottom) leading to the production of VDM $X$ {\it before} EWSB. 
The diagrams with Goldstone bosons identically vanish, leaving only two diagrams with SM fermions.}\vspace{0.2cm}
\label{fig:bewsb}
\end{figure}

Since we are interested in the freeze-in production of the DM, hence we look for all such number changing processes with at least one DM particle in the final state. 
The processes that produce VDM, can easily be cooked up from interactions introduced in the preceding section and vertices collected in Appendix ~\ref{sec:vertex}. 
We classify all the DM number changing processes on the basis of their occurrences before electroweak (EW) symmetry breaking (EWSB) 
i.e. for thermal bath temperature $T>T_\text{EW}\simeq 160~\rm GeV$ or after EWSB, i.e. for $T<T_\text{EW}$. 
Due to the presence of the $\Phi X B$ vertex the VDM $X$ can always be produced from the decay of the scalar $\Phi$ that maintains 
thermal equilibrium with the SM bath via the portal interaction $\lam_{H\Phi}\left|H\right|^2 \Phi^2$. 
This decay channel, shown in Fig.~\ref{fig:phi-decay}, is always present before and after EWSB as $m_\Phi>m_X$, which is anyway our prime 
assumption for the stability of the VDM. After EWSB, the decay occurs to $Z,\gamma$. Apart from the decay, we also have 
four $2\to2$ annihilation channels with one DM in the final state as shown in Fig.~\ref{fig:bewsb} before EWSB~\footnote{All such processes with a pair of DM in the final state are suppressed by $\sim 1/\Lambda^2$ and hence sub-dominant.}. 
The processes include two t-channel and two s-channel graphs including Goldstone bosons ($\phi^{0,\pm}$).
Note that, before EWSB all the SM states are massless and the Goldstone bosons (GB) are propagating degrees of freedom 
as the $SU(2)_L$ scalar has the form:
\bea H=
\begin{pmatrix}
\phi^+\\\phi^0
\end{pmatrix}.
\label{eq:higgs}
\eea

The dark sector fields $\{\Phi,X\}$, on the other hand, are massive due to $U(1)_X$ breaking, which occurs at much higher energy scale. 
Due of the presence of totally anti-symmetric rank four Levi-Civita symbol in the interaction vertex $\Phi X B$ and the momenta dependent interaction vertices for the 
GB's, all the processes involving Goldstone bosons in $t$-channel and $s$-channel identically become zero 
at the level of amplitude squared. Therefore, all those processes with GB's drop out leaving only the $\Phi\to X,B$ decay channel, along with the two 
$2\to2$ annihilation diagrams $f\Phi\to fX,~ff\to X\Phi$ (the top right and bottom left diagram of Fig.~\ref{fig:bewsb}) for DM production via freeze-in before EWSB. 
However, as we shall see in subsequent sections, the decay before EWSB is sub-dominant as compared to the annihilation processes for large reheat temperature ($T_\text{RH}>>m$).

\begin{figure}[htb!]
$$
\includegraphics[scale=0.35]{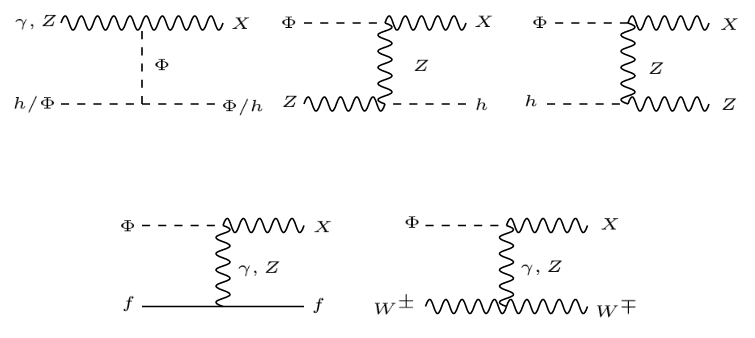}\\
\includegraphics[scale=0.32]{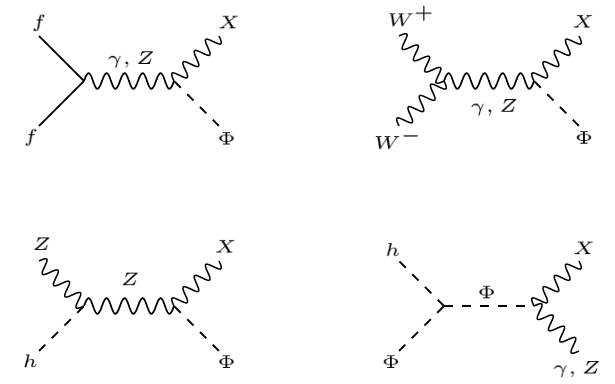}
$$
\caption{Annihilations via $t$-channel (left panel) and $s$-channel (right panel) leading to the production of $X$ {\it after} EWSB.}\vspace{0.2cm}
\label{fig:aewsb}
\end{figure}

Once the EW symmetry is broken, the GB's are no more individual physical degrees of freedom, instead they become longitudinal polarizations of the charged and neutral SM gauge bosons with $v_h=246~\rm GeV$. Also, the physical gauge bosons can be obtained in the mass basis by rotating the weak basis as:
\bea
\begin{pmatrix}
B_\mu \\ W_{3\mu} 
\end{pmatrix}=\begin{pmatrix}
c_w & -s_w\\ s_w & c_w
\end{pmatrix}\begin{pmatrix}
A_\mu \\ Z_\mu
\end{pmatrix}\label{eq:rot-gauge},
\eea
where $c(s)_w$ is the (co)sine of the Weinberg angle. Thus, after EWSB, the decay corresponds to $\Phi\to X,\gamma(Z)$, while all $2\to2$ annihilation channels 
giving rise to DM final states are shown in Fig.~\ref{fig:aewsb}. Due to massive propagator contributions after EWSB, the decay becomes more relevant for the determination of the DM yield, as we will demonstrate and discuss later. 
The decay widths and squares of the annihilation processes appearing before and after EWSB are collected in Appendix~\ref{sec:app-amp-bewsb}. 
%The portal coupling $\lam_{H\Phi}$, however, can be as large as $\lesssim 4\pi$ as a result of which processes involving $h\Phi\Phi$ vertices may start dominating over decay after EWSB. But, as we shall show, for $\lambda_{H\Phi}\lesssim\mathcal{O}(1)$ still allows us to consider only decay to be the dominant DM production channel after EWSB. Thus, for computing DM yield and its relic abundance thereof, we consider contributions from both $2\to2$ annihilation channels and $1\to2$ decay channel before EWSB, while after EWSB we consider only $1\to2$ decay to be the prime production channel modulo $\lam_{H\Phi}\lesssim\mathcal{O}(1)$. 

%%%%%%%%%%%%%%%%%%%%%%%%%%%%%%%%%
\subsection{Boltzmann equations for DM production}
\label{sec:BEQ}
%%%%%%%%%%%%%%%%%%%%%%%%%%%%%%%%%%

The key for freeze-in DM production is to assume that DM was not present in the early universe. In case it is produced via a decay as $\Phi \to BX$, the Boltzmann equation (BEQ) for the number density of $X$ can be written as:
\bea
\dot{n_X}+3 H n_X = \int d\Pi_X d\Pi_B d\Pi_\Phi \left(2\pi\right)^4 \delta^4\left(p_X+p_B-p_\Phi\right) \left|\mathcal{M}\right|_{\text{D}}^2 f_\Phi,
\label{eq:beq-decay}
\eea
where $d\Pi_j=\frac{d^3 p_j}{2E_j\left(2\pi\right)^3}$ are Lorentz invariant phase space elements, and $f_i$ is the phase space density of the $i^\text{th}$ particle with corresponding number density being:
\bea
n_i = \frac{g_i}{\left(2\pi\right)^3}\int d^3p f_i, 
\label{eq:num-density}
\eea
where $g_i$ is the number of internal DOFs. It is important to note that we assume negligible abundance of $X$ as the initial condition, also we disregard  Pauli-blocking/stimulated emission effects, i.e. we assume $1 \pm f_i\approx 1$. 
Indeed it has been assumed that $\Phi$'s are in equilibrium with the thermal bath (SM).

Similarly, the BEQ for DM production via generic annihilation process  $i,j\to X,k$ (with one DM in the final state) 
reads~\cite{Kolb:1990vq}:
\bea
\dot{n_X}+3 H n_X = \sum_{i,j,k}\int d\Pi_X d\Pi_i d\Pi_j d\Pi_k \left(2\pi\right)^4 \delta^4\left(p_X+p_k-p_i-p_j\right) \left|\mathcal{M}\right|_{i,j\to X,k}^2 f_i f_j.
\label{eq:beq-ann}
\eea
The BEQ in Eq.~\eqref{eq:beq-ann} can be rewritten as an integral over the CM energy as~\cite{Edsjo:1997bg,Hall:2009bx}:
\bea
\dot{n_X}+3 H n_X \approx \frac{T}{512\pi^6}\sum_{i,j,k}\int_0^\infty ds d\Omega P_{ij} P_{Xk} \left|\mathcal{M}\right|_{i,j\to X,k}^2 \frac{1}{\sqrt{s}} K_1\left(\frac{\sqrt{s}}{T}\right),
\label{eq:beq-ann-1}
\eea
where $P_{ab}=\frac{1}{2\sqrt{s}}\sqrt{s-(m_a+m_b)^2}\sqrt{s-(m_a-m_b)^2}\to \frac{\sqrt{s}}{2}$ in the limit $m_{a,b}\to 0$. 

Next we define the yield $Y_X\equiv n_X/s$, as a ratio of DM number density $n_X$ and the comoving entropy density in the visible sector $s$.
The BEQ corresponding to the decay in terms of the yield $Y_X$ can be written in the differential form as: 
\bea
\begin{split}
-s(T) H(T) T \frac{dY_X^\text{D}}{dT} &= \frac{g_\Phi m_\Phi^2 \Gamma_{\Phi\to X,B}}{2\pi^2} T K_1\left(m_\Phi/T\right),
\end{split}
\label{eq:yld-temp-decay}
\eea
where we have defined:
\bea
\Gamma_{\Phi\to X,B}=\int \frac{1}{2m_\Phi}\frac{\left|\mathcal{M}\right|_{\Phi\to X,B}^2}{g_\Phi} \left(2\pi\right)^4\delta^4\left(p_X+p_B-p_\Phi\right)d\Pi_X d\Pi_B
\eea
as the decay width of $\Phi$. It is possible to express Eq.~\eqref{eq:yld-temp-decay} in terms of the dimensionless quantity $x\equiv m_X/T$ and the reaction density $\gamma_D$ for decay as:
\bea
\begin{split}
x H s \frac{dY_X^\text{D}}{dx} &= \gamma_D,
\end{split}
\label{eq:yld-x-decay}
\eea
where $\gamma_D$, called reaction density, is defined in Appendix~\ref{sec:reac-den}. 

For the case of annihilation one can similarly write:
\bea
\begin{split}
-s(T) H(T) T \frac{dY_X^{\text{ann}}}{dT} &= \frac{T}{512\pi^6}\sum_{i,j,k}\int_0^\infty ds d\Omega P_{ij} P_{Xk} \left|\mathcal{M}\right|_{i,j\to X,k}^2 \frac{1}{\sqrt{s}} K_1\left(\frac{\sqrt{s}}{T}\right).
% \\& \implies Y_X^\text{ann} = -\int_{T_{max}}^{T_{min}}\frac{1}{s.H}\frac{1}{512\pi^6}\int_0^\infty ds d\Omega \left(\frac{\sqrt{s}}{2}\right)^2 \left|\mathcal{M}\right|_{i,j\to X,k}^2 \frac{1}{\sqrt{s}} K_1\left(\frac{\sqrt{s}}{T}\right) 
\end{split}
\label{eq:beq-ann-yld}
\eea
Note that the sum over $i,j,k$ indicates all the possibilities of producing DM following Figs.~\ref{fig:bewsb}, and \ref{fig:aewsb}.
Again, in terms of the reaction density defined in Appendix~\ref{sec:reac-den}, one can express the yield due to annihilation as:
\bea
\begin{split}
x H s \frac{dY_X^\text{ann}}{dx} &= \gamma_\text{ann}.
\end{split}
\label{eq:yld-x-ann}
\eea

Following Eq.~\eqref{eq:yld-temp-decay} and Eq.~\eqref{eq:beq-ann-yld} the total yield due to decay and due to annihilation can be written as:
\bea
\begin{split}
Y_X^\text{total} &= Y_X^\text{D}+Y_X^\text{ann},\\&= \int_{T_\text{min}}^{T_\text{max}} dT\frac{m_\Phi^2 \Gamma_{\Phi\to X,B}}{2\pi^2}\frac{K_1\left(m_\Phi/T\right)}{s(T) H(T)} \\&+\frac{1}{512\pi^6}\sum_{i,j,k}\int_{T_\text{min}}^{T_\text{max}}\frac{dT}{s(T) H(T)}\int_{s=0}^\infty ds d\Omega \left(\frac{\sqrt{s}}{2}\right)^2 \left|\mathcal{M}\right|_{i,j\to X,k}^2 \frac{1}{\sqrt{s}} K_1\left(\frac{\sqrt{s}}{T}\right). 
\label{eq:totyld1}
\end{split}
\eea

The maximum temperature available to the process is what we call reheat temperature $T_\text{RH}$. Also, we note that the processes after EWSB, are different from those before. Therefore, taken all such processes together, the yield at temperature $T_0$ can finally be written as:
\bea\begin{split}
Y_X^\text{total} \left(T_0\right)& = \Bigg\{\int_{T_\text{EW}}^{T_\text{RH}} dT\frac{m_\Phi^2 \Gamma_{\Phi\to X,B}}{2\pi^2}\frac{K_1\left(m_\Phi/T\right)}{s(T) H(T)} \\
&+\frac{1}{512\pi^6}\sum_{i,j,k}\int_{T_\text{EW}}^{T_\text{RH}}\frac{dT}{s(T) H(T)}\int_{0}^\infty ds d\Omega \left(\frac{\sqrt{s}}{2}\right)^2 \left|\mathcal{M}^{\text{bEWSB}}\right|_{i,j\to X,k}^2 \frac{1}{\sqrt{s}} K_1\left(\frac{\sqrt{s}}{T}\right)\Bigg\}\\
&+\Bigg\{\int_{T_\text{0}}^{T_\text{EW}} dT\frac{m_\Phi^2 \Gamma_{\Phi\to X,\gamma(Z)}}{2\pi^2}\frac{K_1\left(m_\Phi/T\right)}{s(T) H(T)} \\
&+\frac{1}{512\pi^6}\sum_{i,j,k}\int_{T_\text{0}}^{T_\text{EW}}\frac{dT}{s(T) H(T)}\int_{0}^\infty ds d\Omega \left(\frac{\sqrt{s}}{2}\right)^2 \left|\mathcal{M}^{\text{aEWSB}}\right|_{i,j\to X,k}^2 \frac{1}{\sqrt{s}} K_1\left(\frac{\sqrt{s}}{T}\right)\Bigg\},      
    \end{split}
    \label{eq:totyld2}
\eea
where the first parenthesis corresponds to contribution before EWSB while the second one describes the after EWSB production and $\mathcal{M}^{\text{(a)bEWSB}}$ stands for the amplitude for processes appearing (after) before EWSB. Note that for the annihilation processes we have considered the massless approximation which makes the expression less complicated. One can, equivalently, express the BEQ in a more general manner in terms of the reaction densities utilising Eq.~\eqref{eq:yld-x-decay} and Eq.~\eqref{eq:yld-x-ann} as:
\bea
xHs\frac{dY_X}{dx}=\gamma_\text{ann}+\gamma_\text{D}\label{eq:beq-x-1}.
\eea
This is rather more common and convenient way of parametrization. In Sec.~\ref{sec:relic-dm} we will be using these reaction densities to compare DM yield before and after the EWSB.

%%%%%%%%%%%%%%%%%%%%%%%%%%%%%%%%%%
\subsection{UV limit and limitations}
\label{sec:uv-limits}
%%%%%%%%%%%%%%%%%%%%%%%%%%%%%%%%%%

In this section, we demonstrate the difference between massless and massive limit of DM production cross-section 
and therefore we will be able comment on limitations of the Ultra Violet (UV) freeze-in advocated in \cite{Elahi:2014fsa,Barman:2020plp}.
Here we limit ourself to the period before EWSB so all the SM particles are massless. 
The masses of the dark sector $\Phi$ and $X$ are assumed to be of the same order, typically 
$m_\Phi \sim m_X \sim m \sim {\cal O}(1\text{TeV})$.
Hereafter the ``massless limit'' refers to zero-mass approximations, i.e. both  SM and dark masses are zero.
Our task in this section is to estimate size of mass effects, i.e. contributions to the yield that depend on the dark masses.
Since we limit ourself to the temperatures above $T_\text{EW}$ therefore Eq.~\eqref{eq:totyld2} can be simplified~\footnote{The contribution from $\Phi$ decays are negligible here.} as:
\bea\begin{split}
Y_X(T_\text{EW}) &= 
\frac{1}{4 \cdot 512\pi^6}\int_{T_\text{EW}}^{T_\text{RH}}\frac{dT}{s(T) \, H(T)}\\&
\int_{0}^\infty ds \left\{ \int d\Omega \left|\mathcal{M}^{\text{bEWSB}}[s,\cos\theta]\right|_{i,j \to X,k}^2 \right\}
\sqrt{s} K_1\left(\frac{\sqrt{s}}{T}\right).
    \end{split}\label{eq:UV}     
\eea
Here onward we shall refer to $\mathcal{M}^{\text{bEWSB}}$ as $\mathcal{M}$.

For strictly massless case the integration over $s$ can be performed analytically, so the result reads
\bea
Y_{m=0}=A I(\theta_{min})\frac{45T_{RH}}{512\pi^7}\frac{16 M_{pl}}{1.66 g_{\star s}\sqrt{g_{\star\rho}}},
\label{eq:delY1}
\eea
% \frac{1}{512\pi^5}\frac{1}{s(T).H(T)}\int_0^\infty ds \Bigg\{\frac{d\Omega}{4\pi}\left|\mathcal{M}\right|_{m=0}^2\Bigg\}\sqrt{s}K_1\left(\frac{\sqrt{s}}{T}\right)\\&=\frac{A.I(\theta_{min})}{512\pi^5} \frac{45 M_{pl}}{2\pi^2}\frac{1}{1.66 g_{\star s}\sqrt{g_{\star\rho}}}\frac{1}{T^5}\int_0^\infty ds s^{3/2}K_1\left(\frac{\sqrt{s}}{T}\right)\\&=A.I(\theta_{min})\frac{45}{2\times 512\pi^7}\frac{32 M_{pl}}{1.66 g_{\star s}\sqrt{g_{\star\rho}}}\\ &\implies Y_\text{massless}=A.I(\theta_{min})\frac{45T_\text{RH}}{2\times 512\pi^7}\frac{32 M_{pl}}{1.66 g_{\star s}\sqrt{g_{\star\rho}}}.
% \Biggl(\frac{A.I. M_{pl}}{4\times 512\pi^6}\frac{1}{\frac{2\pi^2}{45}(1.66)g_{\star s}\sqrt{g_{\star\rho}}}\Biggr)\frac{1}{T^5}\int_0^{\infty} \underbrace{ds s^{3/2} K_1\left(\frac{\sqrt{s}}{T}\right) }_{=32T^5}\\&=\Biggl(\frac{A.I. M_{pl}}{4\times 512\pi^6}\frac{32}{\frac{2\pi^2}{45}(1.66)g_{\star s}\sqrt{g_{\star\rho}}}\Biggr)=1.8184\times 10^{-9}. 
where $A$ contains all the couplings and constant factors that arise in the computation of 
$\left|\mathcal{M}^{m=0}[s,\cos\theta]\right|_{i,j \to X,k}^2$, and it is defined through the following relation:
\bea
\left|\mathcal{M}^{m=0}[s,\cos\theta]\right|_{i,j \to X,k}^2=A s m(\cos\theta)
\label{eq:Adef},
\eea
with $m(\cos\theta)$ containing all the angular dependance of the amplitude squared while
\beq
I(\tm) \equiv \frac12  \int_{-1}^{\cos\tm} d\cos\theta m(\cos\theta),
\label{eq:Idef}
\eeq
where $\tm$ is an angular cutoff necessary to avoid singularities that appear in the forward direction for t-channel diagrams, in the following $\tm=10^{-2}$ will be adopted.

\begin{figure}[htb!]
$$
\includegraphics[scale=0.32]{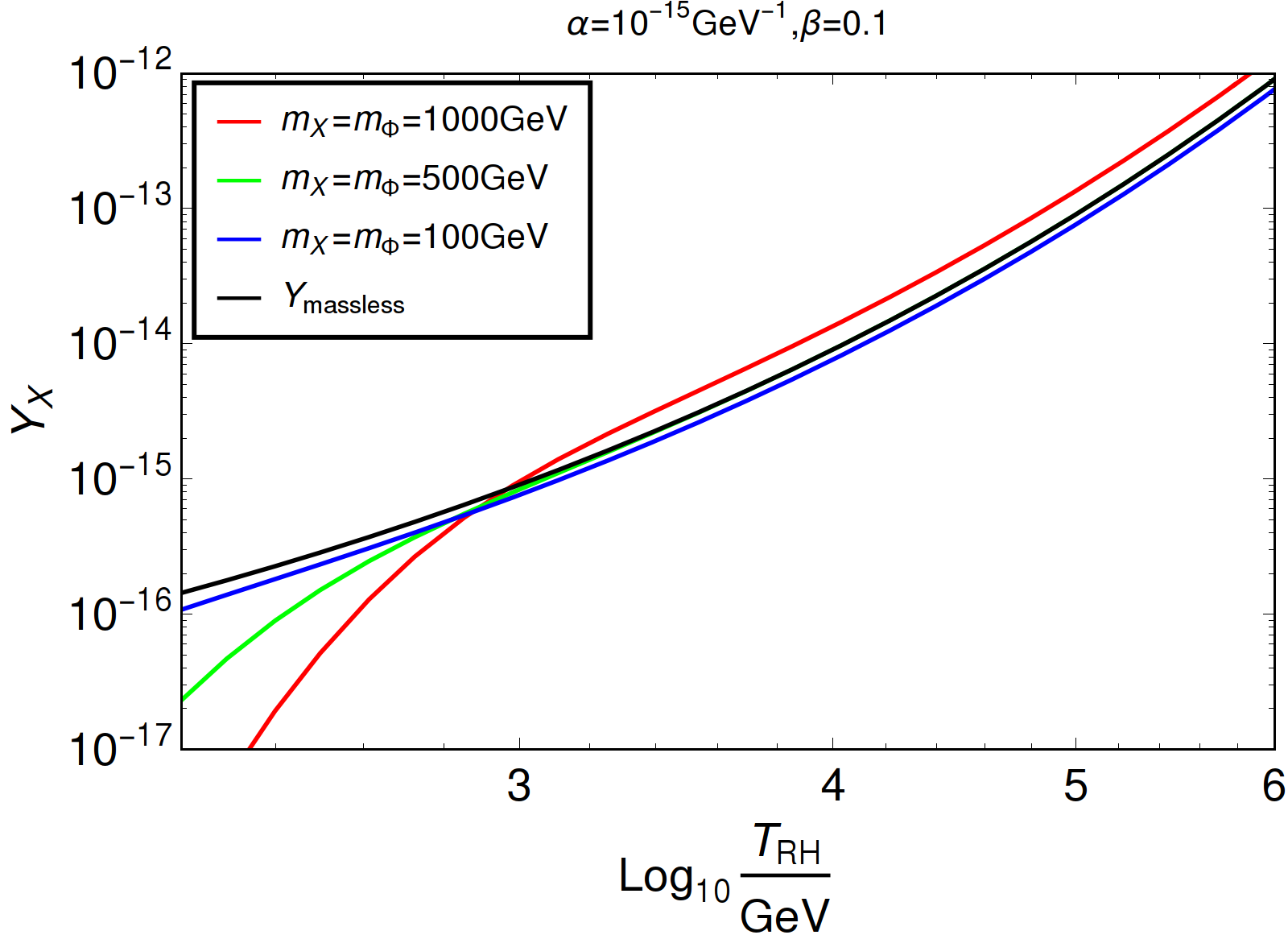}~~
\includegraphics[scale=0.32]{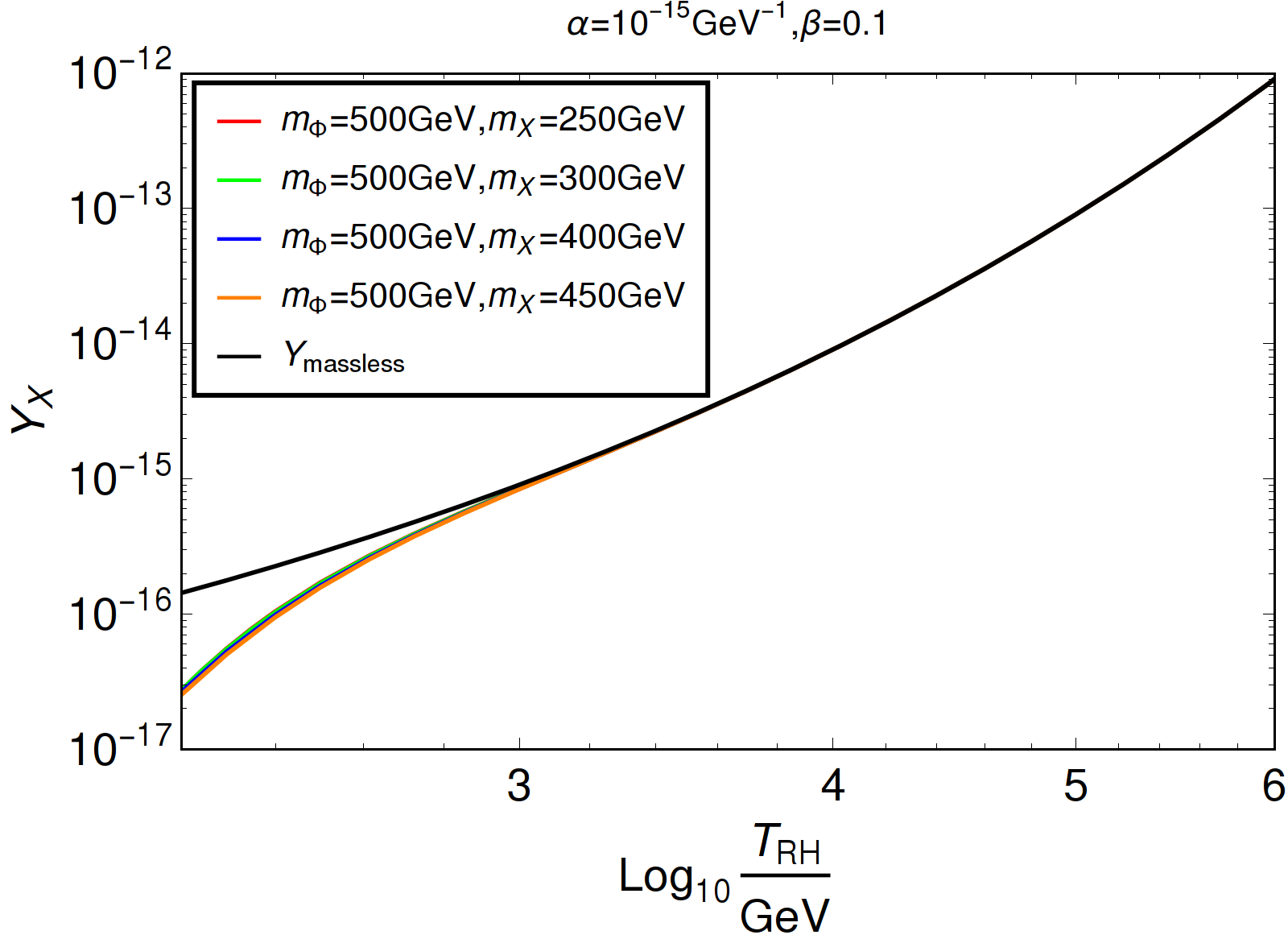}
$$
$$
\includegraphics[scale=0.32]{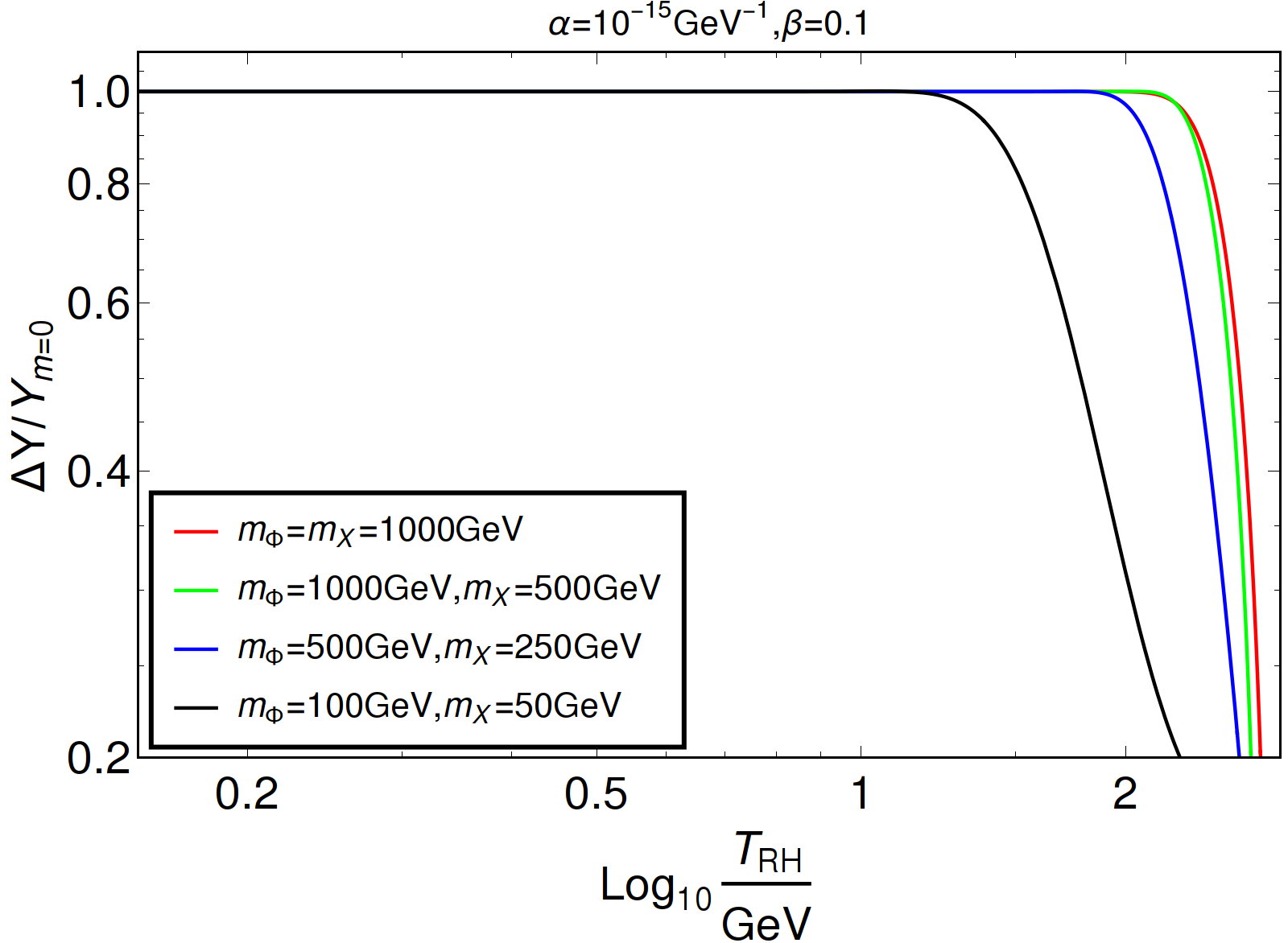}
$$
\caption{Top panels: Comparison of the yield at $T_\text{EW}$ in the massive case versus that in the massless case with respect to the reheat temperature, 
where the red, green and blue curves correspond to the massive cases while the black curve represents the massless case.  Bottom panel: Variation of $\Delta Y/Y_{m=0}$ with the reheat temperature for different choices of 
the masses shown in different colors. In all the plots ``massless'' refers to $\to m=0$.}
\label{fig:compare}
\end{figure}

Now, we would like to estimate the difference between the massless and massive limit. For that, let us define: 
\bea
\Delta Y(T)\equiv Y(T)|_{m=0}-Y(T)|_{m\neq 0}. 
\eea
Also, note that,
\beq
\left|\mathcal{M}^{m\neq 0}[s,\cos\theta]\right|_{i,j \to X,k}^2 = \left|\mathcal{M}^{m=0}[s,\cos\theta]\right|_{i,j \to X,k}^2 
+{\cal O}\left(\frac{m^2}{s}\right).
\eeq
% Then we obtain:
% 
% \beq
% \Delta Y(T) \approx \frac{1}{512\pi^5}\int_{T}^{T_\text{RH}}\frac{dT^\prime}{s(T^\prime) \, H(T^\prime)}
% \int_0^{\sm} ds \left\{ \int \frac{d\Omega}{4\pi} \left|\mathcal{M}^{m=0}[s,\cos\theta]\right|_{i,j \to X,k}^2 \right\}
% \sqrt{s} K_1\left(\frac{\sqrt{s}}{T}\right)
% \eeq
% Hereafter we assume $\sm=m^2$. 
%
%Let's parametrize the matrix element as follows
%\beq
%\left|\mathcal{M}^{m=0}[s,\cos\theta]\right|_{i,j \to X,k}^2
%= A s\frac{5+\cos\theta-\cos 2\theta}{1-\cos\theta}
%\eeq
%where
%$$
%A\equiv \frac{g_1^2}{128}(\alpha^2+\tilde\alpha^2)(Y_D^2+Y_T^2)
%$$
We assume here that the terms ${\cal O }\left(\frac{m^2}{s}\right)$ are negligible, note however that the mass dependance remains in $\sm=m^2$.  Then, for processes with amplitude squared of the form given in Eq.~\eqref{eq:Adef} we find: 
\beq
\Delta Y(T_\text{EW}) \approx A \frac{45}{512\pi^7}\frac{I(\tm)\mpl }{1.66\cdot (g_\star)^{1/2}g_\star^s}\int_{T_\text{EW}}^{\trh} dT^\prime\;\int_0^\frac{m}{T^\prime}
dx \, x^4 K_1(x),
\label{eq:dely1}
\eeq
where $x\equiv \sqrt{s}/T$.
As an example, let's consider a $t$-channel process $\Phi f \to X f$, where $f$ stands for the SM fermions. Then we find $I(\tm) = 1/2  \int_{-1}^{\cos\tm} d\cos\theta (5+\cos\theta-\cos 2\theta)/(1-\cos\theta) \approx 75$. For $\trh \gg m,T_\text{EW}$ we can estimate $\Delta Y(\tew)$ as follows
\bea
\Delta Y (\tew)& \approx A I(\theta_{min})\frac{45}{512\pi^7}\frac{M_{pl}}{1.66 g_{\star s}\sqrt{g_{\star\rho}}} 4.65m.
\label{eq:diff}
\eea
Note that $\Delta Y(\tew)$ is linear in $m$ as it obviously should vanish in the limit $m\to 0$. We should also note that Eq.~\eqref{eq:diff} is process independent and can be written in this particular form whenever the matrix element squared can be expressed 
in the form of Eq.~\eqref{eq:Adef}.

Now, let us find out the condition under which $\Delta Y$ becomes as large as $Y_\text{massless}$, as that will dictate the  condition for which massless limit can be no longer adopted for the UV freeze-in scenario. This can simply be found out by equating Eq.~\eqref{eq:diff} and Eq.~\eqref{eq:delY1}: 
\bea\begin{split}
\Delta Y&= A I(\theta_{min})\frac{45}{512\pi^7}\frac{M_{pl}}{1.66 g_{\star s}\sqrt{g_{\star\rho}}}\Biggl( m\int_0^{T_\text{RH}/m} dug(u) \Biggr)\lsim Y_{m=0}\\&\implies m\lsim 3.5 ~T_\text{RH}.  
    \end{split}
\eea

% \implies  m\int_0^{T_\text{RH}/m} dug(u)\lsim 16T_\text{RH}\implies 4.6m\lsim16T_\text{RH}
This implies that, as long as the masses involved in the freeze-in process are approximately less that three times of the reheat temperature, 
one can overlook the masses and the yield can be computed in the massless limit. 
This, in other words, justifies the fact that for large reheat temperature the massless limit is a 
good approximation for obtaining the UV freeze-in yield. In Fig.~\ref{fig:compare}, we demonstrate the difference between the yield 
obtained for massive and massless case of DM production for all the processes before EWSB put together. 
We plot {\sb $Y_X(T=T_{EW})$} as a function of $T_\text{RH}$ and see that massive case sharply differs from the massless 
case (black line) at small $T_\text{RH}$, while they exactly merge as the reheat temperature becomes large. 
In the bottom panel we show the same feature in terms of $\frac{\Delta Y}{Y_{m=0}}$ for different choices of $m_{X,\Phi}$. 

%%%%%%%%%%%%%%%%%%%
\section{DM relic abundance via freeze-in}
\label{sec:relic-dm}
%%%%%%%%%%%%%%%%%%%

As described in details in the last section, within the freeze-in paradigm, the DM yield is controlled by the annihilation and/or decay of SM as well as dark sector particles. Following Eq.~\eqref{eq:beq-x-1} we write down the BEQ governing the DM yield as~\cite{Duch:2017khv}:
\bea
xHs\frac{dY_X}{dx}=\gamma_\text{ann}+\gamma_\text{D}\label{eq:beq-x},
\eea
where $Y_X=n_X/s$. $H$ is the Hubble parameter given by $H=1.66\sqrt{g_{\star\rho}}T^2/M_{pl}$ and $x=m_X/T$ is a dimensionless quantity to parametrize the temperature of the thermal bath. As mentioned earlier, the $\gamma$'s denote the so-called reaction density for different particles annihilating (decaying) to the DM. The detail expressions of the reaction densities for $2\to2$ annihilations and $1\to2$ decays are given in Appendix.~\ref{sec:reac-den}. In order to compute the DM yield, we solve Eq.~\eqref{eq:beq-x} with the initial condition $Y_X\approx 0$ at large $T$ {\it i.e.,} small $x$ in accordance with the usual FIMP set-up {\footnote{The zero intial abundance of the DM could be a result of reheating itself when the inflaton decays preferentially to the visible sector without reheating the hidden sector, or may be due to some other mechanism.}}. By solving Eq.~\eqref{eq:beq-x} one can obtain the total DM yield $Y_X$ at the present epoch {\it i.e.,} $Y_X(T_0)$. The relic abundance of $X$ at present temperature can then be obtained via:
\bea
\Omega_X h^2 = \left(2.75\times 10^8\right) \left(\frac{m_X}{\text{GeV}}\right) Y_X(T_0).
\label{eq:relicX}
\eea
We must also remind that the PLANCK~\cite{Aghanim:2018eyx} allowed relic density reads:
\beq
\Omega h^2 = 0.11933\pm 0.00091,
\eeq
which we will use to find the relic density allowed parameter space of the model.

Since in our case the connection between the dark and the visible sector proceeds via a non-renormalizable interaction, 
the DM abundance is usually characterized by UV freeze-in~\cite{Elahi:2014fsa,Barman:2020plp} limit, where the 
DM abundance is sensitive to the reheat temperature $T_\text{RH}$ of the universe and NP scale $\Lambda$ only. 
This is in sharp contrast to the Infra-Red or IR freeze-in scenario where the two sectors communicate via 
renormalizable operators, and the DM abundance is set by the IR physics {\it i.e.,} the yield becomes maximum 
at low temperature, typically at $T\sim m_X$~\cite{Hall:2009bx}. Now, the reheat temperature $T_\text{RH}$ is very loosely bounded. 
Typically, the lower bound on $T_\text{RH}$ comes from the measurement of light element abundance during 
Big Bang Nucleosynthesis (BBN), which requires $T_\text{RH}\gtrsim 4.7~\rm MeV$~\cite{deSalas:2015glj}. 
The upper bound, on the other hand, comes from (a) cosmological  gravitino problem~\cite{Moroi:1993mb,Kawasaki:1994af} 
in the context of supersymmetry, that demands $T_\text{RH}\lesssim 10^{10}~\rm GeV$ to prohibit gravitino over production 
and (b) simple inflationary scenarios that require $T_\text{RH}\sim 10^{16}~\rm GeV$~\cite{Kofman:1997yn,Linde:2005ht} 
for a successful inflation. The reheat temperature, thus, can be regarded as a free parameter for our analysis. 
As we have already shown in the preceding section, when reheat temperature drops 
close to the masses involved in the annihilation or DM production process, massive kinematics start 
playing a key role in the yield and UV limit can not be trusted. Therefore our analysis will be divided into two regimes, depending on the 
scale of the reheat temperature $(T_{\text{RH}})$:
\begin{itemize}
 \item $T_{\text{RH}}>>m_i$, where the reheat temperature lies way above different masses that appear in the model. 
 \item $T_{\text{RH}}\gtrsim m_i$, where the reheat temperature lies close to the masses in the theory. 
%  Note that, this assumption does not violate the scale hierarchy in Eq.~\eqref{eq:hierarchy}.
\end{itemize}
In the following sub-sections we will consider the above two cases.
It is important to point out that our calculations of cross-sections have all been done analytically (see Appendices) and then the 
relic density is found out by solving BEQ in {\tt Mathematica} numerically. 
% 
% One can express Eq.~\eqref{eq:beq-x} as a function of temperature in a more explicit form as:
% 
% \bea\begin{split}
% Y_X^\text{total} \left(T\right)& = \Bigg\{\int_{T}^{T_\text{RH}} dT\frac{m_\Phi^2 \Gamma_{\Phi\to X,B}}{2\pi^2}\frac{K_1\left(m_\Phi/T\right)}{s(T).H(T)}+\\&\frac{1}{512\pi^6}\int_{T}^{T_\text{RH}}\frac{dT}{s(T).H(T)}\int_{0}^\infty ds d\Omega \left(\frac{\sqrt{s}}{2}\right)^2\left|\overline{\mathcal{M}^{\text{bEWSB}}}\right|_{i,j\to X,k}^2 \frac{1}{\sqrt{s}} K_1\left(\frac{\sqrt{s}}{T}\right)\Bigg\}\\&+\Bigg\{\int_{T}^{T_\text{EW}} dT\frac{m_\Phi^2 \Gamma_{\Phi\to X,\gamma(Z)}}{2\pi^2}\frac{K_1\left(m_\Phi/T\right)}{s(T).H(T)}+\\&\frac{1}{512\pi^6}\int_{T}^{T_\text{EW}}\frac{dT}{s(T).H(T)}\int_{0}^\infty ds d\Omega \left(\frac{\sqrt{s}}{2}\right)^2 \left|\overline{\mathcal{M}^{\text{aEWSB}}}\right|_{i,j\to X,k}^2 \frac{1}{\sqrt{s}} K_1\left(\frac{\sqrt{s}}{T}\right)\Bigg\},      
%     \end{split}
%     \label{eq:totyld2}
% \eea
% 
% where the first parenthesis corresponds to contribution before EWSB and the second one indicates after EWSB scenario.

%%%%%%%%%%%%%%%%%%%%%%%%%%%%%
\subsection{Large reheat temperature: $T_\text{RH}>>m$}
\label{sec:large-trh}
%%%%%%%%%%%%%%%%%%%%%%%%%%%%%

In this sub-section the reheat temperature is assumed to be much larger than the dark sector masses. We would like to mention that computing the reaction densities, 
we consider both $2\to 2$ annihilation channels and $\Phi \to X B$ decay channel before EWSB, while after EWSB we only take into 
account the $\Phi\to X Z(\gamma)$ decays for reasons elaborated soon. 
Also note that, all the SM fields are massless before EWSB but the dark sector fields ($m_\Phi,m_X$) are massive irrespective of the era, thanks 
to $U(1)_X$ breaking at very high scale. First we compare reactions densities for annihilation and the decay as a function of 
$x \equiv m_X/T$. Often we use a dimensionless variable $r=\frac{m_X}{m_\Phi}<1$ to illustrate scan results.

\begin{figure}[htb!]
$$
\includegraphics[scale=0.32]{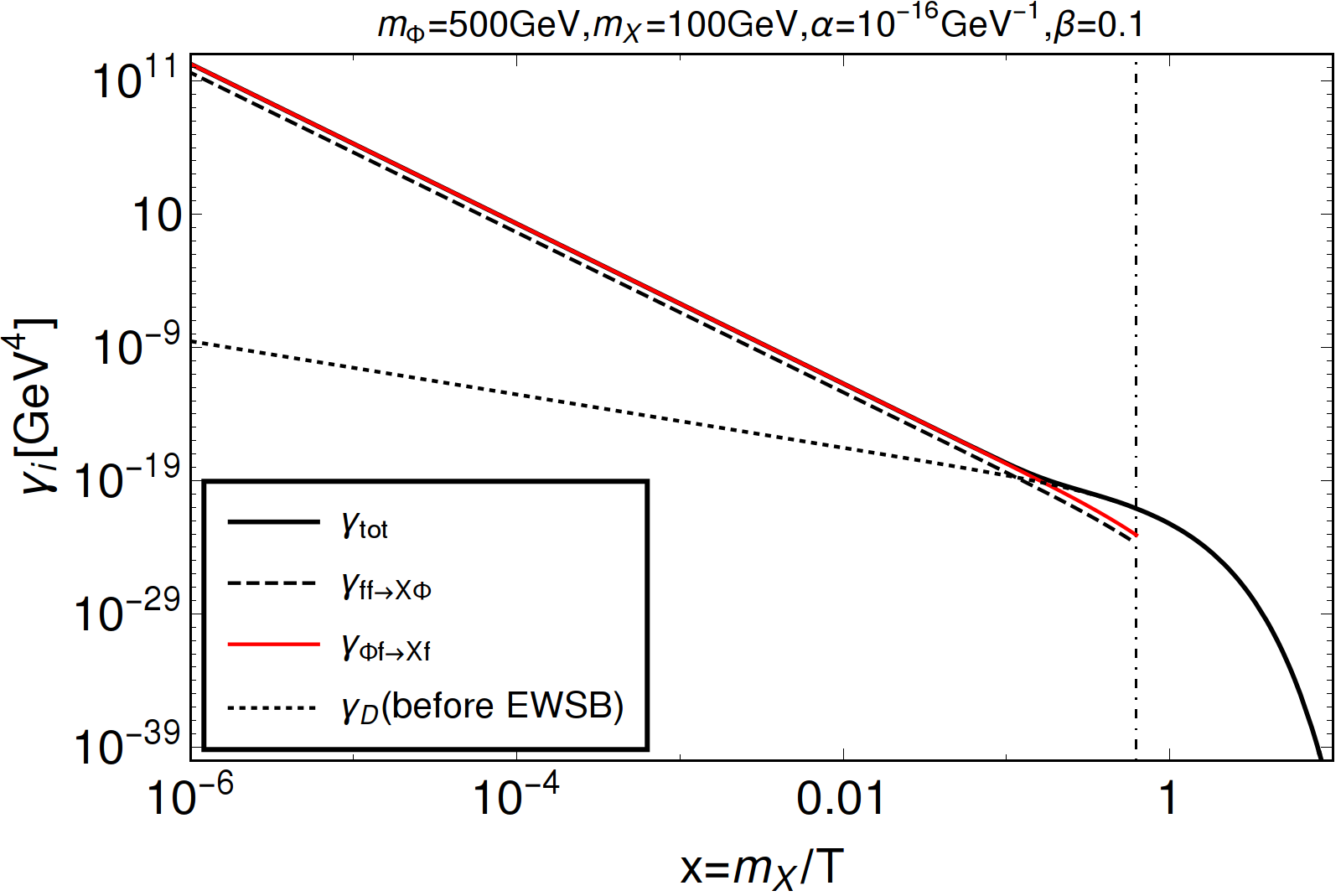}~~
\includegraphics[scale=0.32]{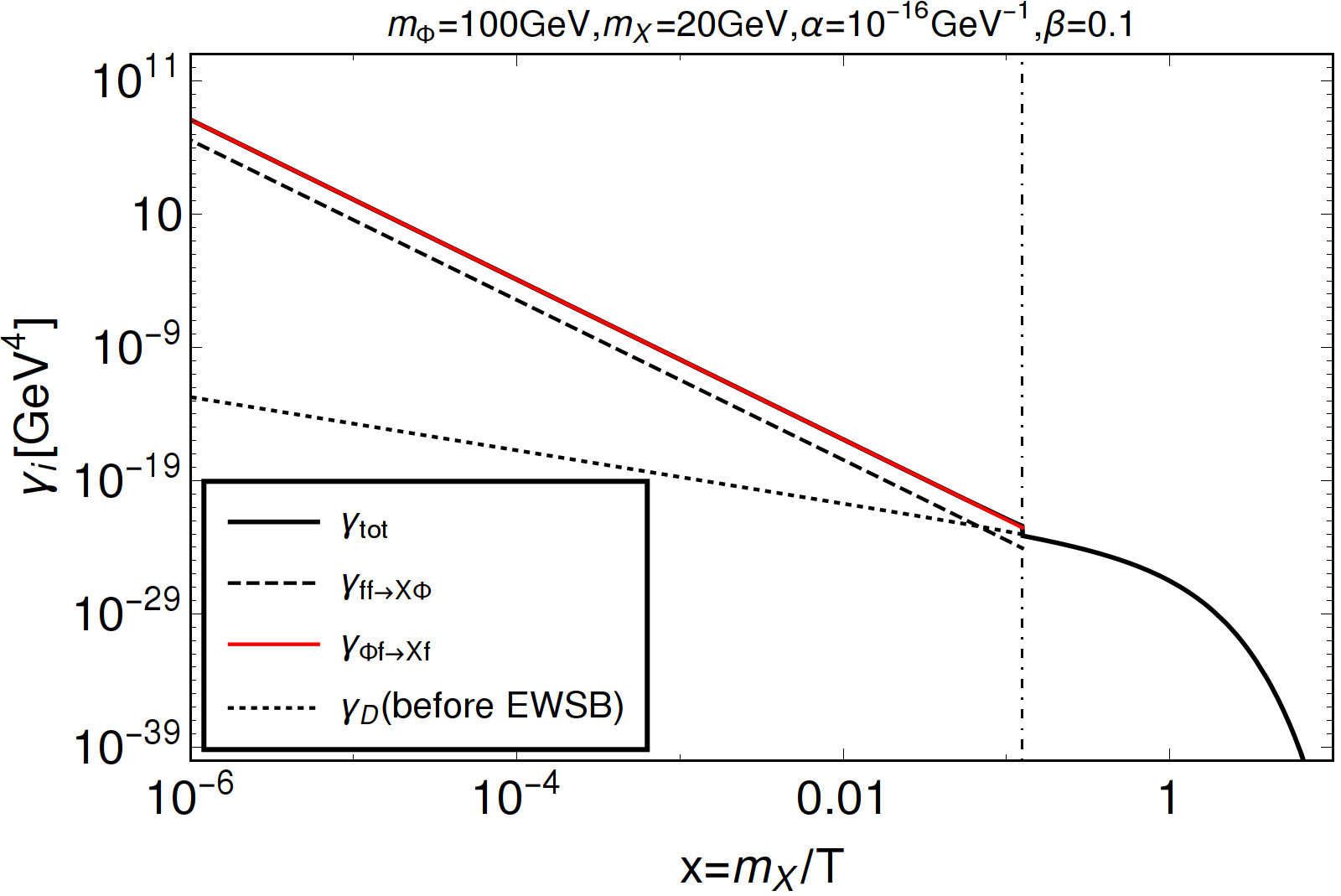}
$$
\caption{Comparison of reaction densities ($\gamma_\text{ann,D}$) for annihilation and the decay.  
Before EWSB we show the $s$-channel $\bar f f \to X \Phi$ process (black dashed) and $t$-channel $\Phi f \to X f$ (red) and 
also the decay contribution (black doted), while after EWSB we plot the density for the decay (solid black) only. 
The solid black line (hidden beneath the red line before EWSB) shows both contributions from the annihilation and the decay together.
In the left panel we choose $m_\Phi=500~\text{GeV}, ~ m_X=100~\text{GeV}$, while in the right, we choose $m_\Phi=100~\text{GeV},~m_X=20~\text{GeV}$. 
In both cases we have $\alpha=10^{-16}~\text{GeV}^{-1},\beta=0.1$. The vertical dashed-dotted lines shows $x$ corresponding to EWSB. }
\label{fig:reac-den1}
\vspace{0.2 cm}
\end{figure}

In Fig.~\ref{fig:reac-den1} we show individual contributions of annihilations and the decay to the reaction densities as a function of $x$ 
keeping the DM mass, the $\Phi$ mass, and the coupling $\alpha(\tilde{\alpha})$ fixed. The vertical dashed-dotted line represents EWSB $x_{EW}=m_X/T_{EW}$. 
Here we clearly see that the reaction densities due to the $s$-channel and $t$-channel annihilation processes are much larger than that due to decay before EWSB, 
while after EWSB the reaction density falls to a very small value. 

The suppression of the decay before EWSB can be understood comparing the analytical forms of reaction densities for annihilation (in massless limit) and decay as follows:
\beq
q\equiv\frac{\gamma_\text{ann}}{\gamma_\text{D}} 
\sim \frac{\int_0^\infty dss^{3/2}\sigma_{ann}(s)K_1\left(\sqrt{s}/T\right)}{\alpha^2(1+\beta^2) m_\Phi^5 K_1\left(m_\Phi/T\right)}
\sim \left(\frac{T}{m_\Phi}\right)^5 \frac{1}{K_1\left(m_\Phi/T\right)},  
\eeq
where $\sigma_{ann}(s)\propto \alpha^2(1+\beta^2)\times \text{const.}$ was assumed. It then follows that for 
$T >> m_\Phi \implies q \gg 1$ and for $T \sim m_\Phi \implies q \sim 1$.
We however, would like to caution the reader that above formula is not strictly valid at $T$ close to EWSB when massive kinematics 
become important for computing annihilation cross-sections as elaborated in Sec.~\ref{sec:uv-limits}. 

\begin{figure}[htb!]
$$
\includegraphics[scale=0.32]{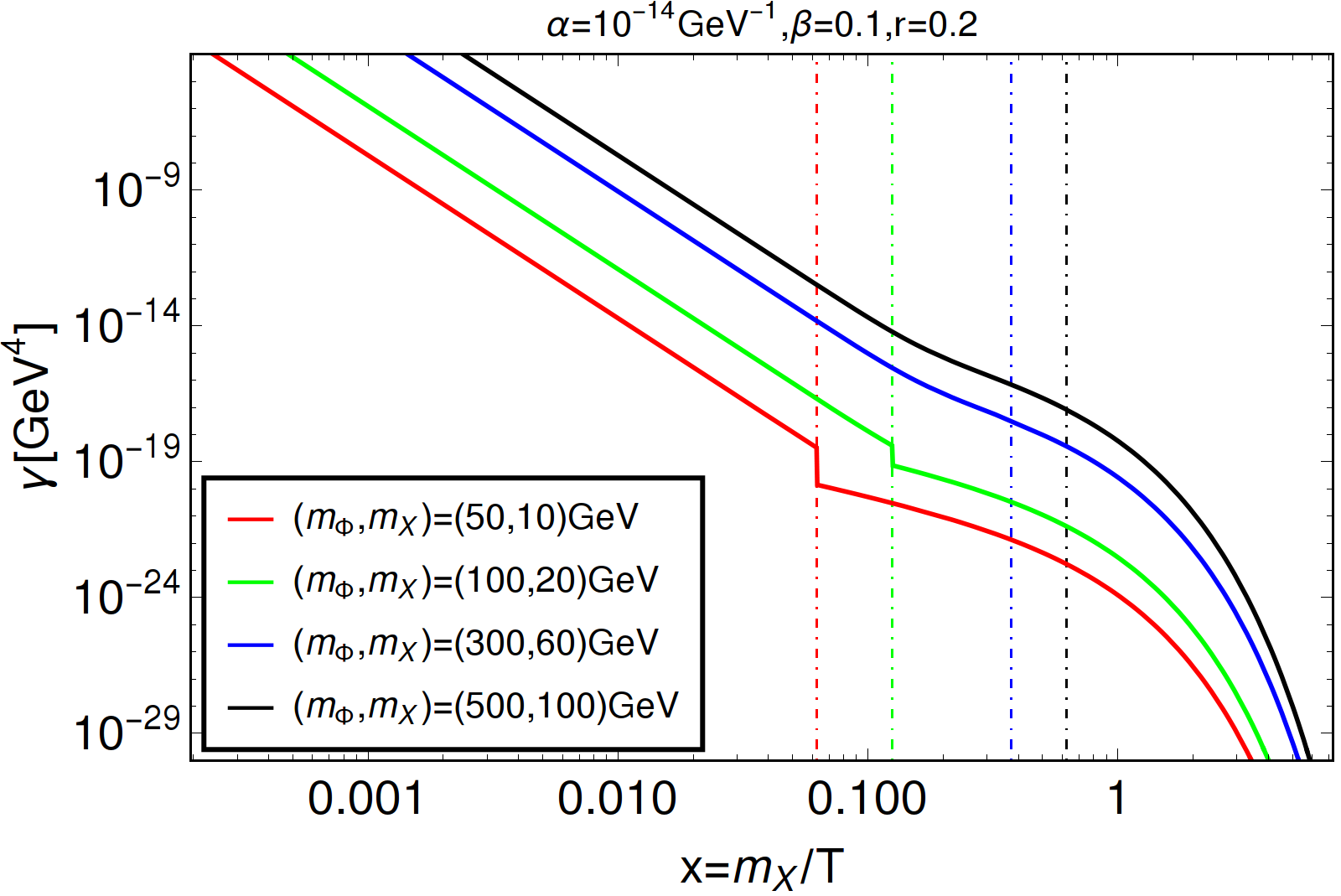}~~
\includegraphics[scale=0.32]{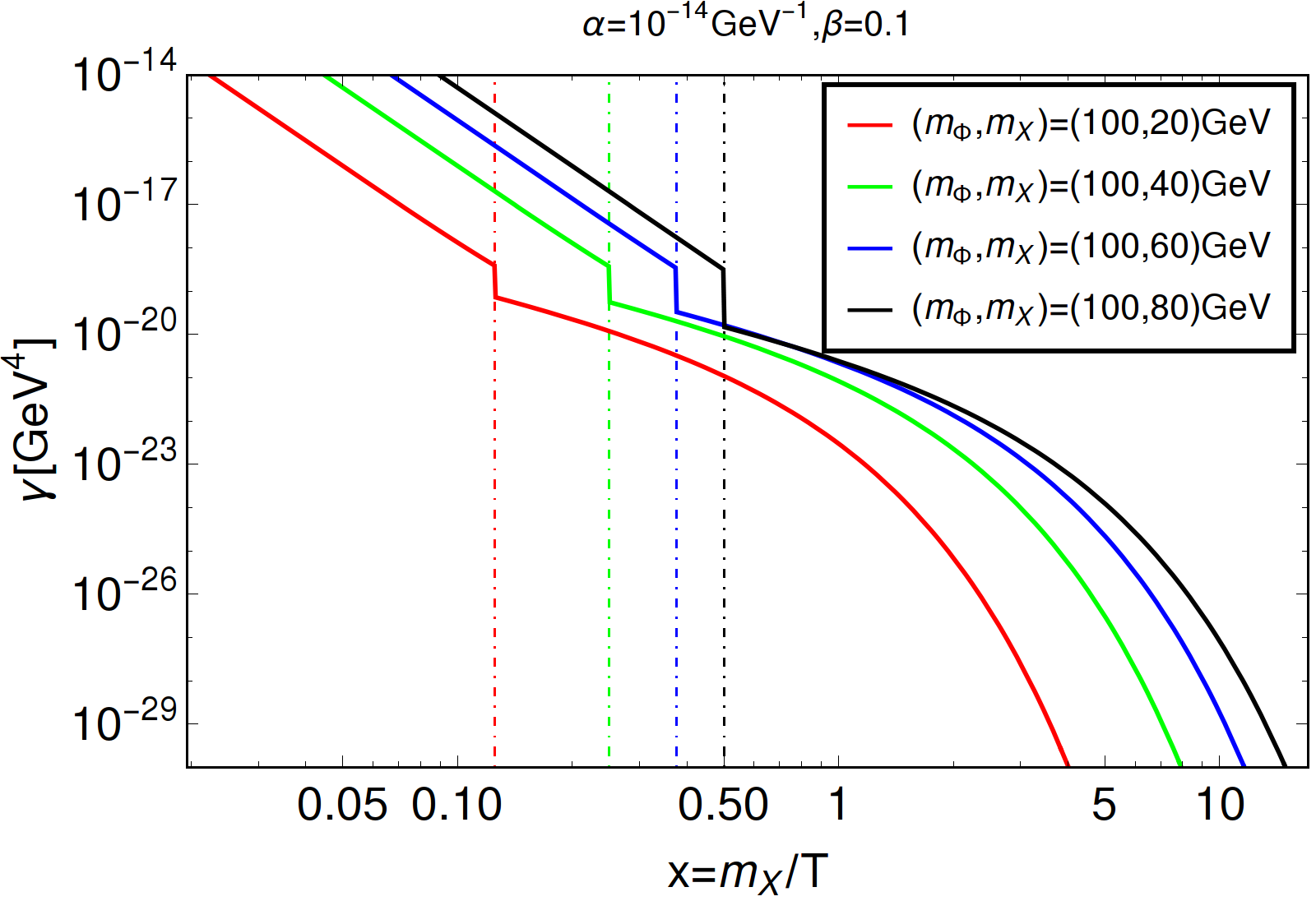}
$$
\caption{Total reaction density over the whole range of temperature before and after EWSB taking into account both annihilation 
and decay before EWSB, but only decay after EWSB. We consider all SM states to be massless before EWSB, 
while dark sector particles are massive. The dashed-dotted vertical lines correspond to $x_{EW}=\frac{m_X}{T_{EW}}$ for each choice of DM mass. 
In the left panel we show densities for different choices of $m_\Phi=\{50,100,300,500\}~\rm GeV$ (in red, green, blue, black, respectively) 
for fixed $r=\frac{m_X}{m_\Phi}=0.2$, while in the right one they are shown for different $r=\{0.2,0.4,0.6,0.8\}$ 
(in red, green, blue, black, respectively) for fixed $m_\Phi=100~\rm GeV$. In both cases we choose $\alpha=10^{-14}~\text{GeV}^{-1},\beta=0.1$.}
\label{fig:reac-den2}
\vspace{0.2 cm}
\end{figure}

This implies that the DM is dominantly produced before EWSB, while the yield accumulated after EWSB is negligibly small. 
This is a typical feature of UV freeze-in where the maximum yield production happens at high temperature. 
An evolution of the total reaction density taking into account all annihilation processes occurring before EWSB together with the decay after EWSB for 
different sets of $\{m_\Phi,m_X\}$ is shown in Fig.~\ref{fig:reac-den2}. It is clear from the discussion above that decay contribution before EWSB is very small, while 
the contribution from decay is dominant over annihilation processes after EWSB. Note here that the vertical 
dashed-dotted lines in different colors represent EWSB at $x_{EW}=m_X/T_{EW}$ corresponding to those $m_X$ values chosen for illustration.

Also note that for $m_\Phi \lesssim 100~\rm GeV$, the reaction densities in two regimes before and after EWSB can be distinctively identified 
by a step, while that for larger $m_\Phi$ they are continuous. This is because with temperature dropping, the $2 \to 2$ production cross-sections drops 
and around $T_{EW}$ the decay contribution dominates over them, resulting a continuous curve before and after EWSB. However, after EWSB, the decay final state 
changes from $\Phi \to X,B$ to $\Phi \to X, \gamma (Z)$. Therefore when $m_\Phi<m_X+m_Z$, one of the decay processes, $\Phi \to X, Z$, become kinematically 
forbidden showing a distinct drop in the reaction density. The values of different parameters chosen for illustration are mentioned in figure inset.

\begin{figure}[htb!]
$$
\includegraphics[scale=0.32]{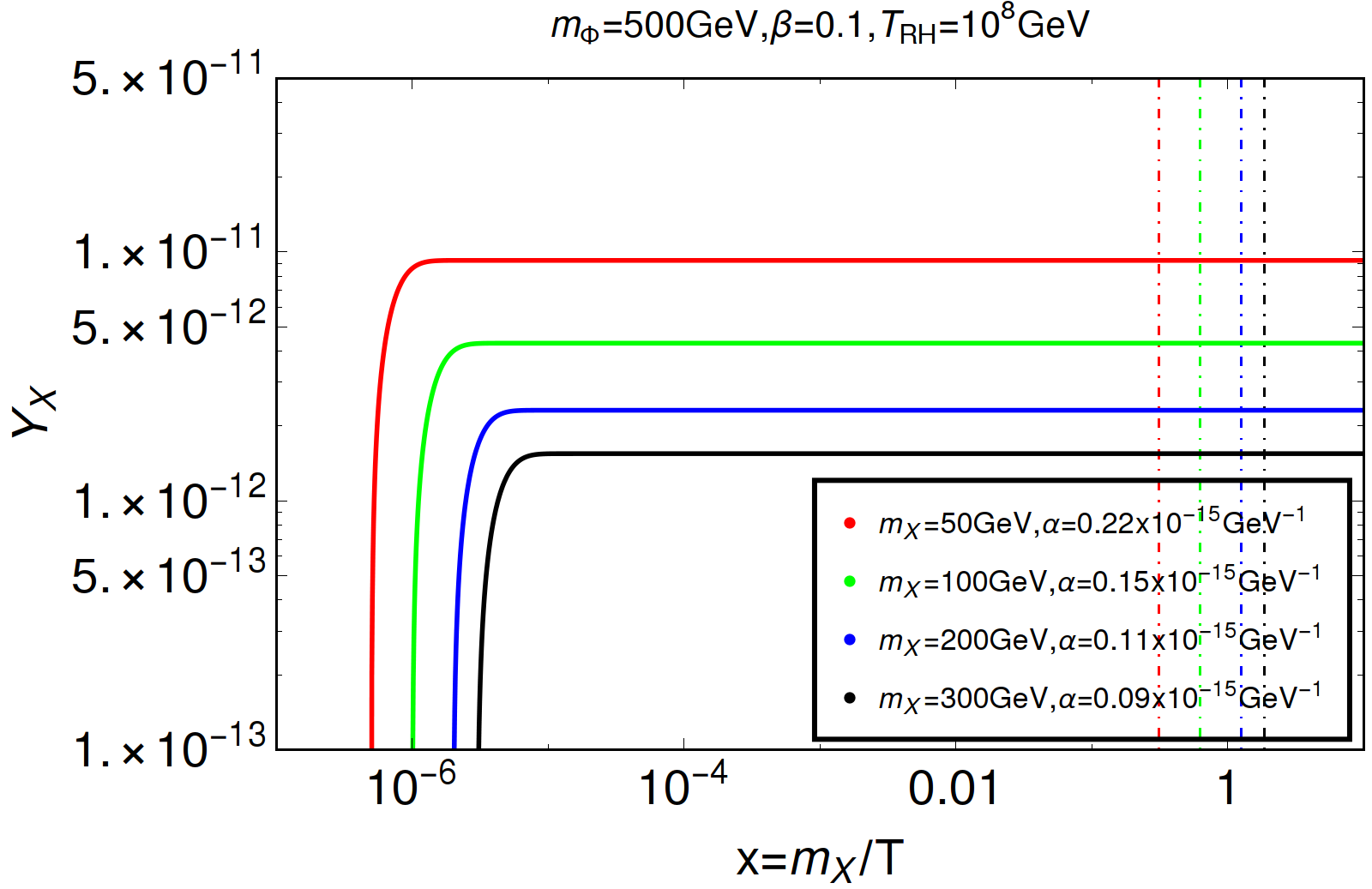}~~
\includegraphics[scale=0.32]{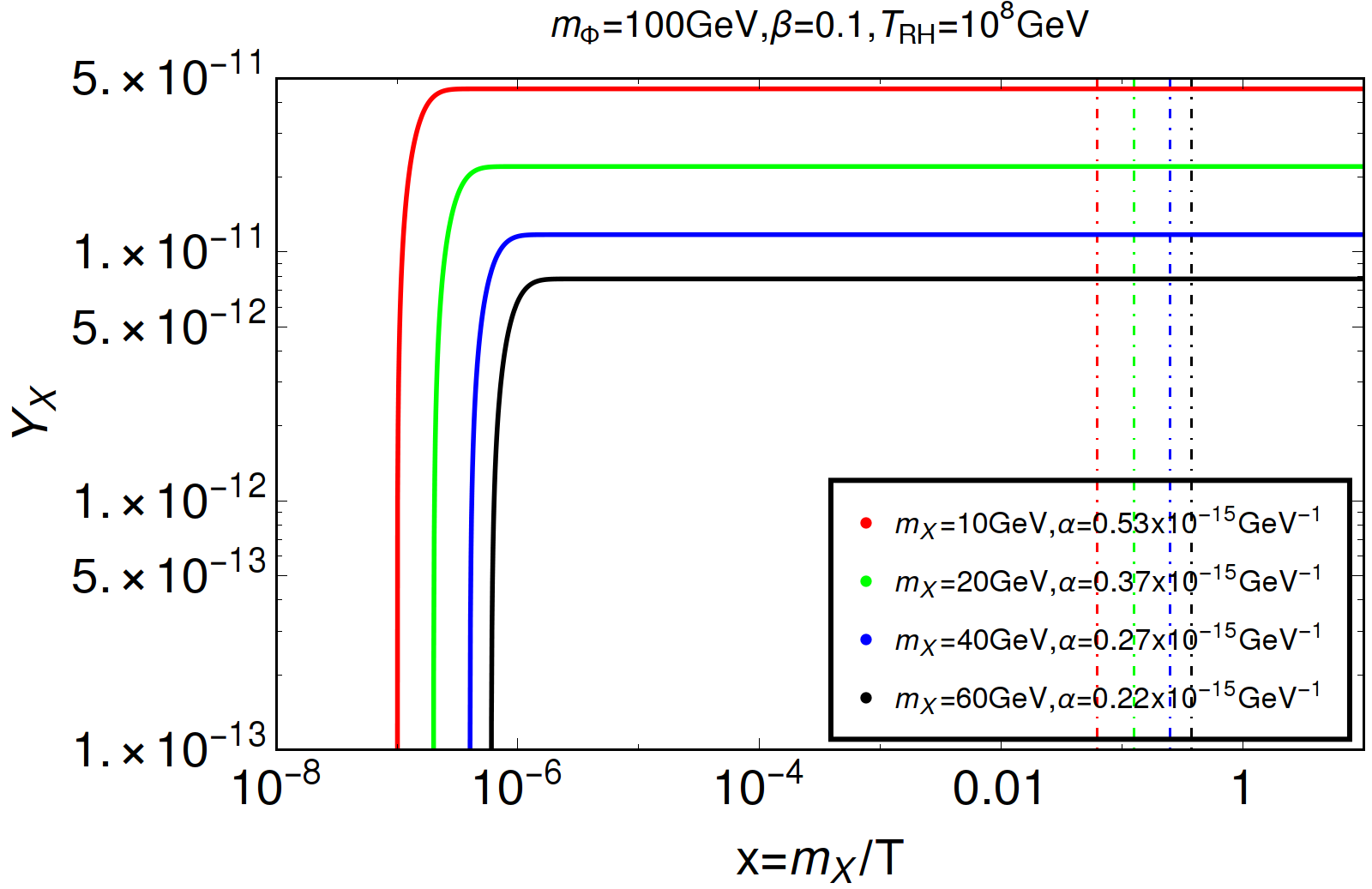}
$$
$$
\includegraphics[scale=0.3]{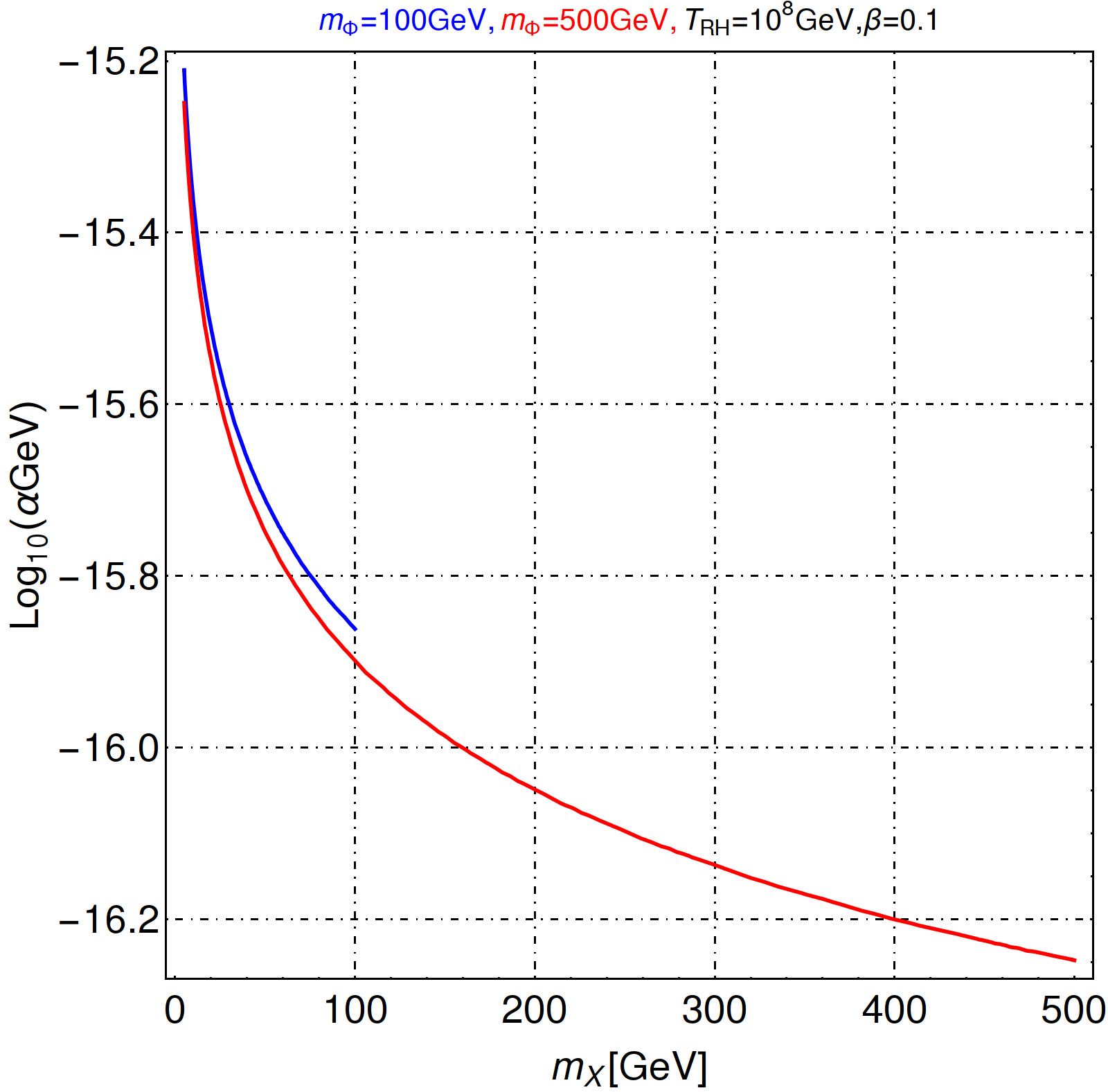}~~
\includegraphics[scale=0.3]{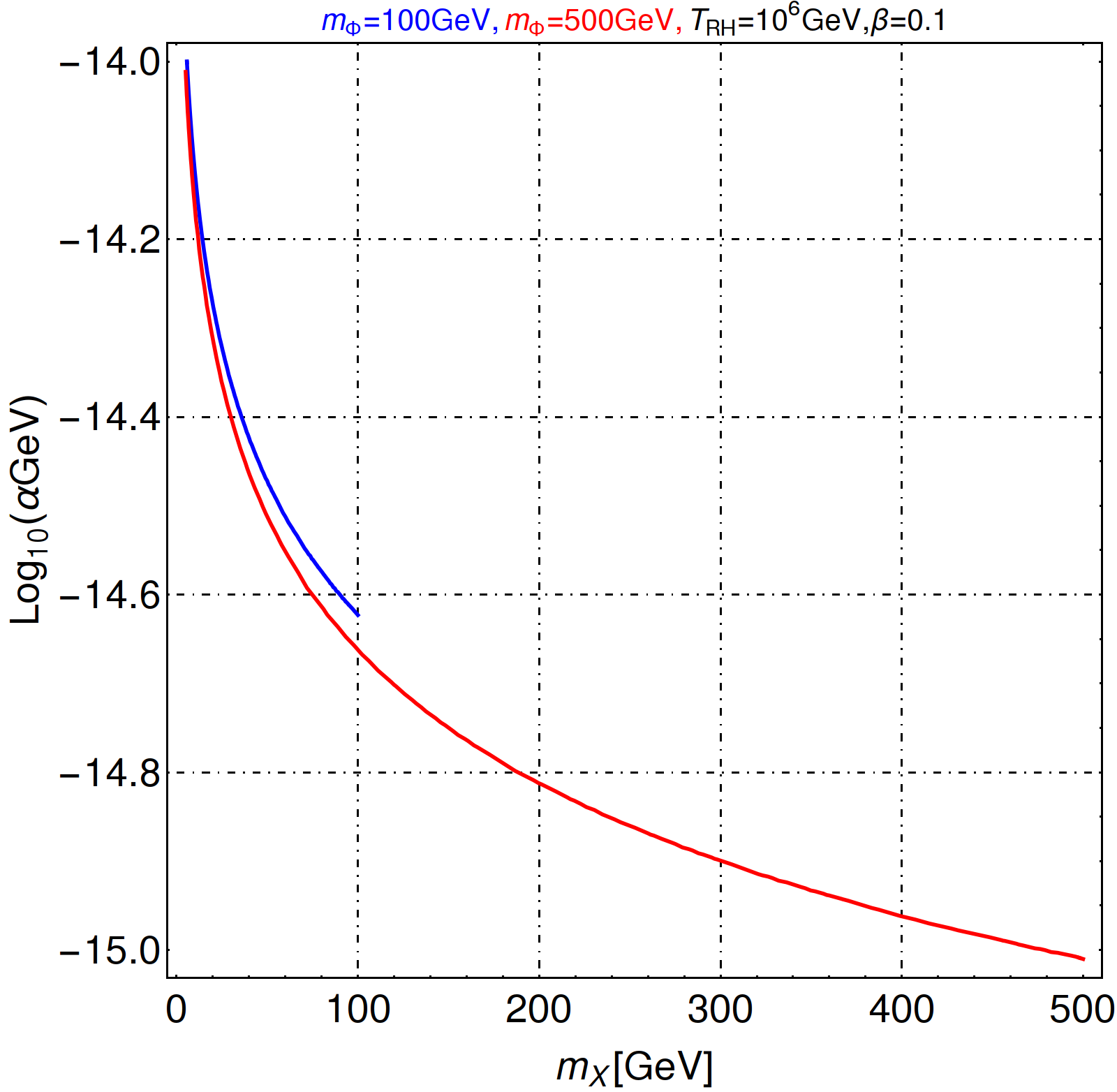}
$$
\caption{Top left: evolution of the DM yield $Y_X$ for parameters which imply correct relic DM abundance as a function of $x=m_X/T$ obtained by solving Eq.~\eqref{eq:beq-x}. Here different colored curves indicate different choices of $r=\frac{m_X}{m_\Phi}=\{0.1,0.2,0.4,0.6\}$ in red, green, blue and black respectively for a fixed $m_\Phi=100~\text{GeV},\alpha=10^{-16}~\text{GeV}^{-1},\beta=0.1$ and reheat temperature $T_\text{RH}=10^8~\rm GeV$. The vertical dashed-dotted lines correspond to $x_\text{EW}=m_X/T_\text{EW}$ for different values of $m_X$. Top right: same as top left but with $m_\Phi=500~\rm GeV$. 
Bottom left:  relic density allowed parameter space in terms of $m_X-\alpha$ where different colored contours correspond to  
$m_\Phi=500~\rm GeV$ (red) and $m_\Phi=100~\rm GeV$ (blue) for a fixed $\beta=\frac{\tilde{\alpha}}{\alpha}=0.1$ and reheat temperature 
$T_\text{RH}=10^8~\rm GeV$. Bottom right: same as bottom left but with $T_\text{RH}=10^6~\rm GeV$.}
\label{fig:rel-highrh}
\vspace{0.2 cm}
\end{figure}
 
The effect of large reheat temperature and large reaction densities at high temperature is reflected in the DM yield shown 
in the top panels of Fig.~\ref{fig:rel-highrh}. In both scans we keep $T_{RH}=10^8$ GeV. 
In the top panel of Fig.~\ref{fig:rel-highrh} we see that the DM yield builds up quickly with $x$ (i.e. with lowering temperature) 
and reaches its maximal value already at very high temperature $T \sim T_\text{RH}$. Then it freezes-in immediately producing an yield that remains constant till $T=T_0\simeq 2.73~\rm K$.
The asymptotic value of the yield, $Y_0\equiv Y(T_0)$, directly implies the PLANCK observed DM relic abundance via (\ref{eq:relicX}). 
As seen from (\ref{eq:relicX}) each choice of the DM mass requires appropriate $Y_0$ what results
in the splitting of the colored curves at large $x$ observed in Fig.~\ref{fig:rel-highrh}. On the other hand each $Y_0$ 
requires the couplings $\alpha, \tilde\alpha$ tuned appropriately, as shown in the legend of Fig.~\ref{fig:rel-highrh}.
The left top panel corresponds to $m_\Phi=500$ GeV, while the top right panel to 
slightly smaller value $m_\Phi=100$ GeV. The vertical dashed lines show the locations of EWSB, although its effect on the final yield is invisible. 
In the bottom panels of Fig.~\ref{fig:rel-highrh} we show contours corresponding to the central value of the PLANCK observed relic abundance 
($\Omega h^2\simeq 0.1199$) in the $\alpha-m_X$ plane for fixed $m_\Phi=500\gev$ and $100\gev$ and two different values of $\trh$. 
The left and right lower panels correspond to the reheat temperature $\trh=10^8$ and $10^6\gev$, respectively.
The relic abundance is obtained following Eq.~\eqref{eq:relicX}. 
Since $\Omega_X \propto m_X$, we see for larger DM mass smaller $\alpha$ is required to compensate for the over abundance. Note, that in each panel 
the kinematical condition $m_\Phi\gtrsim m_X$ is obeyed. 
As expected, for the same DM mass, growing $\trh$ requires lower couplings. We would finally note that to find yield in such a scenario, the masses in all 
reactions can be safely neglected and the processes after EWSB contributes negligibly.

\begin{figure}[htb!]
$$
\includegraphics[scale=0.35]{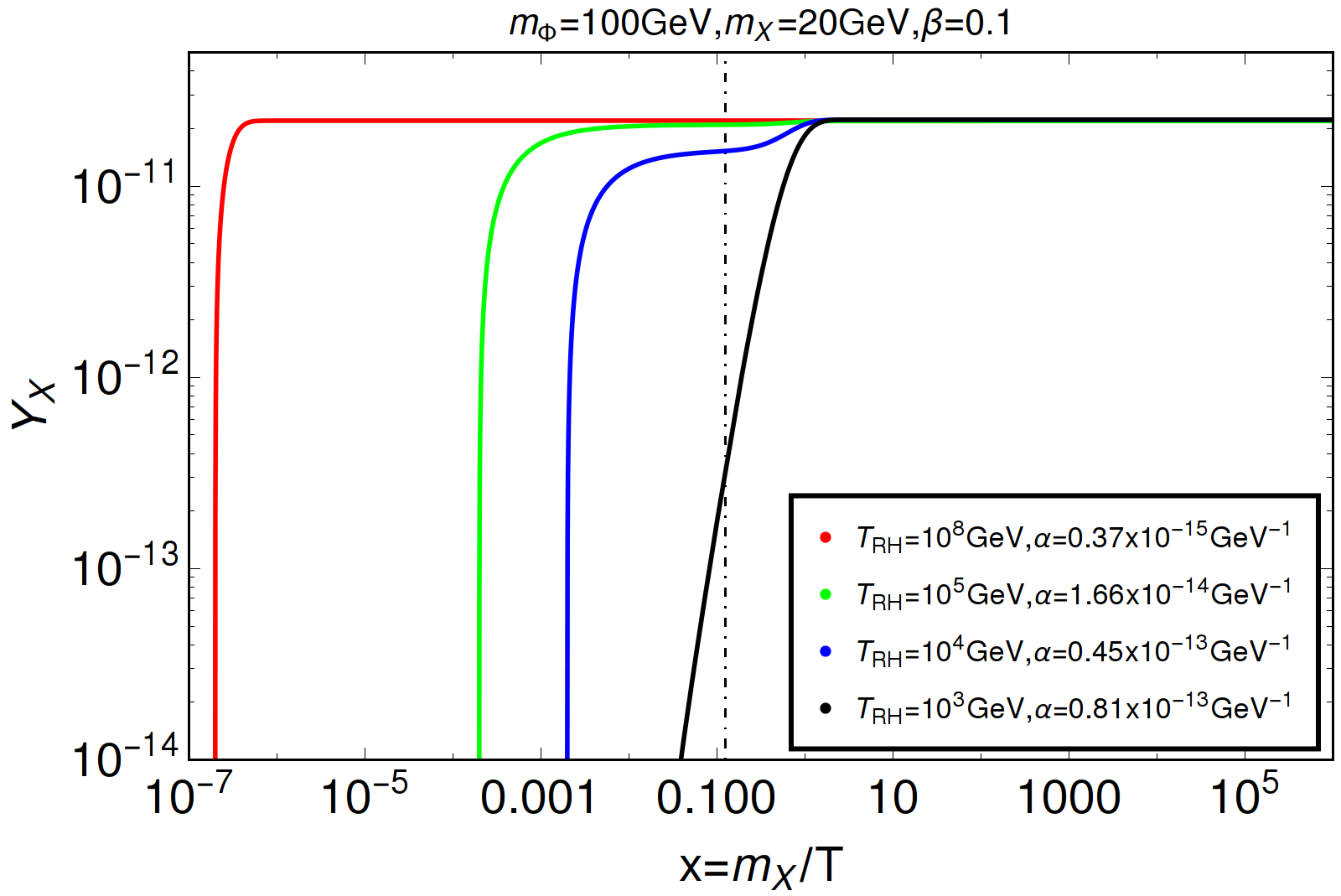}~~
\includegraphics[scale=0.35]{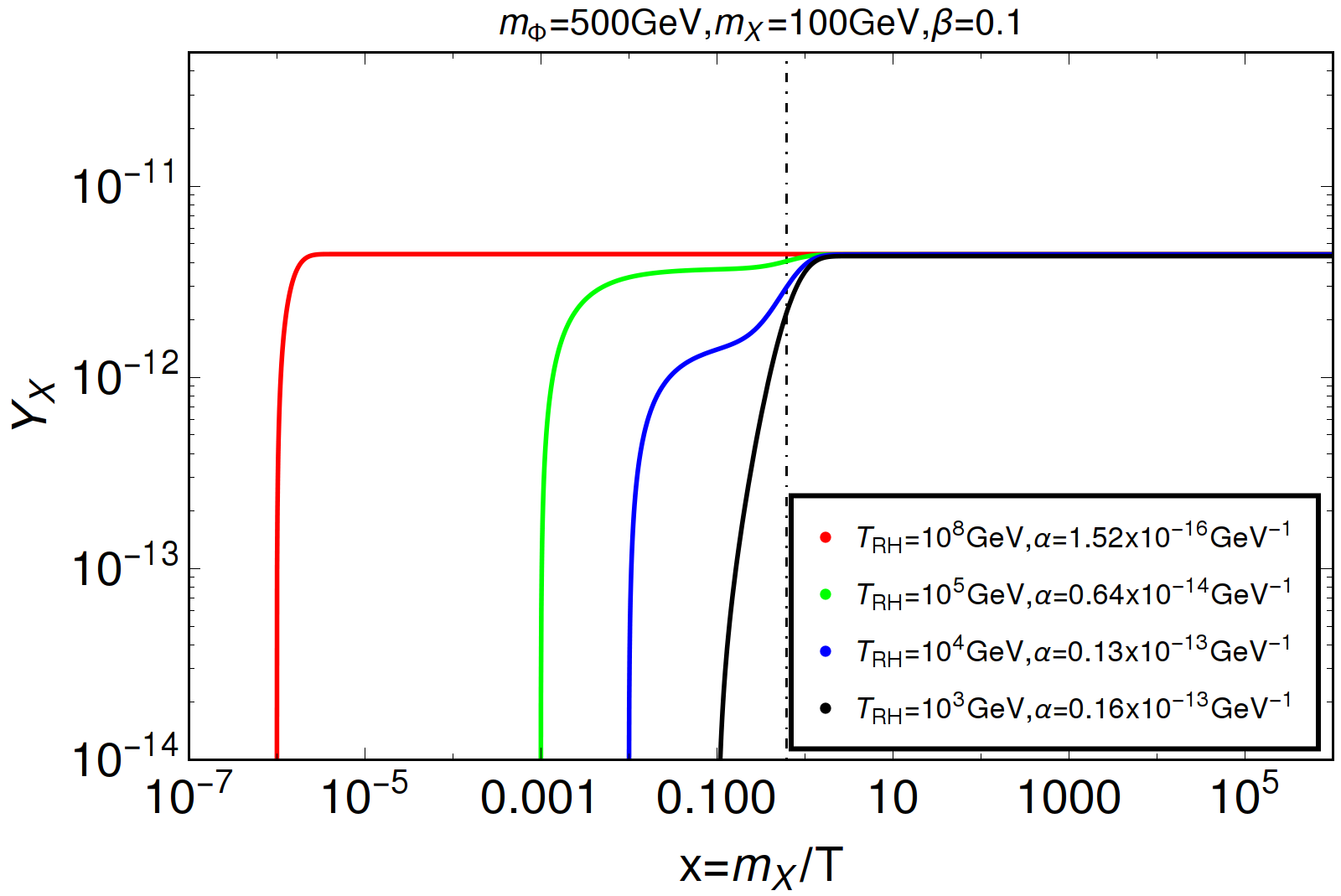}
$$
$$
\includegraphics[scale=0.42]{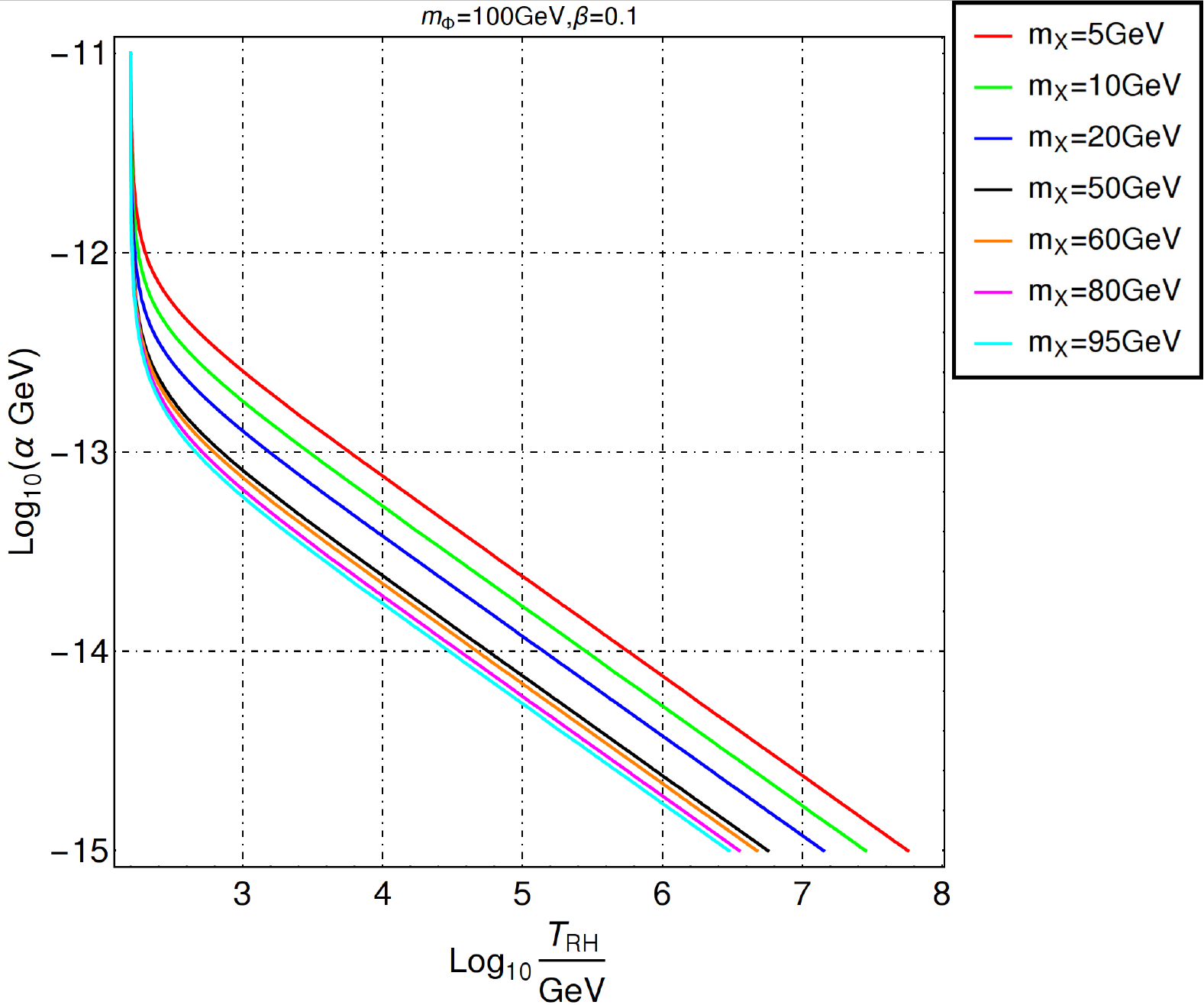}~~~
\includegraphics[scale=0.42]{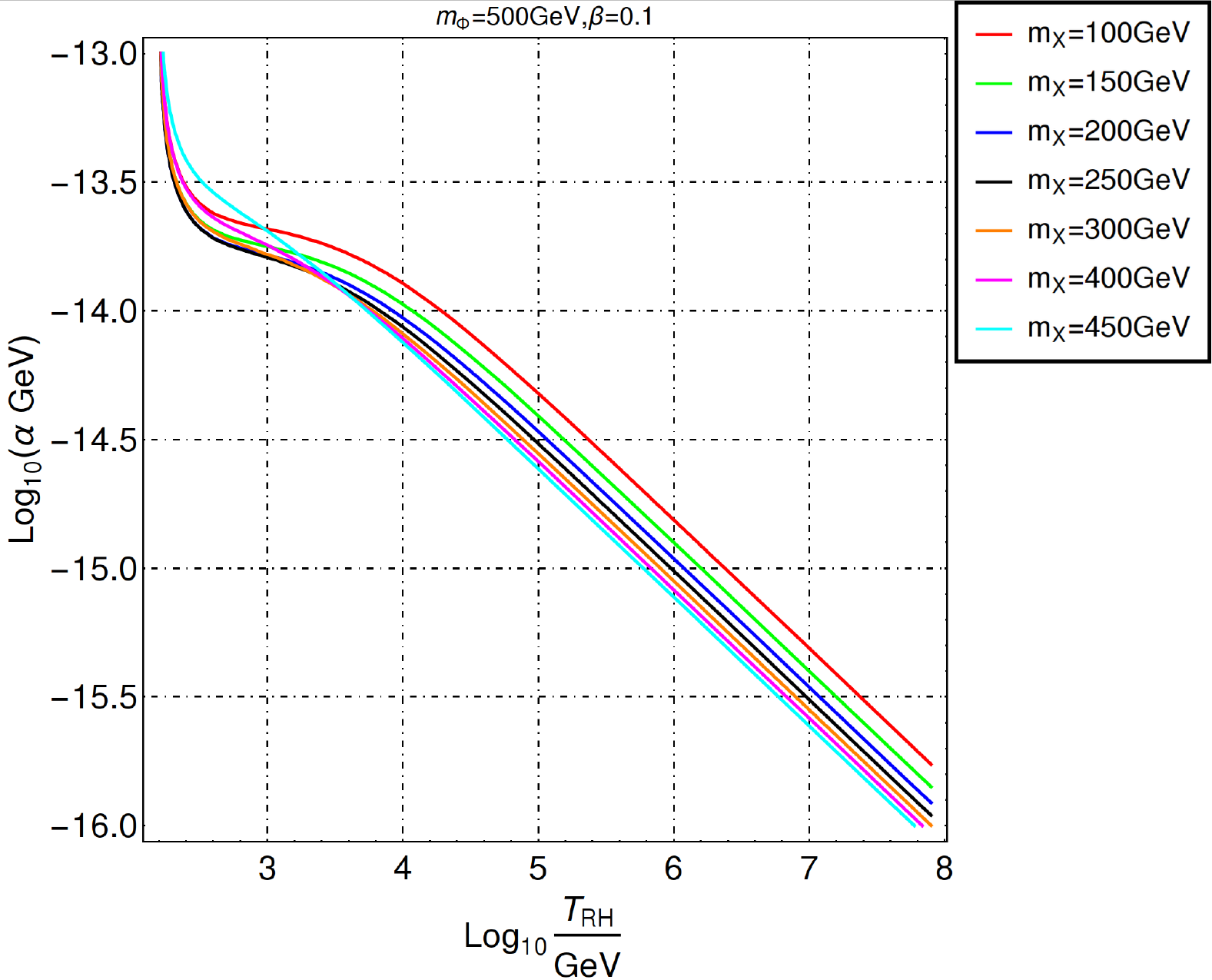}
$$
$$
\includegraphics[scale=0.42]{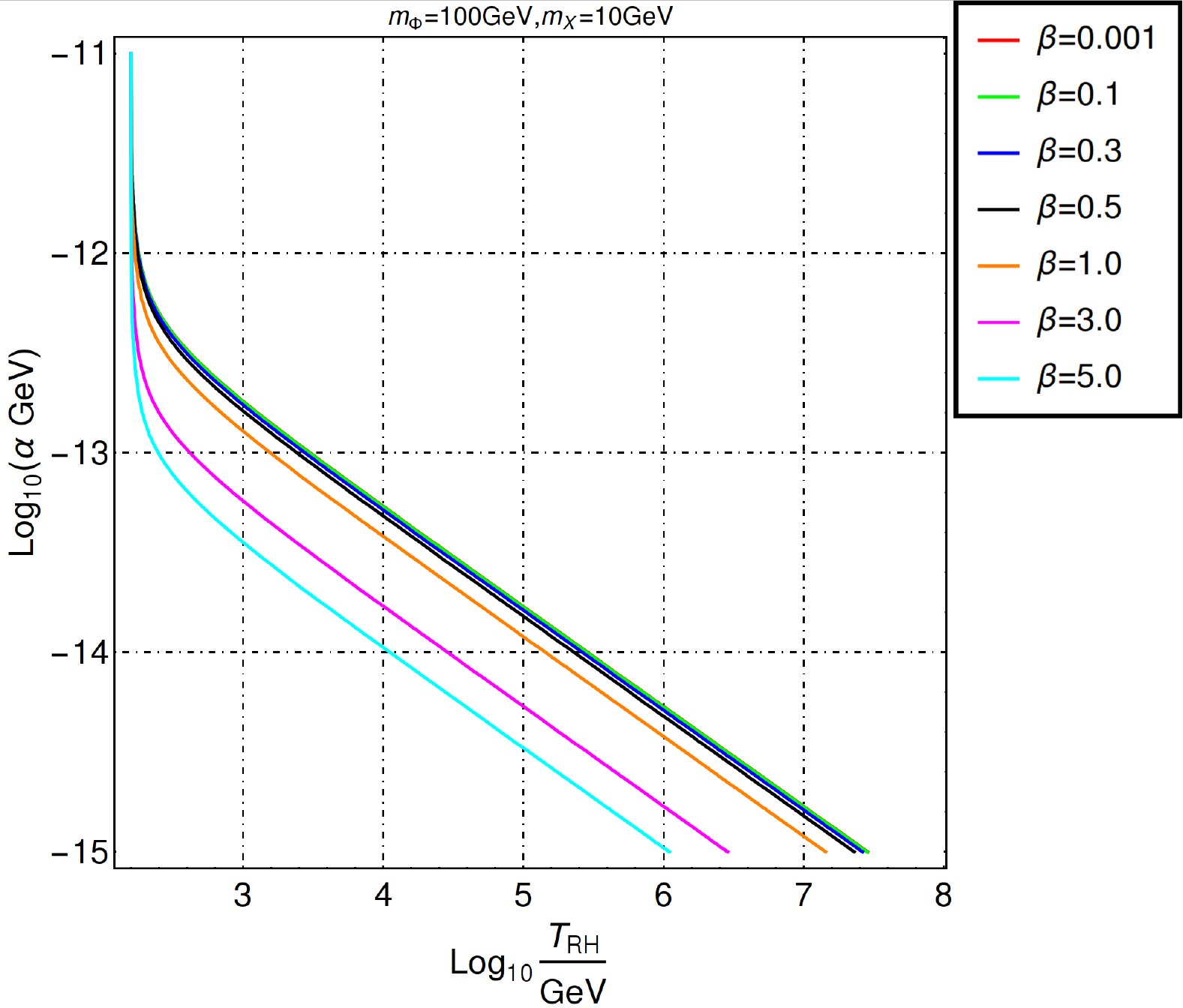}~~~
\includegraphics[scale=0.42]{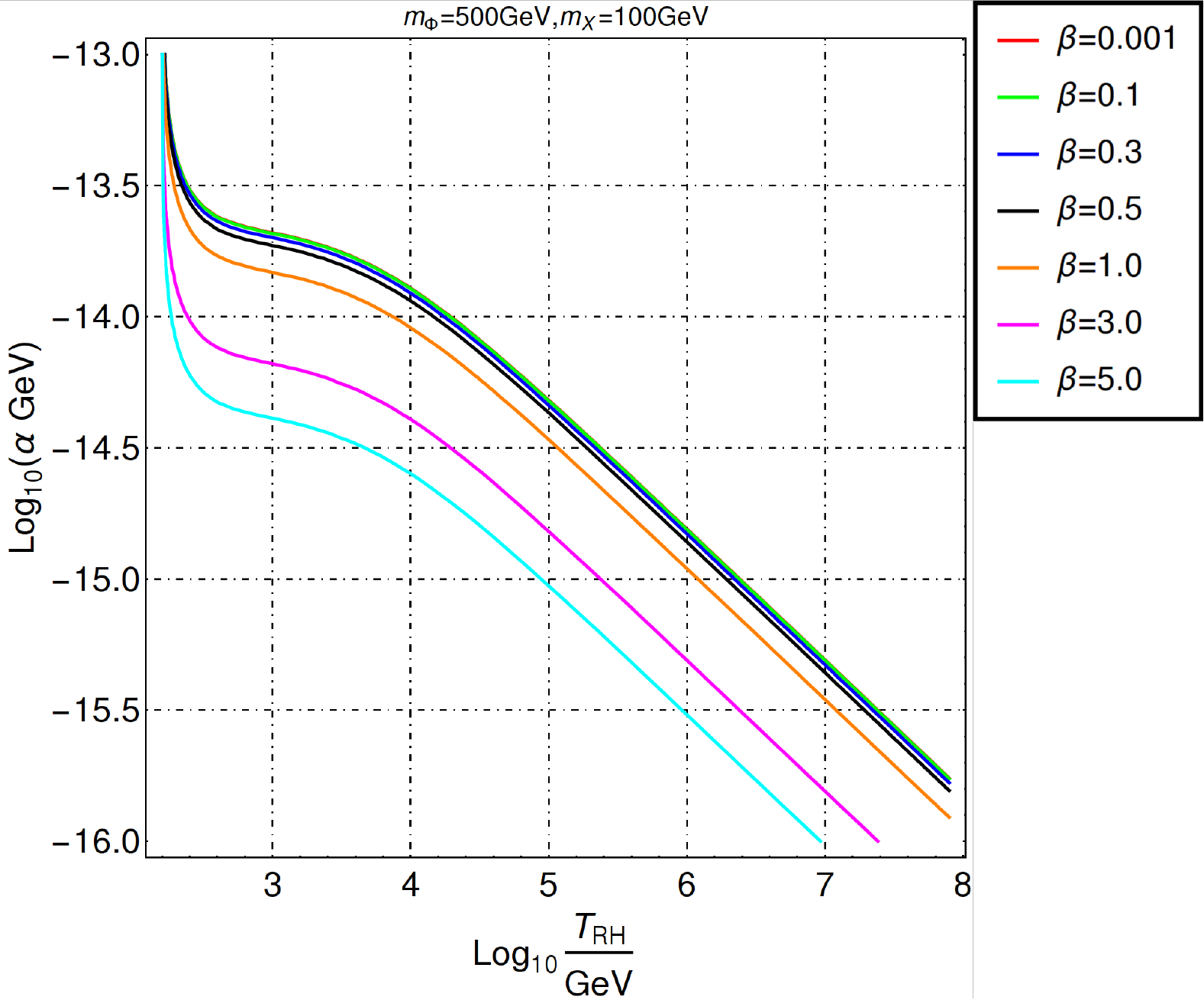}
$$
\caption{Top panel: DM yield as a function of temperature for different choices of the reheat temperature $T_\text{RH}:\{10^5,10^2,10,1\}~\rm TeV$ 
shown in red, green, blue and black curves respectively. Here we have considered both 
annihilation and decay channels before EWSB where SM particles are massless while the dark particles are massive. In the post-EWSB scenario we have considered 
only the decays with massive states. The parameters chosen correspond to right relic abundance. In each of these plots the vertical dashed-dotted lines denote
EWSB ($x_{EW}$). Middle left: Relic density allowed parameter space in terms of $\alpha-T_\text{RH}$ where different colored contours correspond to 
different $m_X$ for a fixed $m_\Phi=100~\rm GeV$. Middle right: Same as middle left panel, but for a different fixed $m_\Phi=500~\rm GeV$. Bottom left: Same as middle panel but for different choices of $\beta:\{0.001,0.1,0.3,0.5,1.0,3.0,5.0\}$ for $m_\Phi=100~\text{GeV}$ and $m_X=10~\text{GeV}$. Bottom right: Same as bottom left with $m_\Phi=500~\text{GeV}$ and $m_X=100~\text{GeV}$}
\label{fig:yld-trh}
\vspace{0.2 cm}
\end{figure}

DM production via annihilation or decay processes can also be compared (at $T_0\sim 0\gev$) by naive dimensional analysis as advocated in 
\cite{Hall:2009bx, Cheung:2010gj}: 
\beq
\frac{Y_X^{ann}}{Y_X^\text{D}}\sim \frac{\sigma M_{pl} T_{FI}}{\Gamma_\Phi M_{pl}/T_{FI}^2}\sim\frac{\alpha^2 M_{pl} T_{RH}}{\left(\alpha^2 m_\Phi^3\right) M_{pl}/m_\Phi^2}
\sim\frac{T_{RH}}{m_\Phi}, 
\label{eq:compare}
\eeq
where $T_{FI}$ denotes the characteristic freeze-in temperature scale at which the yield reaches the constant value (see e.g the plateau in upper panels of Fig.~\ref{fig:rel-highrh}) 
for DM production via annihilation process or decay, which are not quite the same. 
For decays, the freeze-in temperature can be assumed to be the mass of the decaying particle, i.e. $T_{FI}\sim m_\Phi$, which is used in the second step of the above analysis. On the other hand, for DM yield produced via annihilation process, the freeze-in temperature can be assumed to be the highest temperature available for the process, i.e. 
$T_{FI}\sim T_{RH}$. Therefore, for $T_{RH}>>m_\Phi$ ensures that annihilation contribution dominates over decay contribution in the final yield for UV limit. This brings us to an asymptotic 
formula of the yield in UV limit, which can simply be written as:
\bea
Y_X(x=0) \sim \alpha^2 M_{pl} T_{RH},
\eea
neglecting the decay contribution and is validated in Fig.~\ref{fig:yld-trh}, as we explain in a moment.

% as argued in Sec.~\ref{sec:uv-limits} to match the predictions of UV freeze-in.

% Due to the large reheat temperature the major part of the DM production takes place before EWSB, while the contribution from the processes after EWSB is negligible and can be ignored (as $T_{\text{EW}}<<T_\text{RH}$). Now, all the SM particles have zero mass above the EW transition temperature $T_{\text{EW}}\simeq 160~\rm GeV$ and only the dark sector particles $\{\Phi,X\}$ are considered to be massive. The expressions for spin-averaged squared amplitudes before EWSB are given in Appendix.~\ref{sec:app-amp-bewsb}. For the freeze-in yield, we consider both scattering and decay before EWSB, while after EWSB we only consider decay. However, the after EWSB contribution, as mentioned above, is negligible and can be overlooked. In Fig.~\ref{fig:reac-den1} we show the variation of $\gamma$'s due to annihilation and decay before and after the EWSB with $x=m_X/T$ following Eq.~\eqref{eq:beq-x}. In Fig.~\ref{fig:rel-highrh} we show the yield corresponding to the parameters that give rise to observed DM relic abundance. Here we see that the yield immediately freezes in at $T\sim T_\text{RH}$, which shows the UV nature of this scenario.  

%%%%%%%%%%%%%%%%%%%%%%%%%%%%%
\subsection{Low reheat temperature: $T_\text{RH}\gtrsim m$}
\label{sec:low-trh}
%%%%%%%%%%%%%%%%%%%%%%%%%%%%%

This case is more interesting. It turns out that when the reheat temperature drops down to $\sim\text{TeV}$ scale, then processes that take place after EWSB are relevant and contribute significantly to final yield.  
After EWSB all the SM states also become massive, and hence in order to get meaningful results, all masses shall be kept.
As we shall see, in such a case the IR freeze-in starts showing up, i.e. the freeze-in takes place at a temperature $T\sim m_X$.

\begin{figure}[htb!]
 $$
  \includegraphics[scale=0.3]{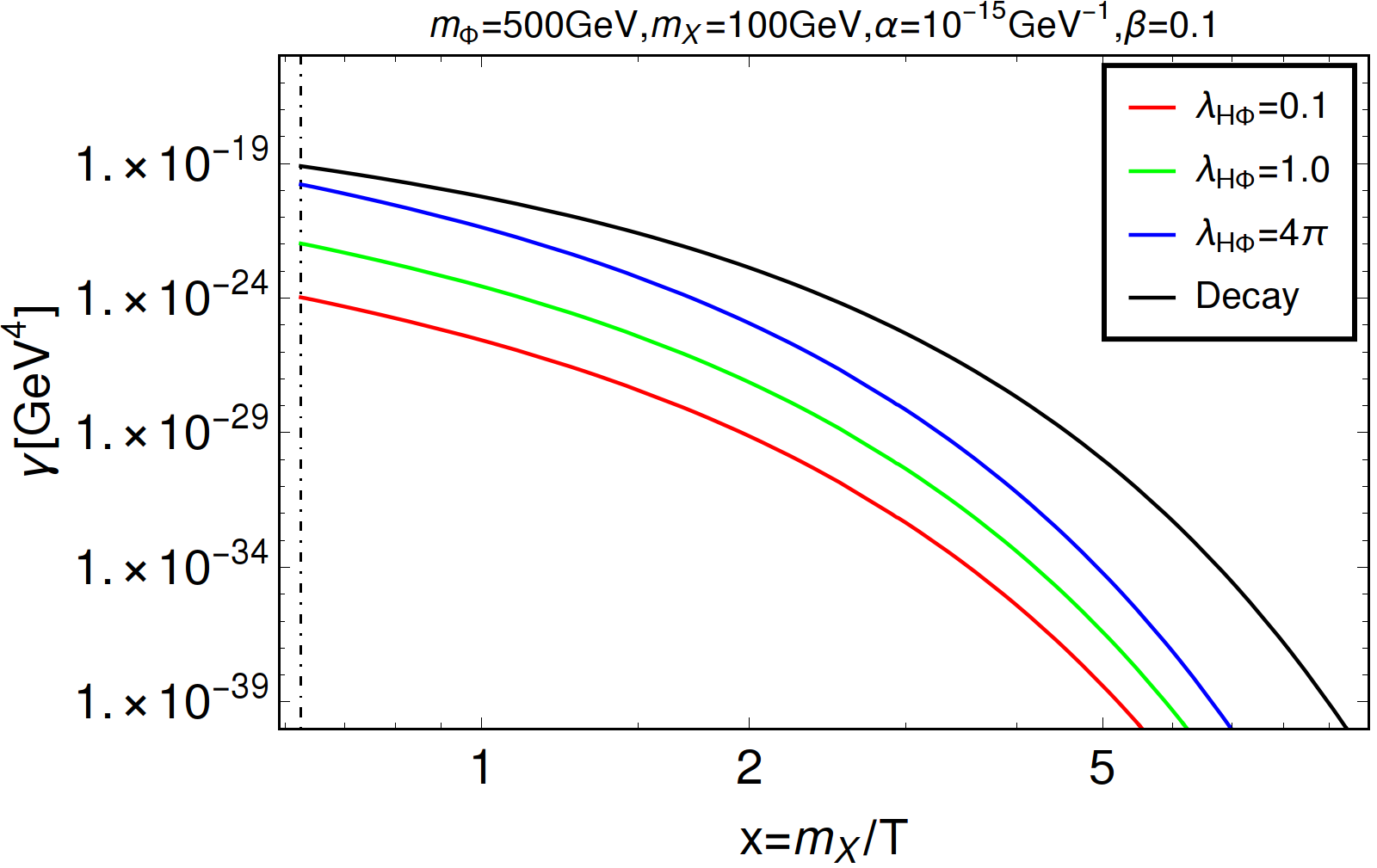}~~
  \includegraphics[scale=0.3]{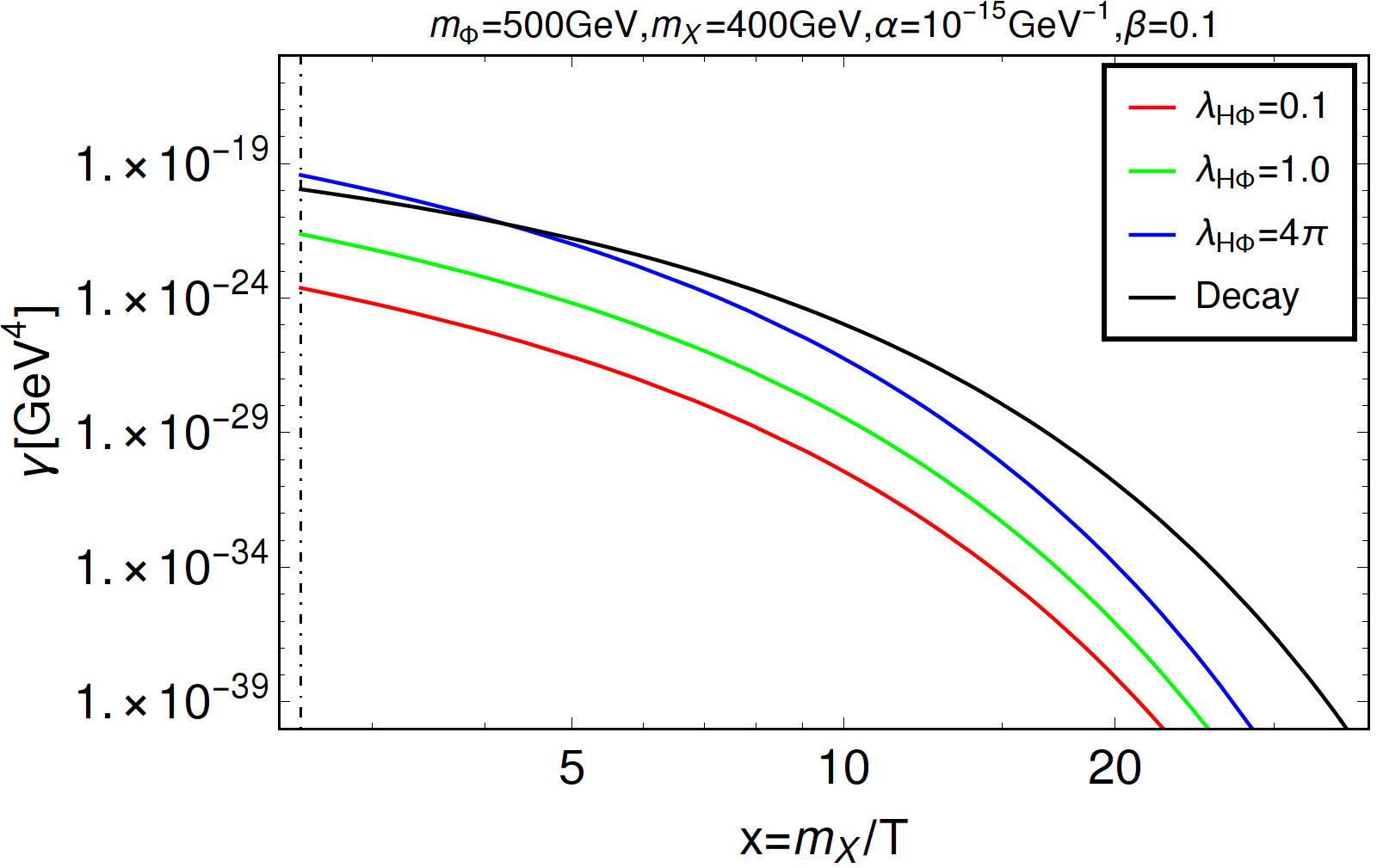}
 $$
 $$
 \includegraphics[scale=0.3]{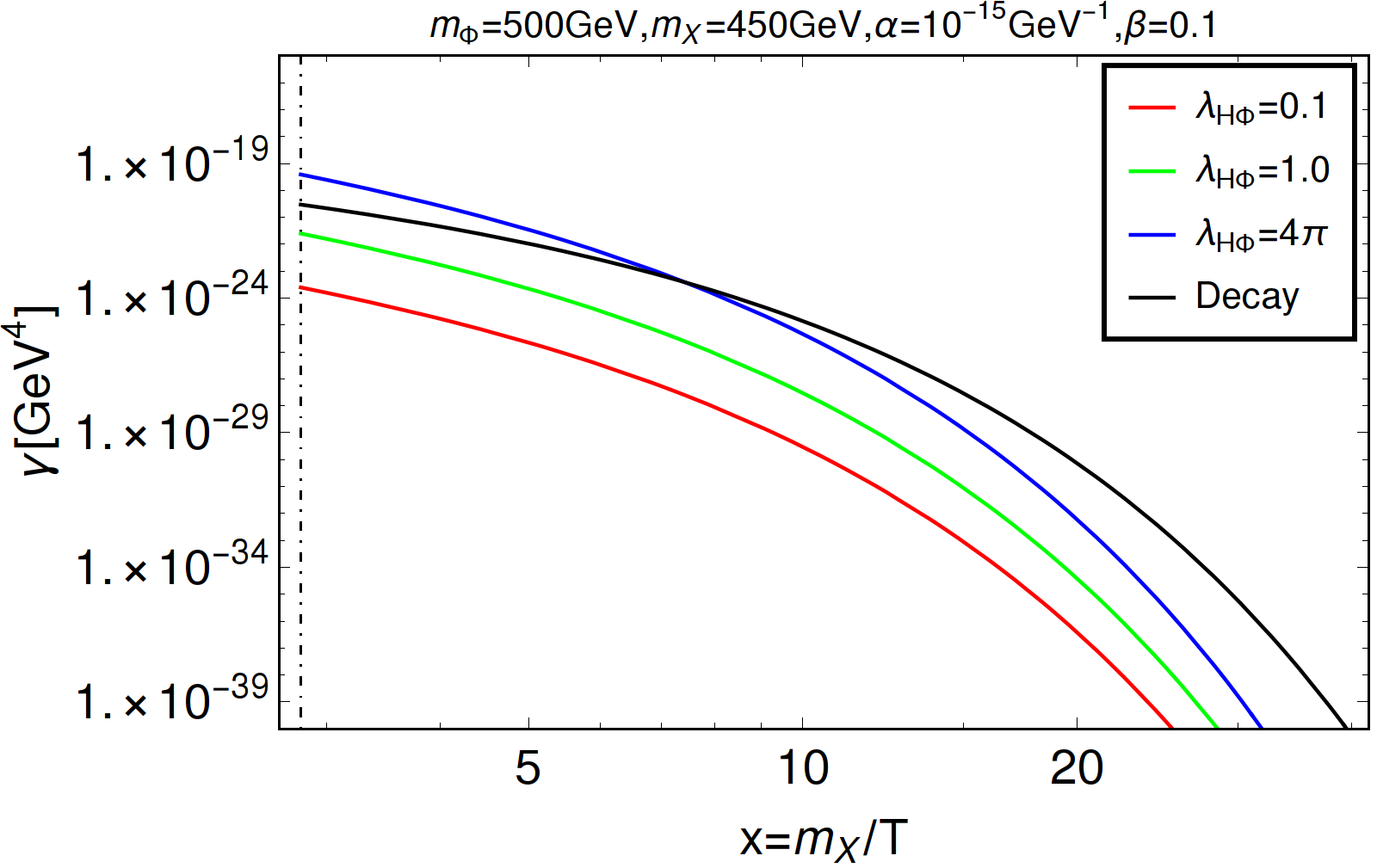}~~
 \includegraphics[scale=0.3]{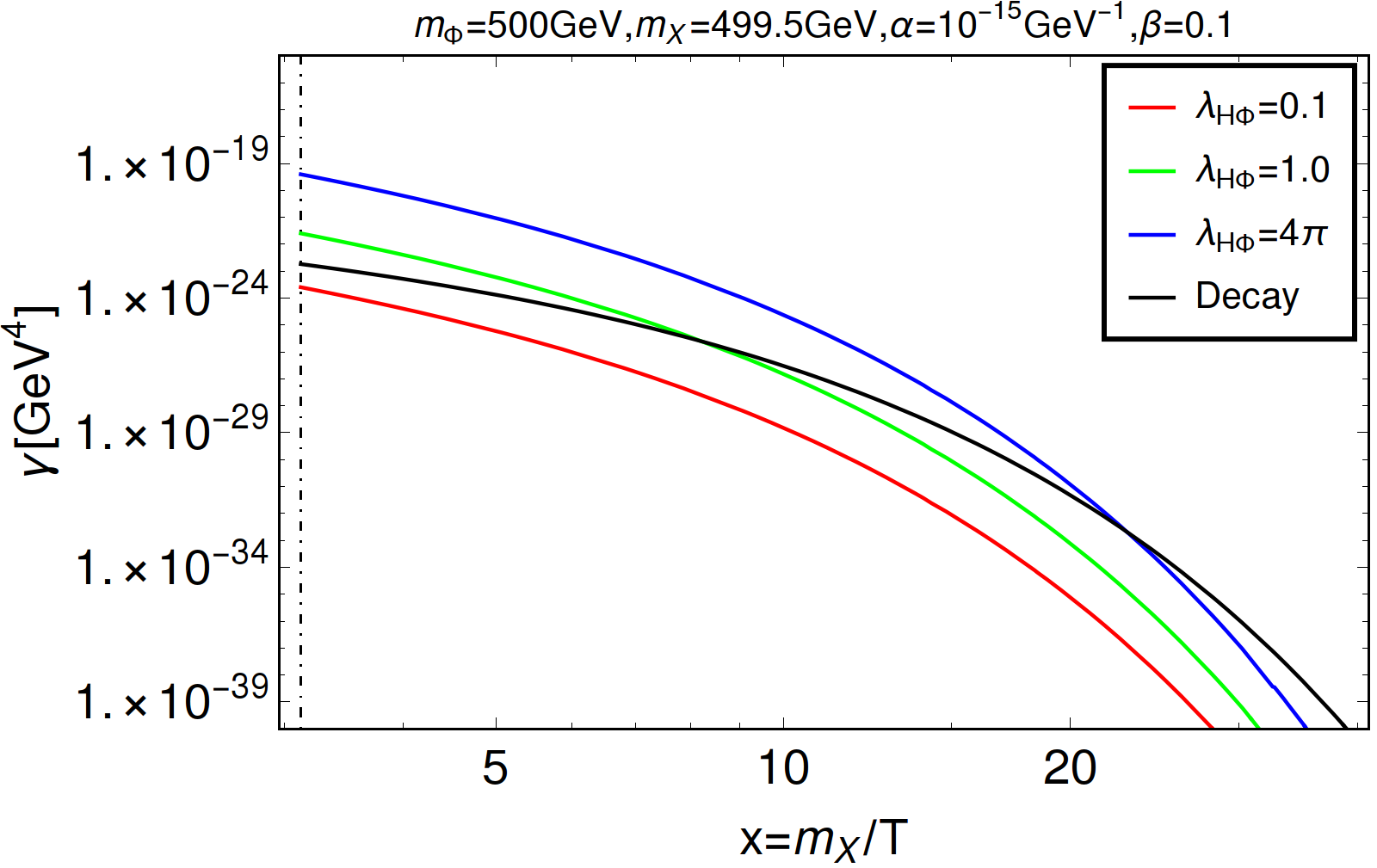}
 $$
 $$
 \includegraphics[scale=0.3]{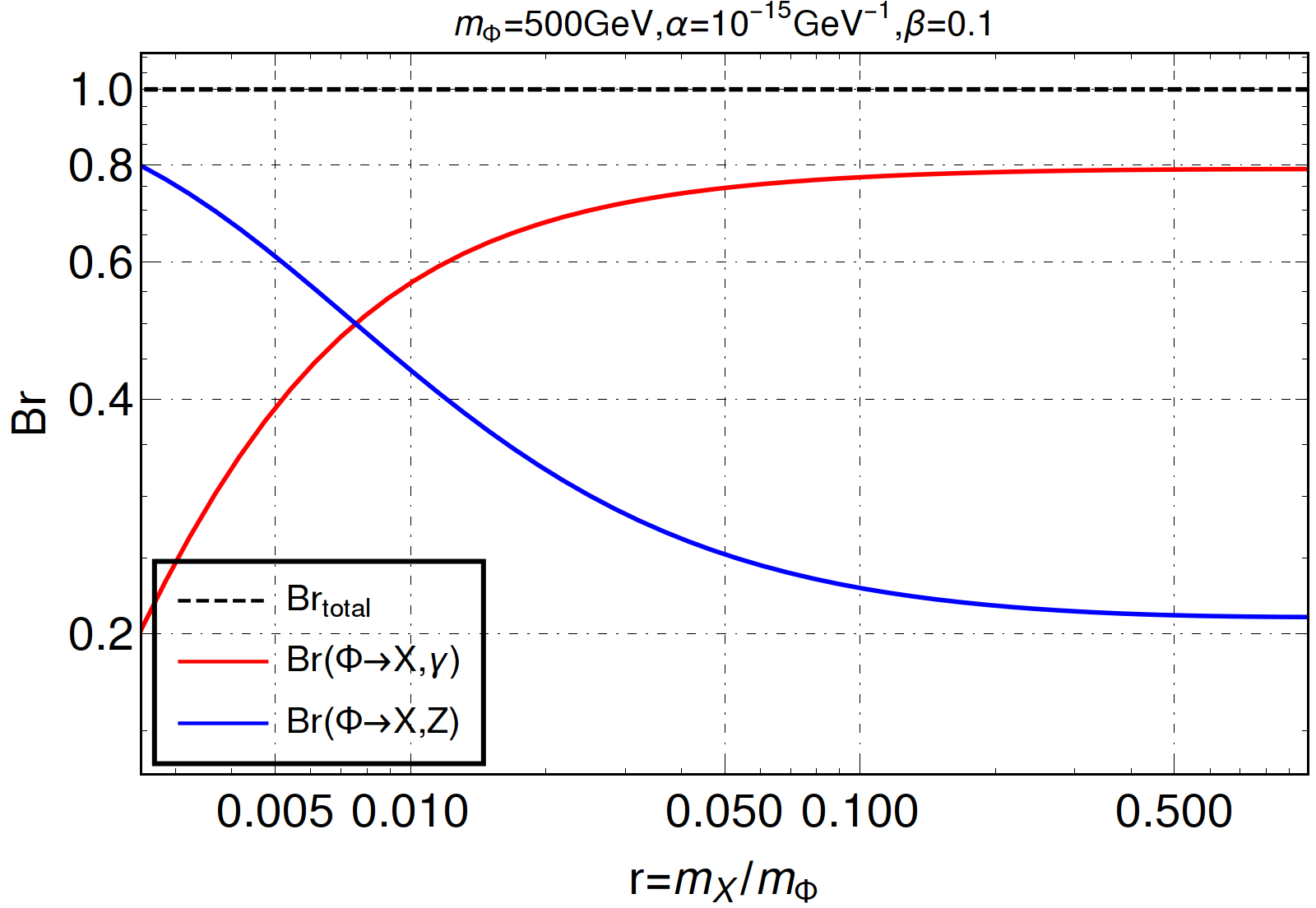}
 $$
 \caption{Top panel: Comparison of reaction densities due to annihilation (colorful lines) and decay (shown in black) after EWSB. Contributions of two annihilation channels are shown: $h,\Phi\to X,Z$ via $Z$ boson mediation in $t$-channel and $h,\Phi\to X,\gamma(Z)$ via $\Phi$ mediation in $s$-channel. All states are assumed to be massive. The amplitudes for $h,\Phi\to X,\gamma(Z)$ channel are proportional to the portal coupling $\lambda_{H\Phi}$ that is being varied in the plots.
The choice of parameters is specified in the headings. Bottom panel: Decay branching fraction of $\Phi \to X \gamma(Z)$ as function of $r=\frac{m_X}{m_\Phi}$.}
\label{fig:gam-ann-decay}
\vspace{0.2 cm}
\end{figure}

\begin{figure}[htb!]
 $$
 \includegraphics[scale=0.33]{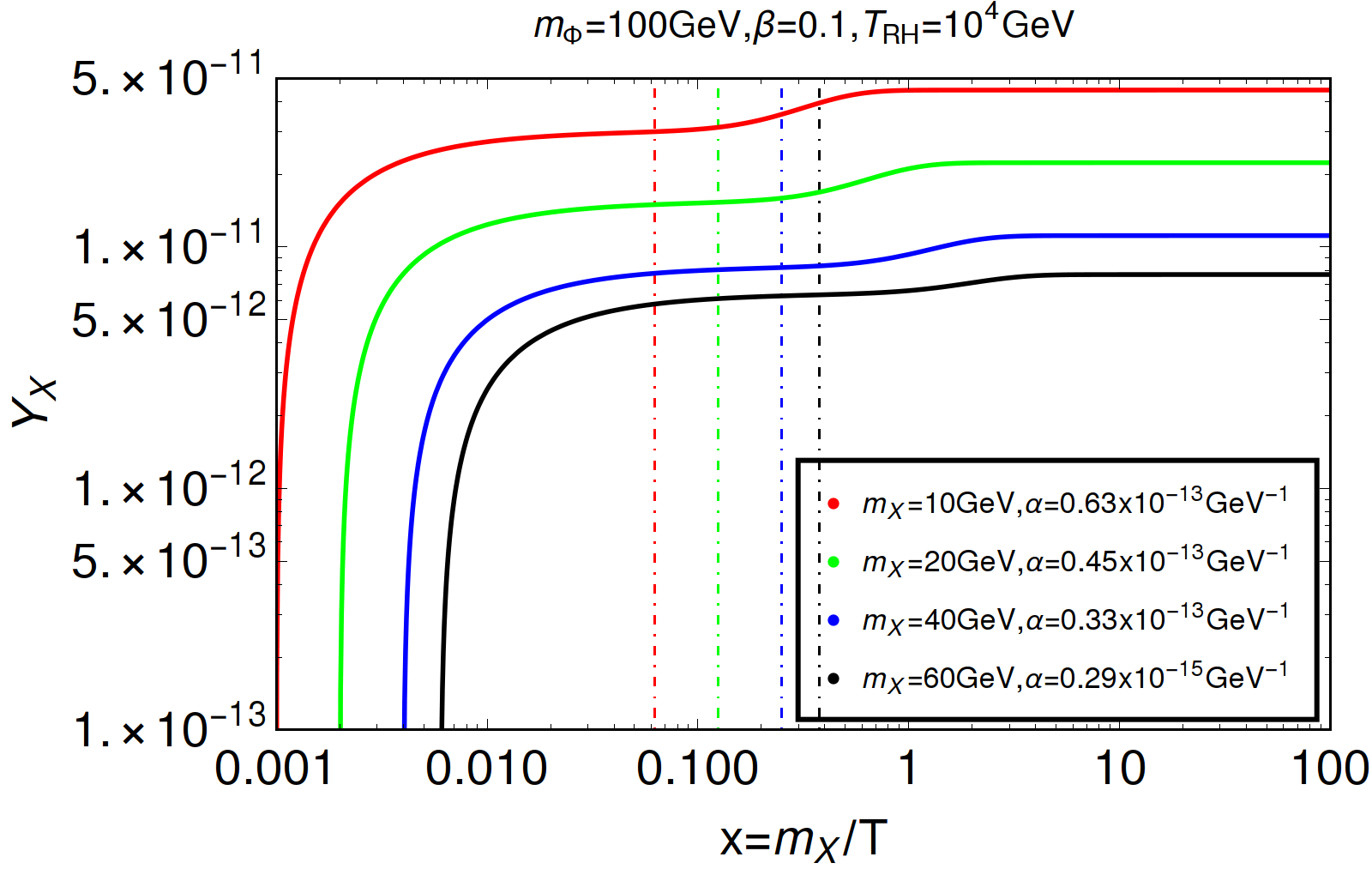}~~
 \includegraphics[scale=0.33]{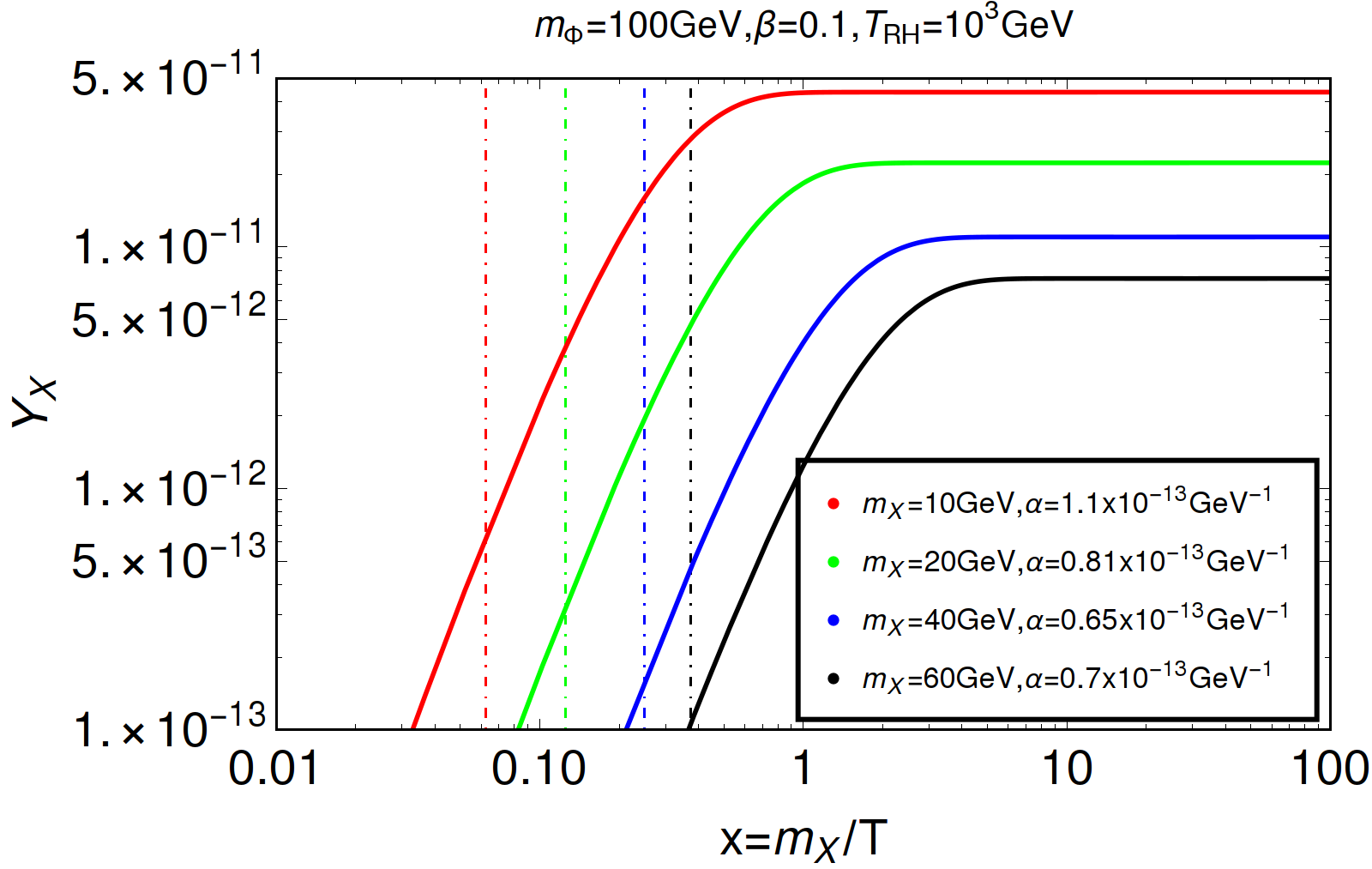}~~
 $$
 $$
 \includegraphics[scale=0.3]{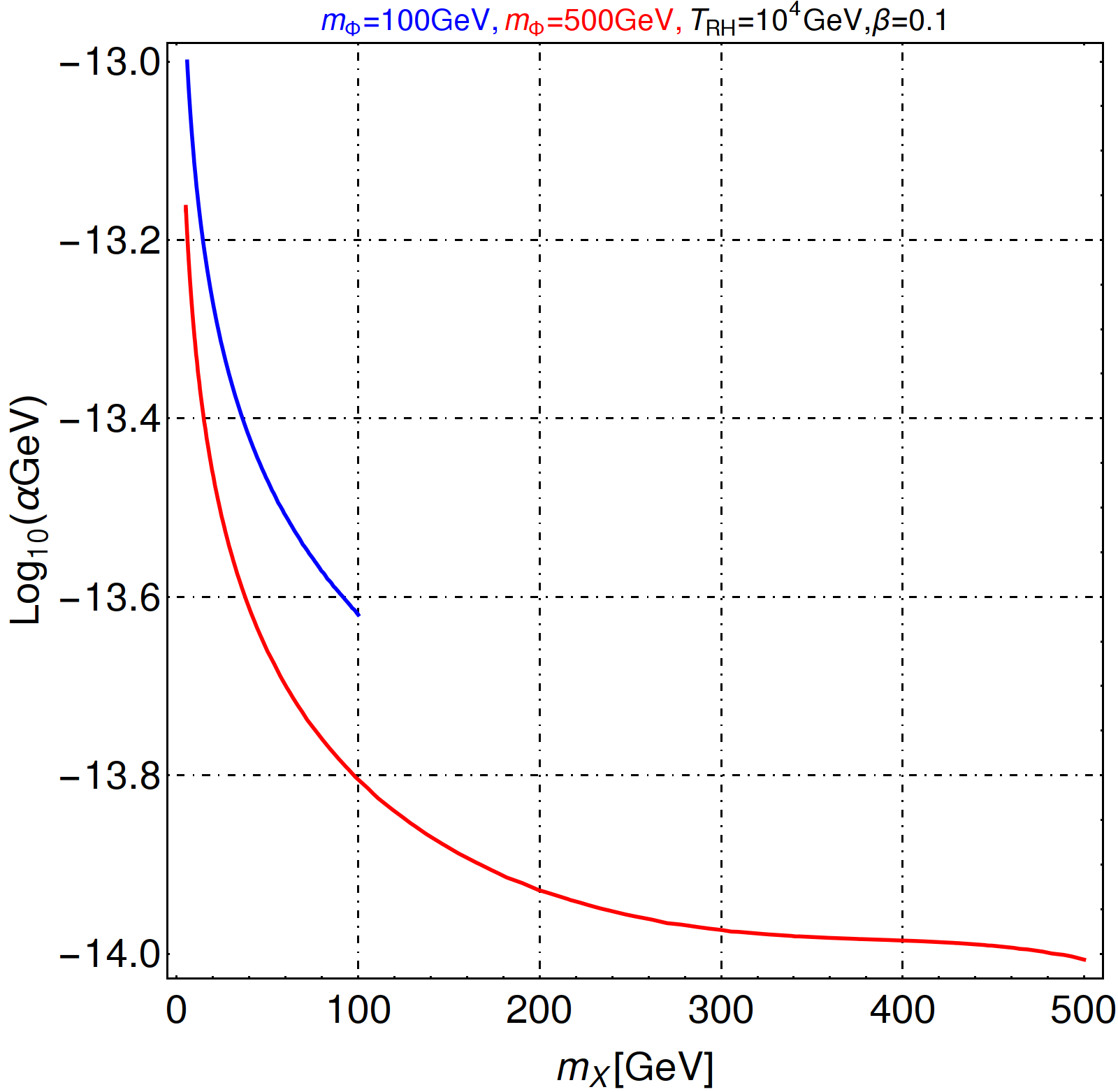}~~
\includegraphics[scale=0.3]{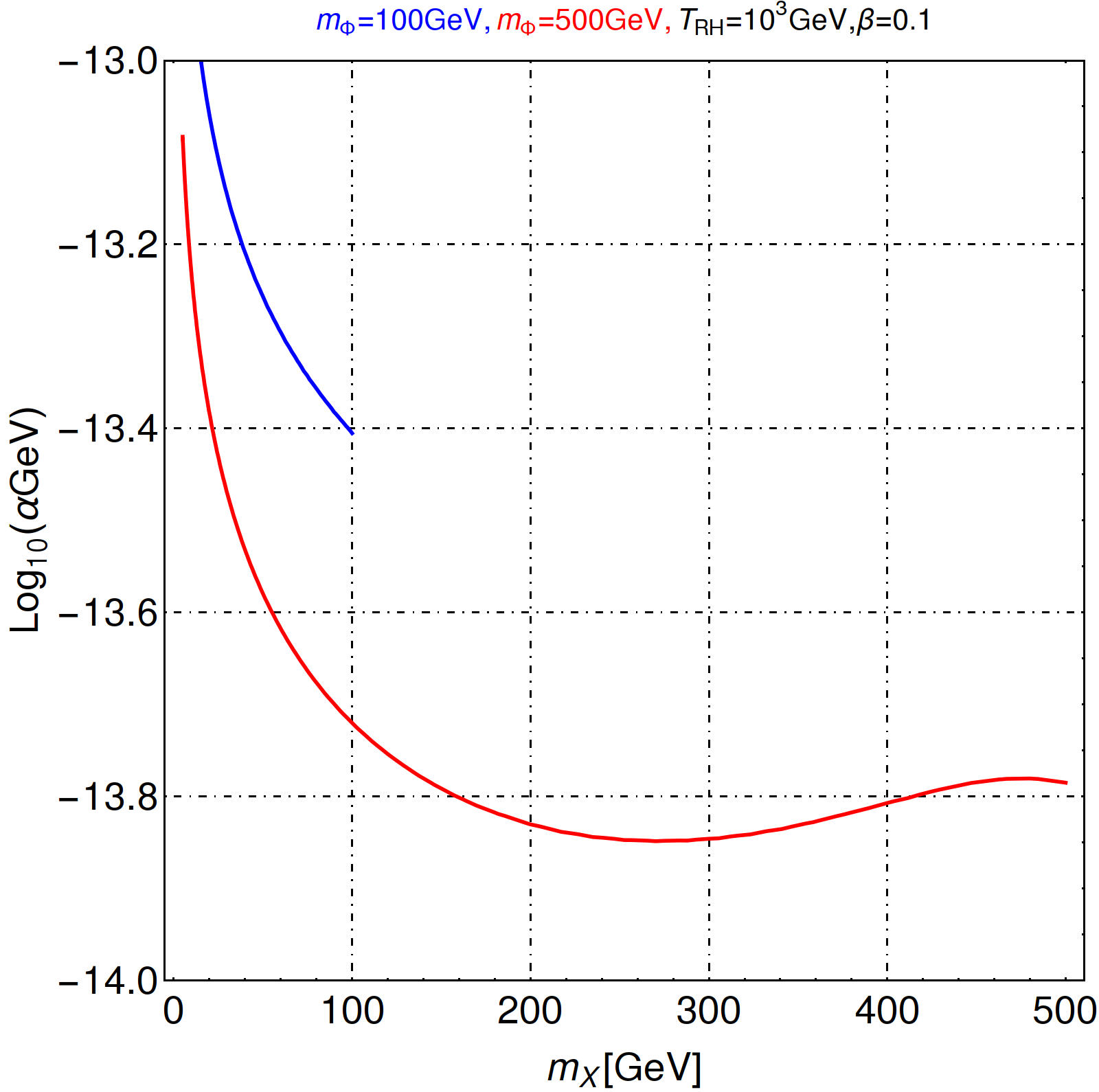}
$$
\caption{Top left: DM yield as a function of $x=m_X/T$ for a reheat temperature $T_\text{RH}=10\tev$ considering both annihilation and decay to be active before EWSB, while keeping only decay after EWSB. We choose $m_\Phi=100\gev$ and $\beta=0.1$ while for each value of $m_X$, $\alpha$ is adjusted to reproduce the central value of the measured relic abundance. Top right: Same as top left but with $\trh=1\tev$. In both of the plots the vertical dashed lines denote
EWSB ($x_{EW}$). Bottom left: Relic density allowed parameter space in $m_X-\alpha$ plane for $\trh=10\tev$ and $\alpha$ adjusted to reproduce the DM abundance. Bottom right: Same as bottom left but with $T_\text{RH}=1~\rm TeV$. }
\label{fig:rel-lowrh}
\vspace{0.2 cm}
\end{figure}

Before solving BEQ, let us estimate first the hierarchy of reaction densities. For illustration, in Fig.~\ref{fig:gam-ann-decay}, we compare reaction densities for the $\Phi$ decay and those for annihilations into $X,\gamma (Z)$ final states. 
For the later final state two annihilation diagrams contribute: (a) $t$-channel annihilation $h,\Phi\to X,Z$ via $Z$ boson mediation and (b) $s$-channel process $h,\Phi\to X,\gamma(Z)$ via $\Phi$ mediation. 
We consider all the states involved in these two processes to be massive. Note that the s-channel amplitude for $h,\Phi\to X,\gamma(Z)$ is proportional to $\lambda_{H\Phi}$, therefore it could be amplified.
We show the reaction density as a function of $x$ for decay as the black curve in all the figures, while we choose three values of $\lambda_{H\Phi}=\{0.1,1,4\pi\}$, shown respectively in red, green and blue, for the annihilation processes. 
It is seen that for small $m_X\simeq 100~\rm GeV$, after EWSB the decay dominates over annihilation even when the portal coupling is close to its limiting perturbative value i.e. $4\pi$. However for $m_X\gtrsim 400~\rm GeV$ and 
$\lambda_{H\Phi}\simeq 4\pi$ the $s$-channel annihilation starts dominating over decay for $x \lsim 7$. 
This is expected since growing DM mass causes phase space suppression of the decay width. 
However, even for $m_X\simeq 499\gev$ for large enough $x$ again the decay dominates over annihilation. 
Therefore, it is fair to conclude that as long as $\lambda_{H\Phi}\lesssim\mathcal{O}(1)$, 
one can safely ignore all the annihilation processes even for large DM mass. 
In the bottom panel of Fig.~\ref{fig:gam-ann-decay} we show variation of $\Phi$-branching ratio to $X$ and photon or $Z$-boson. 
For light $X$, {\sb $\Phi$} decays into $X,Z$ dominate while for $r=m_X/m_\Phi \gsim 0.008$ $\Phi$ decays mostly into $X\gamma$.

% Now, even if all the processes after EWSB (annihilation and decay) have significant contributions in this case, but we can still ignore the contribution due to the annihilation channels compared to the decay, modulo the coupling $\lambda_{H\Phi}\lsim 1.0.$ such that diagrams proportional to $\lambda_{H\Phi}^2$ are sub-dominant. Therefore, in this case, as before we consider both annihilation and decay processes before EWSB with massive new fields and massless SM fields, while after EWSB we only consider the decay. 

% In Fig.~\ref{fig:gam-lowrh} we have shown the variation of the reaction density with $x=m_X/T$ as before for a low reheat temperature. Now, as the abundance can be dimensionally written as: $\Omega_X h^2\sim m_X\alpha^2 T_\text{RH}$, hence decrease in reheat temperature would require a larger $\alpha$ in order to satisfy relic density for a given DM mass, which is evident from Fig.~\ref{fig:rel-lowrh}. Also note from the left panel of Fig.~\ref{fig:rel-lowrh}, the freeze-in in this case takes place at $x\sim\mathcal{O}(1)$, which is a typical signature of IR freeze-in. 

Now let us find relic density in the low $T_\text{RH}$ regime. In the left and right upper panels of Fig.~\ref{fig:rel-lowrh} we have shown 
the DM yield for several values of $m_X$ as a function of $x$ for $\trh=10^4\gev$ and $10^3\gev$, respectively. 
In order to satisfy the relic abundance, for smaller $\trh$, we need larger $\alpha$ as $Y_X\sim T_\text{RH}\alpha^2$.
Note that the left panel already shows formation of IR-like behavior of the yield that is typical for low $\trh$: the ``second slopes'' 
that appear for larger $x$ in the left panel ($\trh=10^4\gev$) reach their corresponding plateaus at the same location 
as in the right panel for $\trh=10^3\gev$. In the right panel, the typical UV-like sudden DM production disappeared leaving only after-EWSB 
IR-production (decays) at much larger $x$. The IR behavior is, of course, more prominent for smaller $\trh$. 
In the lower panels we have shown curves in the $\mx-\alpha$ plane that reproduce proper DM abundance for a fixed $\trh$. 
Comparing with the high $\trh$ regime, Fig.~\ref{fig:rel-highrh}, one can observe that since $Y_X\sim T_\text{RH}\alpha^2$ 
the required $\alpha$ had to be two orders of magnitude smaller than here.

Finally, let us present an approximate formula for the yield for the case of low reheat temperature ($\trh \sim m$). 
Proceeding in a similar way as in the previous subsection, we can estimate the contribution to Yield from annihilation and decay 
via dimensional argument as in Eq.~\ref{eq:compare}. However, we need to remind that now the freeze-in temperatures ($T_{FI}$) are roughly the same for both 
annihilation and decay when $T_{RH}\sim m_\Phi$, resulting in similar decay and annihilation contributions to the yield, i.e.
\beq
\frac{Y_X^{ann}}{Y_X^\text{D}}\sim \frac{\sigma M_{pl} T_{FI}}{\Gamma_\Phi M_{pl}/T_{FI}^2}\sim\frac{\alpha^2 M_{pl} T_{RH}}{\left(\alpha^2 m_\Phi^3\right) M_{pl}/m_\Phi^2}
\sim\frac{T_{RH}}{m_\Phi} \sim 1.   
\eeq
Therefore, the final yield for such a situation can be written as:
\bea
Y_X(x=0) \sim 2\alpha^2 M_{pl} T_{FI} \sim 2\alpha^2 M_{pl} m,
\eea 
where $T_{FI} \sim m$ characterizes dark sector mass.

%%%%%%%%%%%%%%%%%%%%%%%%%%%%%
\subsection{Summary results}
\label{sec:summary}
%%%%%%%%%%%%%%%%%%%%%%%%%%%%%

Effects of varying reheat temperature for DM yield evolution has been shown in the top panels of Fig.~\ref{fig:yld-trh} for two different sets of dark sector masses. 
For $T_\text{RH}=10^8~\rm GeV$ (shown by the red curve) i.e. for the $\trh \gg m$,  we observe yield that follows typical UV freeze-in pattern and becomes maximum at 
$T\sim T_\text{RH}$. With gradual decrease in $T_\text{RH}$, although the characteristic UV freeze-in is still visible at smaller $x$, however the yield also builds up 
at larger $x$ and final freeze-in occurs at $T\sim m_X$, as shown by the blue ($T_\text{RH}=10^4~\rm GeV$) and black curves ($T_\text{RH}=10^3~\rm GeV$). 
For $T_\text{RH}=1~\rm TeV$ the IR characteristic of freeze-in is more prominent. Note that, all the parameters chosen in these plots reproduce the observed relic abundance 
and hence the yields for the same $m_X$ at low $x$ converge to the same asymptotic value, as they indeed must do according to (\ref{eq:relicX}). 
Colorful curves in the upper panels correspond to different values of $\alpha$ adjusted so that in spite of varying $\trh$, the asymptotic value is the same and corresponds to the observed abundance.~\footnote{The initial condition for all our solutions of the freeze-in BEQ assumes no DM at $\trh$, i.e. $Y_X(m_X/\trh)=0$. On the other hand
$\alpha$ is adjusted so that the observed abundance is satisfied, i.e. for the same $m_X$ the curves in upper panels of Fig.~\ref{fig:yld-trh} converge to the same value for $x\to 0$.} 
For large $T_\text{RH}$ one requires a smaller $\alpha$ to obtain the right abundance to compensate the effect of larger integration region. 
We also note that the yield $Y(x \to 0)$ values in these the two upper panels are different due to different choices of DM masses.
In the middle panel of Fig.~\ref{fig:yld-trh} we show curves in the $\alpha-\trh$ space that imply proper DM abundance. 
It is interesting to note that for $T_\text{RH}\lesssim 1~\rm TeV$, the relic density becomes 
independent of the reheat temperature as the IR freeze-in dominates over UV freeze-in. Beyond 1 TeV the effective coupling $\alpha$ must decreases with grow of 
$T_\text{RH}$ for a fixed DM mass in order to satisfy the central value of the PLANCK observed relic abundance as  $Y_X\propto T_\text{RH}\alpha^2$. 
Also for a fixed $T_\text{RH}$, larger DM mass requires smaller $\alpha$ simply because $\Omega_X\propto m_X$ following Eq.~\eqref{eq:relicX}. In the bottom panel of Fig.~\ref{fig:yld-trh}, we illustrate the effect of $\tilde{\alpha}$ by varying 
$\beta=\frac{\tilde{\alpha}}{\alpha}=\{0.001,0.1,0.3,0.5,1.0,3.0,5.0\}$ (shown in different colors) 
on the resulting relic density allowed parameter space. Note that quantities like reaction rates, DM yields or $\Phi$-lifetime depend on $\alpha$ and $\beta$ via 
$\alpha^2+\widetilde\alpha^2=\alpha^2\left(1+\beta^2\right)$.
We have decided to present those quantities for fixed $\beta=0.1$ and various values of $\alpha$. Equivalently the numbers shown in plots could be 
parametrized by $\alpha\left(1+\beta^2\right)^{1/2} =  \alpha_\text{old}\left(1+(0.1)^2\right)^{1/2}$, where $\alpha_\text{old}$ is the parameter
specified in our plots with $\beta=0.1$. Clearly, larger $\beta$ requires smaller $\alpha$ and 
this is what we see in the orange ($\beta=1.0$), magenta ($\beta=3.0$) and cyan ($\beta=5.0$) colored contours. 

%\bg{********************** \it I finish here ***************************}

% the yield decreases as the upper limit of the yield integration becomes lower. For $T_\text{RH}=1~\rm TeV$ we see the effect of IR freeze-in at large $x$, shown by the bump in the black curve. For large reheat temperature the DM freezes in immediately showing its thypical UV freeze-in behaviour, but for $T_\text{RH}\sim m_i$ although the yield builds up quickly at smaller $x$ (dominated by UV freeze-in) but the yield freeze-in near $T\sim m$ showing the effect of IR freeze-in. This nature holds both for $m_\phi=100~\rm GeV$ (shown in the left panel of Fig.~\ref{fig:yld-trh}) and becomes more prominent for $m_\Phi=500~\rm GeV$ (shown in the right panel of Fig.~\ref{fig:yld-trh}) even when $T_\text{RH}=10~\rm TeV$. Here all parameters are chosen arbitrarily just for illustration and they do not follow any constraint.

% Now, if the coupling $\lambda_{H\Phi}\gtrsim\mathcal{O}(1)$ then one can no more avoid the contribution due to the annihilation channels involving $h\Phi\Phi$ vertices after EWSB, although other channels shall still be sub-dominant compared to the decay.

%%%%%%%%%%%%%%%%%%%%%%%%%%%%%%%%%%%%%%%
\section{Signature of the model}
\label{sec:collider}
%%%%%%%%%%%%%%%%%%%%%%%%%%%%%%%%%%%%%%%%

\begin{figure}[htb!]
$$
\includegraphics[scale=0.4]{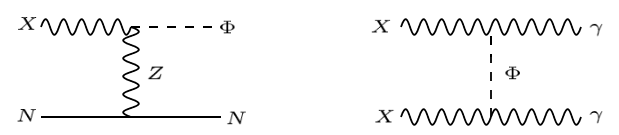}
$$
\caption{Left: Inelastic DM scattering off the nuclei $N$ for direct search of DM; right: DM annihilating to photon final states via $\Phi$ mediation for indirect search of DM.}
\label{fig:dd}
\end{figure}

As we have already demonstrated, the dimensionful $\Phi-B-X$ vertex ($\alpha(\tilde{\alpha})$) is constrained by relic density of DM to be $\alpha(\tilde{\alpha}) \lesssim 10^{-12}~\text{GeV}^{-1}$ for $T_\text{RH}\gtrsim 10^3$ GeV. 
Therefore, assuming $c\sim{\cal O}(1)$~\footnote{As mentioned in ref.~\cite{Macias:2015cna} the Lagrangian (\ref{dim5}) can be generated at the tree-level by integrating out anti-symmetric tensor mediators, then indeed $c\sim{\cal O}(1)$.}, one can conclude that freeze-in of the VDM requires the NP scale $\Lambda\gsim 10^{12}$ GeV at least or higher for larger reheat temperature. If the effective operators (\ref{dim5}) where $n$-loop generated, then we would roughly conclude that $\Lambda \gsim 10^{12-2n}\gev$. Hereafter we assume that the Lagrangian (\ref{dim5}) is indeed tree-level generated.
Therefore, the phenomenology of the model is severely constrained. 
%In this section, we will be noting a few points in this direction. 
First of all, note that there is no $XX-SM$ vertex, therefore no elastic scattering of the DM against nuclei is possible.
So, this model easily avoids stringent constraints from non-observation of DM scattering at direct search experiments. DM can only scatter off nuclei inelastically with $\Phi$ in the final state as shown in left panel of Fig.~\ref{fig:dd}. Since $m_\Phi>m_X$, 
hence such an inelastic scattering is forbidden even if the mass difference $\delta m=m_\Phi-m_X\gtrsim\mathcal{O}(100)~\rm MeV$~\cite{TuckerSmith:2001hy}. On the other hand, due to the presence of $\Phi-X-\gamma$ vertex, the DM pair annihilation 
may give rise to monochromatic X-ray line (right panel of Fig.~\ref{fig:dd}) but such photon flux will be hugely suppressed by $1/\Lambda^2$ and can not account for the, say, galactic-center gamma ray excess as observed. 
Hence this model in its freeze-in realization of DM can not be probed from either direct nor indirect DM search experiments.
\begin{figure}[htb!]
$$
\includegraphics[scale=0.15]{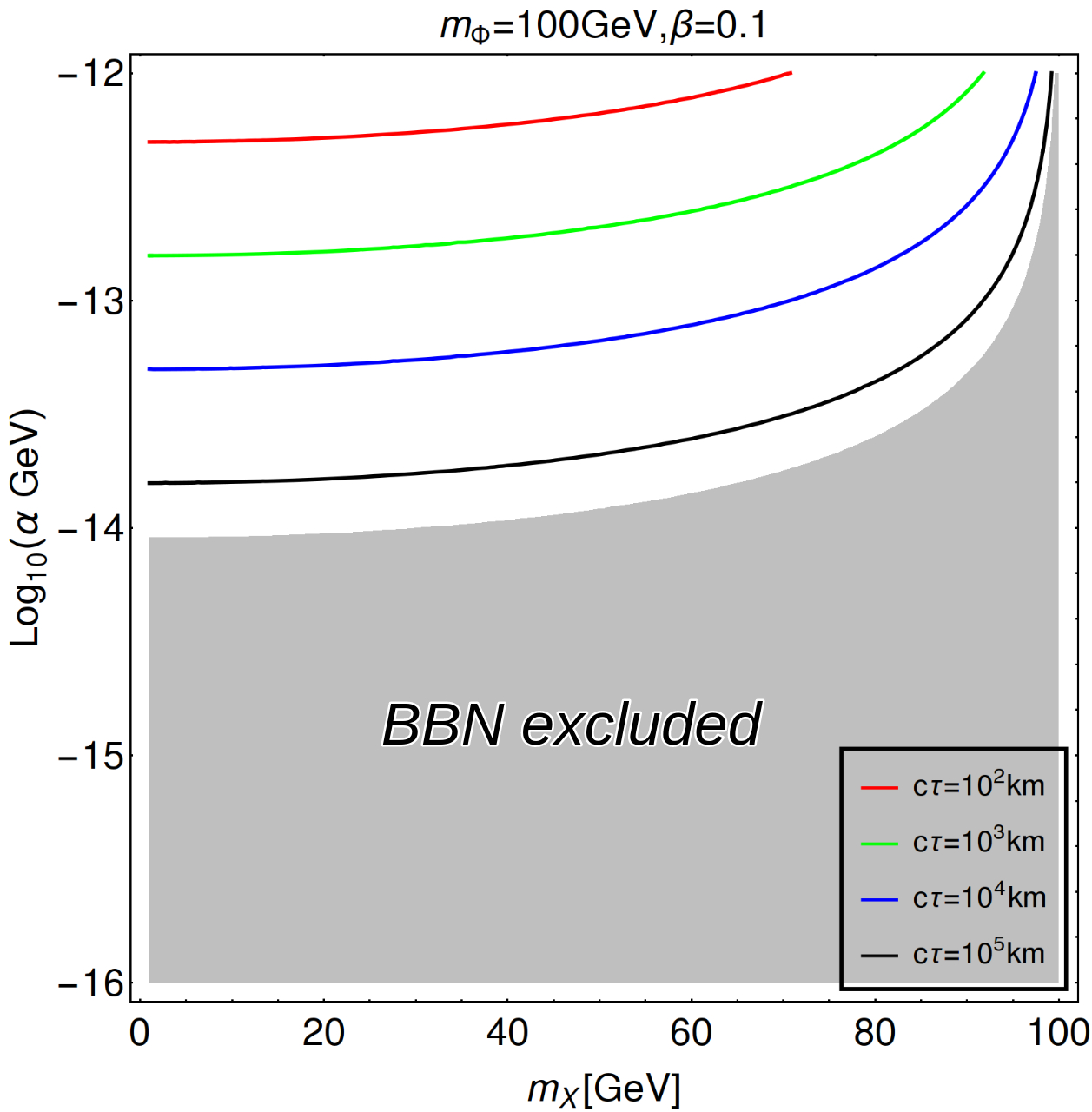}~~
\includegraphics[scale=0.15]{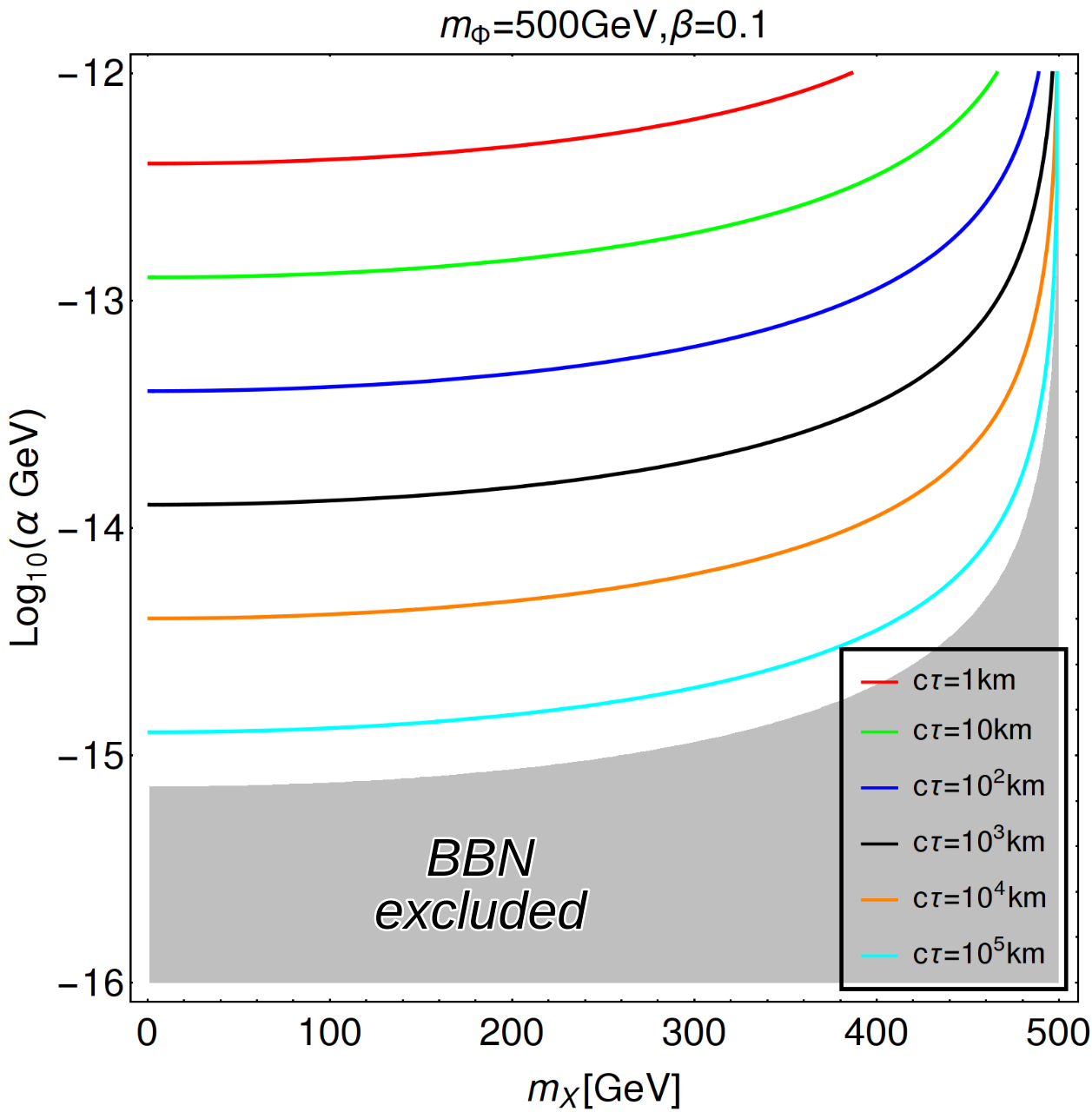}
$$
\caption{The contours show different decay lengths of $\Phi$ for particular choices of the parameters that can produce right relic abundance for the DM.  On left panel, we choose $m_\Phi=100$ GeV, while on the right panel we show the
case for $m_\Phi=500$ GeV. The shaded region at the bottom is excluded by BBN constraint $\tau_\Phi \lsim \tau_\text{BBN} \sim 1~\sec$.}\label{fig:phi-dec-col}
\vspace{0.2 cm}
\end{figure}

\begin{figure}[htb!]
$$
\includegraphics[scale=0.22]{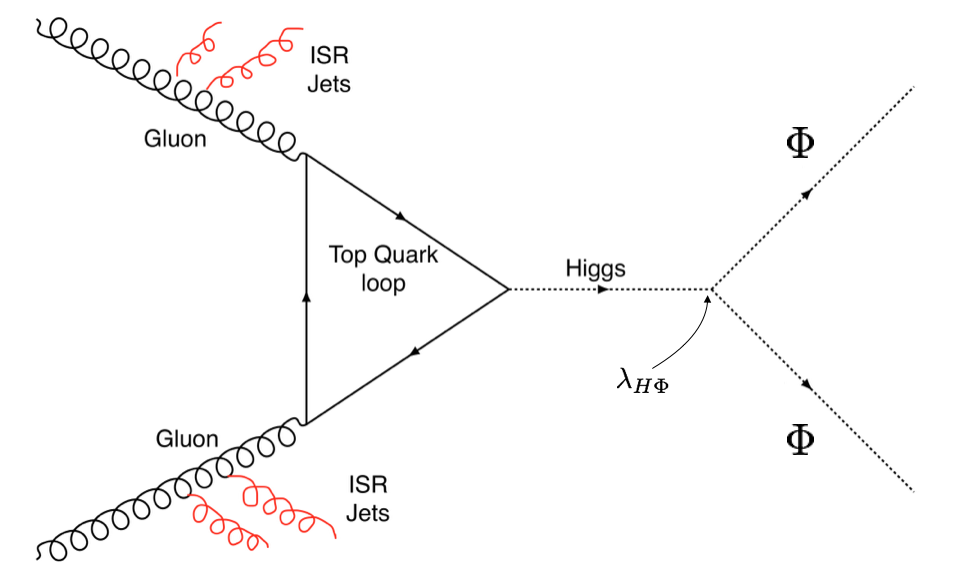}
$$
\caption{$\slashed{E}_T+\text{jets}$ signature that the model can produce at the collider.}
\label{fig:DM-production}
\vspace{0.2 cm}
\end{figure}

However, since the portal coupling $\lam_{H\Phi}$ is unconstrained by existing observations and as argued before this coupling can be as large as $\lam_{H\phi}\sim\mathcal{O}(1)$, hence the non-standard scalar $\Phi$ can be produced at 
the collider via Higgs mediation (see Fig.~\ref{fig:DM-production}). Once these $\phi$'s get produced at colliders they will eventually decay via Fig.~\ref{fig:phi-decay} to DM and SM final states like $Z,\gamma$. 
This is the same channel that also gives rise to DM production via freeze-in. For particular choice of the effective coupling $\alpha$ that gives rise to right relic abundance, the $\Phi$ remain stable over the detector length. 
This is shown in Fig.~\ref{fig:phi-dec-col}, where we see that for $\alpha=10^{-14}-10^{-13}\gev^{-1}$ the average decay length of $\Phi$ is $L_\Phi=c\tau_\Phi\gtrsim 100~\rm km$, where $c=3\times 10^8~\text{ m/sec}$. In such a case the $\Phi$'s basically escapes LHC detector and gives rise to missing energy $(\slashed{E}_T)$, which can be constructed out of the recoil of an initial state radiation (ISR) of a gluon, $\gamma$, $W^\pm,Z,H$ as 
\bea
 \slashed{E}_T = -\sqrt{\left(\sum_{\ell,j} p_x\right)^2+\left(\sum_{\ell,j} p_y\right)^2},
 \eea
where the sum runs over all visible objects that include leptons and jets, and unclustered components. Therefore, the model can finally produce monojet~\footnote{Multijet final states will be infested with huge SM backgrounds.} plus missing energy signal that has extensively been searched at the LHC~\cite{Chala:2015ama,Aaboud:2016uro,Aaboud:2016tnv} as a vanilla DM signal particularly for Higgs portal DM models. 
However, usually, when one produces DM that is connected with the SM via a Higgs portal, then the coupling is tightly constrained from direct search. Therefore, such signals are pretty small and submerged into huge SM background. 
In our case, as mentioned, the coupling ($\lam_{H\phi}$) can be large and can produce a significant number of such mono-X signal events, that may be of interest for next run of LHC. 

It is worth noting that decays of $\Phi$ have to be completed before the onset of Big Bang Nucleosynthesis (BBN)~\cite{Kolb:1990vq,PhysRevD.98.030001}, so that it does not alter the standard BBN picture. 
Therefore here we will require that $\tau_\Phi \lsim \tau_\text{BBN} \sim 1~\sec$ which, for fixed $m_\Phi$ and $m_X < m_\Phi$ puts a lower bound on $\alpha$. This has been illustrated in 
Fig.~\ref{fig:DM-production} where we show, for $m_\Phi=100 \gev$ and $500\gev$, regions allowed by the BBN constraint ($c\tau=10^5$~km). 
Concluding one can see that usually $\alpha \gsim 10^{-15}\gev$ is allowed, while the region $m_X \sim m_\Phi$ is forbidden for any value of $\alpha$.

%\bb{In our case $\Phi$ can decay both into photon  and into fermion final states. As shown in~\cite{Jedamzik:2006xz}, effects of electromagnetically particles on BBN yields due to photodisintegration are operative at comparatively late times $\gtrsim 10^5~\rm sec$. On the other hand, neutrinos decouple from equilibrium at $T\sim 4.3~\rm MeV$ (muon and tau neutrinos) and $T\sim 2.6~\rm MeV$ (electron neutrinos). Thus, if $\Phi$ decays take place around 2.3 MeV, they may lead to an increase of the electron neutrino density. This would increase the rate of reactions $\nu_e,n\to p,e$ and $\overline{\nu_e},p\to n,\overline{e}$ which keep protons and neutrons in thermal equilibrium and hence decrease the $He^4$ abundance. Considering $2\sigma$ upper bound on $He^4$ mass fraction we obtain $\tau_\Phi\lesssim 1~\rm sec~(T\sim 1~\text{MeV})$~\cite{Kawasaki:2000en,Kainulainen:2015sva} (taking into account the change in effective number of neutrino species). However, in our case decay to neutrinos depends on the invisible branching ratio or suppressed in case of off-shell decay with 3-body final state.}

% \bg{. \it{more details and references are needed here.}}

%%%%%%%%%%%%%%%%%%%%%%%%%%%%%%%%%%%%%%%%%%
\section{Summary and Conclusions}
\label{sec:summ}
%%%%%%%%%%%%%%%%%%%%%%%%%%%%%%%%%%%%%%%%%%

A vector boson DM weakly coupled to the visible SM sector via dimension-5 operator has been presented and the parameter space allowed by observed relic DM density 
has been found. The advantage of the model is the absence of tree-level elastic DM scattering against nuclei and a double suppression of present time DM annihilation in, 
for instance, dwarf galaxies. Therefore this scenario easily and naturally satisfies existing experimental constraints.
The model contains a dark sector composed, in a unitary gauge, of a massive vector $X_\mu$, and a real scalar $\Phi$. 
$X_\mu$ is a gauge boson of spontaneously broken extra $U(1)_X$ gauge symmetry. The vector $X_\mu$ and the scalar $\Phi$ are odd under a $Z_2$ 
symmetry introduced to stabilize the DM candidate, $X_\mu$. $\Phi$ is assumed to be heavier than $X_\mu$.
The SM sector is extended by an extra heavy neutral Higgs boson that decouples when its 
mass goes to infinity, as we assume here. The lowest dimensional operator responsible for DM-SM interaction are $1/\Lambda\;X^{\mu\nu}B_{\mu\nu}\Phi$ and 
$1/\Lambda\;\tilde{X}^{\mu\nu}B_{\mu\nu}\Phi$. It has been shown that the model can be formulated in a Stueckelberg-like fashion as a limit of the SM extended by the 
$U(1)_X$ gauge symmetry together with a complex scalar charged under the $U(1)_X$ (needed to spontaneously break the symmetry) and a real scalar $\Phi$.

We have investigated a possibility of DM production via a freeze-in mechanism through decays of $\Phi$ and annihilations including $\Phi$. It turned out to 
be convenient to consider two distinct regimes of the reheat temperature. The first one is when the reheat temperature is significantly higher than masses involved in 
the production process. This situation mimics the case of UV freeze-in, when the production happens mostly before EWSB and all processes after EWSB are 
insignificant. However the situation alters, when reheat temperature (which can be thought of a free parameter, being very loosely constrained by BBN) 
drops to lower values close to the mass scale ($m$) typical for the dark sector. It has been shown that UV freeze-in, although advertised to describe the 
case of freeze-in production of DM in EFT formalism, is not fully correct, massive contributions start playing an important role and effects of IR freeze-in i.e. 
DM yield building even up to lower temperature ($T \sim m$) starts showing up.
%Therefore, some important correlations are drawn between reheat temperature and NP scale to satisfy correct relic density. Although, this is analysed in context of the VDM freeze-in that we we discussed, but the departure point of UV limit analysed here can be borrowed in any other EFT context and seems to be model independent. 

In order to predict properly the observed DM abundance, the scale of the dimension-5 operators must be large $\Lambda \sim 10^{12}-10^{16}\gev$ 
depending on the DM mass $m_X$, the reheat temperature $\trh$ and an underlying mechanism for the generation of the relevant effective operators.  
The huge size of $\Lambda$ implies that at the lowest level of perturbative expansion neither elastic scattering off nuclei is allowed nor present time annihilations of DM 
in e.g. centers of galaxies are possible. However, it turns out that LHC collider signals mediated by Higgs boson exchange are possible, 
$gg \to H^* \to \Phi\Phi$. Since the scale of $\Lambda$ required by the DM abundance is large $\Lambda \sim 10^{12}-10^{16}\gev$ the heavier scalar $\Phi$ is 
effectively stable at the detector length scale and hence can produce mono-jet, photon, $Z,W^\pm$ or $H$ events accompanied by missing energy drifted away by pairs of 
$\Phi$ bosons.The signal cross-section could be quite substantial as the portal coupling between $\Phi$ and the SM remains unconstrained.

Finally, we must mention that a freeze-out possibility of the same model can also be thought of. In that case, the phenomenological signatures will become richer.
In contrast to the case considered here the freeze-out scenario implies constraints that are more difficult to satisfy~\cite{Fortuna:2020wwx}.

%%%%%%%%%%%%%%%%%%%%%%%%%%%%%%%%%%%%%%
\vspace*{10mm}  {\bf Acknowledgement}\\
%%%%%%%%%%%%%%%%%%%%%%%%%%%%%%%%%%%%%%
SB would like to acknowledge DST-SERB grant CRG/2019/004078 and WHEPP meeting at IIT Guwahati, where the work was initiated. 
BB and SB would like to thank Sunando Patra for helping out with numerical computations. 
The work of B.G. is supported in part by the National Science Centre (Poland) as a research project, decision no 2017/25/B/ST2/00191.

%%%%%%%%%%%%%%%%%%%%%%%%%%%%%%%%%%%%%%%%%%%%%%%
\appendix
%%%%%%%%%%%%%%%%%%%%%%%%%%%
\section{The parameters of the scalar potential}
\label{parameters_relations}
%%%%%%%%%%%%%%%%%%%%%%%%%%%%
Here we collect useful relations between potential parameters.
\bea
\begin{split}
m_1^2 &= \sec (2 \alpha ) \left(2 \lambda_H v_h^2 \cos ^2\alpha-2 \lambda_{S}v_S^2 \sin ^2\alpha\right),\\ 
m_2^2 &= \sec (2 \alpha ) \left(2 \lambda_Sv_S^2 \cos ^2\alpha-2 \lambda_H v_h^2 \sin ^2\alpha\right),\\ 
m_{\Phi}^2 &= 2 \mu_\Phi^2+\lambda_{H\Phi}v_h^2+\lambda_{S\Phi}v_S^2.
\end{split}
\label{eq:phymass1}
\eea
The couplings, likewise, can be expressed in terms of the physical masses and mixing:
\bea
\begin{split}
\lambda_H=\frac{m_1^2 \cos ^2\alpha+m_2^2 \sin ^2\alpha}{2 v_h^2}, ~\lambda_S=\frac{m_1^2 \sin ^2\alpha+m_2^2 \cos ^2\alpha}{2v_S^2},\\ 
\lambda_{SH}= \frac{\sin \alpha \cos \alpha \left(m_1^2-m_2^2\right)}{v_hv_S},~\lambda_{S\Phi}= \frac{-2 \mu_\Phi ^2+m_3^2-\lambda_{H\Phi}v_h^2}{v_S^2},\\
\end{split}
\eea
with,
\bea
\sin\left(2\alpha\right) = \left(\frac{2 v_h v_S}{m_1^2-m_2^2}\right)\lambda_{SH}.
\label{eq:alpha-mix}
\eea
Now, from Eq.~\eqref{eq:phymass1} we see:
\bea
m_1^2+m_2^2=2\left(\lambda_H v_h^2+\lambda_{S} v_S^2\right).
\label{eq:mh1mh2}
\eea
From (\ref{eq:vsvh}-\ref{eq:mh1mh2}) we find a useful expression for $m_2$:
\beq
m_2^2=\vh^2\frac{2(\lH-\lSM)(4\lH\lS-\lSH^2)}{4\lS(\lH-\lSM)-\lSH^2}
%m_2^2 = \frac{1}{8} v_S^2 \Bigg\{16 \lambda_{S}+\frac{\lambda_H \left(4 \lambda_{S} (\lambda_{SM}-\lambda_H)+\lambda_{SH}^2\right)}{\lambda_{SM}^2 (\lambda_H-\lambda_{SM})^2}\Bigg\}.
% 2 v_h^2 \left(\lambda_H+\frac{4 \lambda_{S} \lambda_{SM} (\lambda_{SM}-\lambda_H)}{4 \lambda_{S} (\lambda_{SM}-\lambda_H)+\lambda_{SH}^2}\right),
\label{m2_use}
\eeq

%%%%%%%%%%%%%%%%%%%%%%%%%%%%%%%%%%
\section{Relevant vertices}
\label{sec:vertex}
%%%%%%%%%%%%%%%%%%%%%%%%%%%%%%%%%

Adopting the Lagrangian of the model in Eq.~\eqref{Ltot}, one finds relevant vertices and propagators collected 
in the table~\ref{tab:vert-fac}. Here the notation have usual meaning, for example, $g_{1,2}$ are the gauge couplings corresponding to 
$U(1)_Y$ and $SU(2)_L$ gauge groups, respectively. $c_v$ and $c_a$ are defined as: 
$c_v^f=T_{3L}-2 \sin^2\theta_w Q_f$ and $c_a^f=T_{3L}$, where $T_{3L}$ is the $SU(2)_L$ isospin quantum number and $Q_f$ is the charge of the SM fermion $f$ concerned.

\FloatBarrier
\begin{table}[htb!]\scriptsize
\begin{center}
\begin{tabular}{|c|c|c|c|c|c|c|c|c|c|c|}
\hline\hline
Vertex&Vertex factors\\
\hline\hline
&\\
\begin{minipage}{.15\textwidth}
      \includegraphics[scale=0.25]{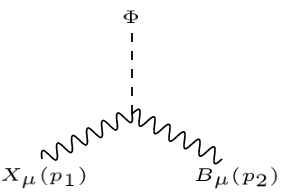}
    \end{minipage}& $i\tilde{\alpha}\varepsilon_{\mu\nu\rho\sigma}p_1^\rho p_2^\sigma+i\alpha\left(\eta_{\mu\nu}p_1.p_2-p_{1_\nu} p_{2_\mu} \right)$  \\\cline{2-2}
    &\\
\begin{minipage}{.2\textwidth}
      \includegraphics[scale=0.25]{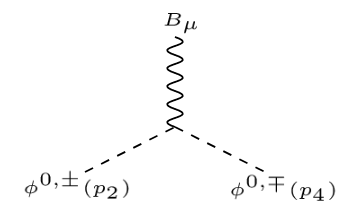}
    \end{minipage}& $-\frac{ig_1}{2}\left(p_2-p_4\right)_\mu$  \\\cline{2-2}
    &\\
\begin{minipage}{.2\textwidth}
      \includegraphics[scale=0.26]{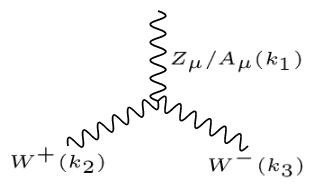}
    \end{minipage} & $-ie\left[\left(k_1-k_2\right)_a g_{bc}+\left(k_2-k_3\right)_b g_{ac}+\left(k_1-k_3\right)_c g_{ab}\right]$ \\
 & $-ig_2c_w\left[\left(k_1-k_2\right)_a g_{bc}+\left(k_2-k_3\right)_b g_{ac}+\left(k_1-k_3\right)_c g_{ab}\right]$ \\    
 &\\
 \hline\hline
 &\\
 $f\overline{f}B_\mu$ &$
\frac{ig_1}{2}\gamma^\mu\frac{1}{2}\begin{cases}
\left(Y_L^{\ell}+Y_R^{\ell}\right)-\gamma_5\left(Y_L^{\ell}-Y_R^{\ell}\right) & \\
\left(Y_L^Q+Y_R^Q\right)-\gamma_5\left(Y_L^Q-Y_R^Q\right)   & 
\end{cases}$ \\
&\\
$h\Phi\Phi$ & $2i\lambda_{H\Phi}v_h$ \\ 
$hZZ$ & $\frac{iv_h}{4}\left(g_1^2+g_2^2\right)$ \\
&\\
$f\overline{f}Z_\mu$ & $
-i\frac{g_2}{c_w}\gamma_\mu\frac{1}{2}\begin{cases}
\left(c_v^{\ell}-c_a^{\ell}\gamma_5\right)& \\
\left(c_v^Q-c_a^Q\gamma_5\right)& 
\end{cases}$  \\
&\\
$f\overline{f}\gamma$ & $-ieQ_f\gamma_\mu$ \\
&\\
\hline
\hline 
Propagator & $R_\xi$ gauge Feynman rules \\ [0.5ex] 
\hline\hline
&\\
\begin{minipage}{.2\textwidth}
      \includegraphics[scale=0.4]{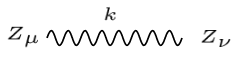}
    \end{minipage}&$\frac{i}{k^2-m_Z^2}\Biggl[-g_{\mu\nu}+\left(1-\xi\right)\frac{k_\mu k_\nu}{k^2-\xi m_Z^2}\Biggr]$\\
    &\\
\begin{minipage}{.2\textwidth}
      \includegraphics[scale=0.4]{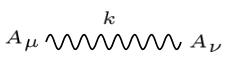}
    \end{minipage}&$\frac{i}{k^2}\Biggl[-g_{\mu\nu}+\left(1-\xi\right)\frac{k_\mu k_\nu}{k^2}\Biggr]$\\
&\\
\hline
\hline
\end{tabular}
\end{center}
\caption {Relevant DM-SM interaction vertices and vertex factors along with SM propagators. All momenta are assumed flowing towards the vertices. Here $\ell$ stands for 
SM leptons and $Q$ stands for the SM quark generations. $Y_{L,R}^{\ell(Q)}$ is the hypercharge for left and right-chiral leptons (quarks): $Y_L^\nu=-1,Y_L^e=-1,Y_R^e=-2;Y_L^Q=1/3,Y_R^u=4/3,Y_R^d=-2/3$.
}\vspace{0.1cm} 
\label{tab:vert-fac}
\end{table}\FloatBarrier

%%%%%%%%%%%%%%%%%%%%%%%%%%%
\section{Reaction densities}
\label{sec:reac-den}
%%%%%%%%%%%%%%%%%%%%%%%%%%%
For a $2\to2$ annihilation channel the reaction density is defined as:
\bea\begin{split}
\gamma\left(a,b\to1,2\right)&=\int\prod_{i=1}^4 d\Pi_i \left(2\pi\right)^4 \delta^{(4)}\biggl(p_a+p_b-p_1-p_2\biggr)f_a^{eq}f_b^{eq}\left|\mathcal{M}_{a,b\to1,2}\right|^2\\&=\frac{T}{32\pi^4}g_a g_b \int_{s_{min}}^\infty ds~\frac{\biggl[\bigl(s-m_a^2-m_b^2\bigr)^2-4m_a^2 m_b^2\biggr]}{\sqrt{s}}\sigma\left(s\right)_{a,b\to1,2}K_1\left(\frac{\sqrt{s}}{T}\right)\label{eq:gam-ann},    
    \end{split}
\eea
where $a,b(1,2)$ are the incoming (outgoing) states and $g_{a,b}$ are corresponding degrees of freedom. Here $f_i^{eq}\approx\exp^{-E_i/T}$ is the Maxwell-Boltzmann distribution. The Lorentz invarint 2-body phase space is denoted by: $d\Pi_i=\frac{d^3p_i}{\left(2\pi\right)^3 2E_i}$. The amplitude squared (summed over final and averaged over initial states) is denoted by $\left|\mathcal{M}_{a,b\to1,2}\right|^2$ for a particular $2\to2$ scattering process. The lower limit of the integration over $s$ is $s_{min}=\text{max}\biggl[\left(m_a+m_b\right)^2,\left(m_1+m_2\right)^2\biggr]$.

For a $1\to2$ decay process the reaction density is given by:
\bea\begin{split}
\gamma\left(a\to1,2\right)&=\int\sum_{i=1}^3 d\Pi_i\left(2\pi\right)^4 \delta^{(4)}\biggl(p_a-p_1-p_2\biggr)f_a^{eq}\left|\mathcal{M}_{a\to1,2}\right|^2\\&=\frac{g_a}{2\pi^2}m_a^2 \Gamma_{a\to1,2}TK_1\left(\frac{m_a}{T}\right)
\label{eq:gam-dec}.      
    \end{split}
\eea
%%%%%%%%%%%%%%%%%%%%%%%%%%%%%%%%%%
\section{Expressions for squared amplitudes before EWSB}
\label{sec:app-amp-bewsb}
%%%%%%%%%%%%%%%%%%%%%%%%%%%%%%%%%%
\subsection{t-channel annihilation before EWSB}
%%%%%%%%%%%%%%%%%%%%%%%%%%%%%%%%%%%%%%%%
The spin averaged amplitude squared for $f\Phi\to fX$ process is given by:
\bea\begin{split}
\left|\overline{\mathcal{M}}\right|^2_{f\Phi\to fX} &=\frac{g_1^2N_c}{128s\left(s-m_X^2\right)\left(s-m_\Phi^2\right)\big(\cos\theta-1\big)}\left(\left[Y_L^f+Y_R^f\right]^2+\left[Y_L^f-Y_R^f\right]^2\right)\\&\Biggl[\alpha^2\Bigl(1+\beta^2\Bigr)\Bigg\{4 \cos\theta  \left(s-m_X^2\right) \left(m_\Phi^2-s\right) \left[3s^2+s\left(m_\Phi^2-5m_X^2\right)+m_X^2m_\Phi^2\right]\\&+\cos2 \theta\left(s-m_X^2\right)^2 \left(s-m_\Phi^2\right)^2+s\Bigl(2 m_X^2 m_\Phi^4-22 m_X^4 m_\Phi^2\Bigr)\\&+s^2\Bigl(27 m_X^4-4 m_X^2 m_\Phi^2+3 m_\Phi^4\Bigr)+s^3\Bigl(10 m_\Phi^2-14 m_X^2\Bigr)-5s^4\Bigg\}\\&+\alpha^2\Bigl(1-\beta^2\Bigr)\Bigg\{16 m_X^2 s \sin ^2\left(\frac{\theta }{2}\right) \left(m_X^2-s\right) \left(m_\Phi^2-s\right)\Bigg\}\Biggr],
    \end{split}\label{eq:bewsb-t}
\eea

where $N_c=1(3)$ for the SM leptons (quarks). Also note that all the SM fermions are massless. In the limit $m_\Phi=m_X=0$ this reduces to a relatively simplified form:

\bea
\left|\overline{\mathcal{M}}\right|^2_{f,\Phi\to f,X} =g_1^2 N_c s \alpha^2\left(1+\beta^2\right)\left(\left[Y_L^f+Y_R^f\right]^2+\left[Y_L^f-Y_R^f\right]^2\right)\left(\frac{ 5+12 \cos \theta-\cos2\theta}{128 (1-\cos \theta)}\right).
\eea

Corresponding annihilation cross-section is given by:

\bea\begin{split}
& \sigma\left(s\right)_{f\Phi\to fX}\simeq \frac{6g_1^2 N_c}{25}\alpha^2\left(1+\beta^2\right)\left(\left[Y_L^f+Y_R^f\right]^2+\left[Y_L^f-Y_R^f\right]^2\right).   
    \end{split}
\eea

%%%%%%%%%%%%%%%%%%%%%%%%%%%%%%%%%%%%%%%5
\subsection{s-channel annihilation before EWSB}
%%%%%%%%%%%%%%%%%%%%%%%%%%%%%%%%%%%%%%%%
The spin averaged amplitude squared for $ff\to X\Phi$ process is given by:
\bea\begin{split}
\left|\overline{\mathcal{M}}\right|^2_{f,f\to \Phi,X}&=\frac{g_1^2 N_c}{256 s^3}\left(\left[Y_L^f+Y_R^f\right]^2+\left[Y_L^f-Y_R^f\right]^2\right)\\&\Biggl[\alpha^2\Bigl(1+\beta^2\Bigr)\Bigg\{3 m_X^4 m_\Phi^4-2 m_X^4 m_\Phi^2 s+3 m_X^4 s^2+2 m_X^2 m_\Phi^4 s\\&-4 m_\Phi^2\cos\theta \left(m_X^2-s\right) \left(m_X^2+s\right) \left(m_\Phi^2-s\right)+ \cos2 \theta  \left(s-m_X^2\right)^2 \left(s-m_\Phi^2\right)^2\\&+2 m_X^2 s^3+3 m_\Phi^4 s^2-6 m_\Phi^2 s^3+3 s^4\Bigg\}+8\alpha^2\Bigl(1-\beta^2\Bigr)m_X^2s^3\Biggr],
    \end{split}\label{eq:bewsb-s}
\eea
with $N_c=1(3)$ for SM leptons (quarks). In the limit $m_X=m_\Phi=0$ this reduces to:
\bea
\left|\overline{\mathcal{M}}\right|^2_{ff\to \Phi,X}=\frac{g_1^2 N_c}{64} s \alpha ^2\left(1+\beta^2\right) (\cos2\theta+3) \left(\left[Y_L^f+Y_R^f\right]^2+\left[Y_L^f-Y_R^f\right]^2\right),
\eea

Corresponding annihilation cross-section is given by:

\bea\begin{split}
& \sigma\left(s\right)_{ff\to\Phi X}\simeq \frac{g_1^2 N_c}{1000}\alpha ^2\left(1+\beta^2\right)\left(\left[Y_L^f+Y_R^f\right]^2+\left[Y_L^f-Y_R^f\right]^2\right).   
    \end{split}
\eea

% \bg{Basabendu, Could you add cross-sections here?}

%%%%%%%%%%%%%%%%%%%%%%%%%%%%%%%%%%%%%%%5
\subsection{Decay of $\Phi$}
%%%%%%%%%%%%%%%%%%%%%%%%%%%%%%%%%%%%%%%%
\subsubsection{Before EWSB}
%%%%%%%%%%%%%%%%%%%%%%%%%%%%%%%%%%%%%%%%
The amplitude squared for the $\Phi\to X,B$ decay is given by:
\beq
\left|\mathcal{M}\right|_{\text{D}}^2 = \frac{m_\Phi^4 }{2} \alpha ^2 \left(1+\beta ^2\right) \left(1-r^2\right)^2.
\label{eq:decayamp}
\eeq
The resulting decay width can be written as:
\beq
\Gamma_{\Phi\to X,B} = \frac{m_\Phi^3 }{32\pi}\alpha ^2 \left(1+\beta ^2\right) \left(1-r^2\right)^3,
\label{eq:width-bewsb}
\eeq
%%%%%%%%%%%%%%%%%%%%%%%%%%%%%%%%%%%%%%%%
\subsubsection{After EWSB}
%%%%%%%%%%%%%%%%%%%%%%%%%%%%%%%%%%%%%%%%
After EWSB $\Phi$ decays to photon and $Z$ final states resulting: 
\bea
\Gamma_{\text{total}} = \Gamma_{\Phi\to X,Z}+\Gamma_{\Phi\to X,\gamma}.
\label{eq:width-aewsb}
\eea
The squared amplitude for decay to photon and massive $Z$-boson final state takes the form:
\bea\begin{split}
\left|\mathcal{M}\right|^2_{\text{total}}&=\underbrace{\frac{m_\Phi^4 }{2} \alpha ^2 \left(1+\beta ^2\right) \left(1-r^2\right)^2 c_w^2}_\text{due to photon}\\&+\underbrace{\frac{1}{2} \alpha ^2 s_w^2 \Bigl[ m_\Phi^4\left(1+\beta^2\right) \left(1-r^2-y^2\right)^2-4 \beta ^2 m_X^2 m_Z^2+2 m_X^2 m_Z^2\Bigr]}_\text{due to massive $Z$-boson}
% \\&=\frac{m_\Phi^4 }{2} \alpha ^2 \left(1+\beta ^2\right)\biggl[\left(1-r^2\right)^2-2y^2\left(1-r^2\right) s_w^2+y^4 s_w^2\biggr].     
    \end{split}\label{eq:amp-decay-aewsb}
\eea
Thus, the total decay width after EWSB can be expressed as:
\bea\begin{split}
\Gamma_{\text{total}} &=\frac{\alpha^2m_\Phi^3\left(1+\beta^2\right)}{32\pi}\biggl[c_w^2\left(1-r^2\right)^3+s_w^2 \left(r^4-2 r^2+\left(y^2-1\right)^2\right)\\& \sqrt{1- \left(r-y\right)^2} \sqrt{1- \left(r+y\right)^2}\biggr],    
    \end{split}
    \label{eq:width-aewsb2}
\eea
where $0<r=m_X/m_\Phi\leq1$ and $0<y=m_Z/m_\Phi\leq1$. 

%%%%%%%%%%%%%%%%%%%%%%%%%%%%%%%%%%%%%%%%%%%%%%%%%%%%
\bibliographystyle{JHEP}
\bibliography{Bibliography}

%%%%%%%%%%%%%%%%%%%%%%%%%%%%%%%%%%%%%%%%%%%%%%%%%%%
\end{document}